\documentclass[english]{article}
\usepackage[T1]{fontenc}
\usepackage[latin9]{inputenc}
\usepackage{units}
\usepackage{amsthm}
\usepackage{amsmath}
\usepackage{amssymb}
\usepackage{graphicx}
\usepackage{esint}

\makeatletter

\providecommand{\tabularnewline}{\\}
\newcommand{\lyxdot}{.}

 \theoremstyle{definition}
 \newtheorem*{defn*}{\protect\definitionname}
  \theoremstyle{plain}
  \newtheorem*{thm*}{\protect\theoremname}

\@ifundefined{date}{}{\date{}}
\usepackage{babel}
\usepackage{url}
\usepackage[numbers,sort&compress]{natbib}
\DeclareMathOperator{\tr}{Tr}
\newcommand{\ba}{\bar{a}}
\newcommand{\bb}{\bar{b}}
\newcommand{\normord}[1]{{:}\!\mathrel{#1}\!{:}}
\newcommand{\ket}[1]{\left|#1\right\rangle}
\newcommand{\bra}[1]{\left\langle #1\right|}
\newcommand{\ketp}[1]{\left|#1\right\} }
\newcommand{\brap}[1]{\left\{ #1\right|}
\newcommand{\dprod}[2]{\left\langle #1|#2\right\rangle}
\newcommand{\dprodp}[2]{\left\{ #1|#2\right\} }
\newcommand{\mat}[1]{\mathbf{#1}}

\makeatother

\usepackage{babel}
  \providecommand{\definitionname}{Definition}
  \providecommand{\theoremname}{Theorem}

\begin{document}

\title{Numerical study of the simplest string bit model}

\author{Gaoli Chen%
\thanks{Email: gchen@ufl.edu%
} and Songge Sun%
\thanks{Email: uranussg@phys.ufl.edu%
}}

\maketitle
\begin{center}
\emph{Institute for Fundamental Theory, }
\par\end{center}

\begin{center}
\emph{Department of Physics, University of Florida, Gainesville, Florida
32611, USA}
\par\end{center}
\begin{abstract}
String bit models provide a possible method to formulate a string
as a discrete chain of pointlike string bits. When the bit number
$M$ is large, a chain behaves as a continuous string. We study the
simplest case that has only one bosonic bit and one fermionic bit.
The creation and annihilation operators are adjoint representations
of the $U\left(N\right)$ color group. We show that the supersymmetry
reduces the parameter number of a Hamiltonian from 7 to 3 and, at
$N=\infty$, ensures a continuous energy spectrum, which implies the
emergence of one spatial dimension. The Hamiltonian $H_{0}$ is constructed
so that in the large $N$ limit it produces a world sheet spectrum
with one Grassmann world sheet field. We concentrate on numerical
study of the model in finite $N$. For the Hamiltonian $H_{0}$, we
find that the would-be ground energy states disappear at $N=\left(M-1\right)/2$
for odd $M\leq11$. Such a simple pattern is spoiled if $H$ has an
additional term $\xi\Delta H$ which does not affect the result of
$N=\infty$. The disappearance point moves to higher (lower) $N$
when $\xi$ increases (decreases). Particularly, the $\pm\left(H_{0}-\Delta H\right)$
cases suggest a possibility that the ground state could survive at
large $M$ and $M\gg N$. Our study reveals that the model has stringy
behavior: when $N$ is fixed and large enough, the ground energy decreases
linearly with respect to $M$, and the excitation energy is roughly
of order $M^{-1}$. We also verify that a stable system of Hamiltonian
$\pm H_{0}+\xi\Delta H$ requires $\xi\geq\mp1$.
\end{abstract}
\newpage{}

\section{Introduction}

The idea of string bits, proposed over two decades ago \cite{Thorn:1991fv},
is one approach to formulate string theory. In this formulation, strings
in $D$-dimensional spacetime are chainlike objects comprised of pointlike
entities, string bits, moving in space of $d=D-2$ dimensions. The
dynamics of the string bits is chosen to retain the Galilei symmetry
described by the group $\mathrm{Galilei}\left(d,1\right)$. While
one spatial coordinate is missing and the Lorentz invariance is not
built in $\emph{a priori}$, both of them are regained in the critical
dimension when the number of string bits is large enough. Thereby,
string theory emerges. Since the physics in $\left(d+1\right)$-dimensional
space is described by physics in $d$-dimensional space, the string
bit models provide an implementation of 't Hooft's holography hypothesis
\cite{'tHooft:1990eb,'tHooft:1987,'tHooft:1993gx}.

Such an idea is motivated by the discretization of a continuous string.
Consider a string in lightcone coordinates \cite{Mandelstam:1973jk,Mandelstam:1974hk}, 

\[
x^{\pm}=\frac{x^{0}\pm x^{1}}{\sqrt{2}},\quad\mathbf{x}=\left(x^{2},\cdots,x^{d+1}\right),
\]
 where $\mathbf{x}$ is the transverse coordinates, the Hamiltonian
of the string reads \cite{Goddard:1972ky,Goddard:1973qh}
\begin{equation}
P^{-}=\frac{1}{2}\int_{0}^{P^{+}}d\sigma\,[\mathbf{p}^{2}+T_{0}^{2}\mathbf{x}^{\prime2}],\label{eq:Ham-contious-string}
\end{equation}
where $P^{\pm}$ are the momenta conjugate to $x^{\mp}$ coordinates.
In analogy to (\ref{eq:Ham-contious-string}), a harmonic chain of
$M$ string bits, each of which has mass $m$, is described by the
Hamiltonian
\begin{equation}
H=\frac{1}{2m}\sum_{k=1}^{M}\left[\mathbf{p}_{n}^{2}+T_{0}^{2}\left(\mathbf{x}_{n+1}-\mathbf{x}_{n}\right)^{2}\right].\label{eq:Ham-discrete-chain}
\end{equation}
Under the Galilei transformation $x^{k}\to x^{k}+V^{k}x^{+}$, the
timelike coordinate $x^{+}$ and the mass of each string bit are invariant.
Consequently, $P^{+}=Mm$ can be considered as the Newtonian mass
of the bitchain. For $M\to\infty$, $P^{+}$ behaves like a continuous
variable of which the conjugate can be interpreted as the missing
coordinate $x^{-}$. If the bound states for a many-bit system are
closed linear chains and the excitation energies scale as $1/M$ for
large $M$, Lorentz invariance is regained and leads to a Poincaré
invariant dispersion relation $P^{-}=\left(\mathbf{P}^{2}+\mu^{2}\right)/(2P^{+})$.
It is noteworthy that such bound states can be achieved in the context
of the 't Hooft large $N$ limit \cite{'tHooft:1973jz,Thorn:1979gu}.

However, the Hamiltonian (\ref{eq:Ham-discrete-chain}) for a bosonic
closed string bit chain leads to inevitable instability. The ground
state energy of such a system in the limit $M\to\infty$ is given
by 
\[
E_{G}=\frac{2dT_{0}M}{m\pi}-\frac{\pi dT_{0}}{6Mm}+{\cal O}\left(M^{-3}\right).
\]
The first term can be dropped as the bit number is conserved in string
interaction\cite{Giles:1977mpa}. Because of the negative $\mathcal{O}\left(M^{-1}\right)$
term, a long closed bit chain tends to split into multiple smaller
chains for a lower energy state. This instability issue can be fixed
by introducing supersymmetry\cite{Gliozzi:1976qd,Ramond:1971gb,Neveu:1971rx,Neveu:1971iw,Thorn:1971jc,Neveu:1971iv}.
In supersymmetry, string bits are multiplets with both bosonic and
fermionic degrees of freedom \cite{Green:1980zg,Bardakci:1970nb}.
It turns out that, for models with $d$ bosonic and $s$ fermionic
world sheet degrees of freedom, the ground energy becomes\cite{Thorn:1996fa}
\[
E_{G}=\frac{(s-d)\pi T_{0}}{6Mm}.
\]
It implies that the system is stable for $s>d$ and unstable for $s<d$.
The supersymmetric case $s=d$ gives rise to exact cancellation between
bosonic and fermionic contributions for all $M$.

To set up the dynamics of the superstring bit model, we employ 't
Hooft's large $N$ limit and follow the standard second-quantized
formalism\cite{Bergman:1995wh}. A general superstring bit annihilation
operator is an $N\times N$ matrix denoted by 
\[
\left(\phi_{\left[a_{1}\cdots a_{n}\right]}\right)_{\alpha}^{\beta}\left(\mathbf{x}\right),\quad n=0,\cdots,s,
\]
 where each $a_{i}$ is a spinor index running over $s$ values and
$\alpha,\beta=1,\cdots,N$ are color indices for the adjoint representation
of the color group $SU\left(N\right)$. $\phi$ is bosonic for even
$n$ and fermionic for odd $n$. The square bracket in the subscript
denotes complete antisymmetric relation among $a_{i}$ indices. For
superstring theory, the Poincaré symmetry demands $s=d=8$.

In Ref. \cite{Sun:2014dga}, Thorn and one of us studied the simplest
case of the model with $d=0,\, s=1$, where there are $N^{2}$ bosonic
annihilation operators $\left(a_{\alpha}^{\beta}\right)$ and $N^{2}$
fermionic annihilation operators $\left(b_{\alpha}^{\beta}\right)$,
with corresponding creation operators defined as $\bar{a}_{\alpha}^{\beta}\equiv\left(a_{\beta}^{\alpha}\right)^{\dagger}$
and $\bar{b}_{\alpha}^{\beta}\equiv\left(b_{\beta}^{\alpha}\right)^{\dagger}$.
These operators satisfy the (anti)commutation relations, 
\begin{equation}
\left[a_{\alpha}^{\beta},\bar{a}_{\gamma}^{\delta}\right]=\delta_{\alpha}^{\delta}\delta_{\gamma}^{\beta},\quad\left\{ b_{\alpha}^{\beta},\bar{b}_{\gamma}^{\delta}\right\} =\delta_{\alpha}^{\delta}\delta_{\gamma}^{\beta},\label{eq:commutation-relation}
\end{equation}
 and all others vanishing. With these creation operators, we can build
trace states as follows. Introduce the vacuum state $\ket{0}$ annihilated
by all the $a_{\alpha}^{\beta}$ and $b_{\alpha}^{\beta}$. We can
act on $\ket{0}$ with a sequence of $\bar{a}$ and $\bar{b}$ to
obtain a nonvacuum state with color indices. Finally, we take the
trace of the creation operators to obtain a color-singlet state. Each
creation operator in the trace state is interpreted as a string bit.
Trace states with an even number of $\bar{b}$ are bosonic states,
while those with an odd number of $\bar{b}$ are fermionic states.
To give a few examples, $\tr\bar{a}^{3}\ket{0}$, $\tr\bar{a}^{2}\tr\bar{a}\ket{0}$,
and $\tr\bar{a}\bar{b}^{2}\ket{0}$ are 3-bit bosonic trace states;
$\tr\bar{a}\bar{b}\ket{0}$ and $\tr\bar{a}\tr\bar{b}\ket{0}$ are
2-bit fermionic trace states. Note that, because of the property of
the trace and the anticommutation relation in (\ref{eq:commutation-relation}),
some of such expressions are not a valid trace state, for example,
$\tr\bar{b}\bar{b}\ket{0}=-\tr\bar{b}\bar{b}\ket{0}=0$. Clearly,
the number of trace states increases exponentially as $M$ increases.
In Appendix \ref{app:Counting-Trace-States}, we provide a formula
to count the single trace states and an algorithm to calculate the
number of trace states, including both single and multiple trace states.
In Appendix \ref{app:Trace-States-List}, we list all the different
bosonic trace states from 1 bit to 7 bits. 

The Hamiltonian of the toy model in Ref. \cite{Sun:2014dga} is chosen
to be a linear combination of single trace operators
\begin{equation}
\tr\bar{a}^{2}a^{2},\quad\tr\bar{b}^{2}b^{2},\quad\tr\bar{b}^{2}a^{2},\quad\tr\bar{a}^{2}b^{2},\quad\tr\bar{a}\bar{b}ba,\quad\tr\bar{a}\bar{b}ab,\quad\tr\bar{b}\bar{a}ba,\quad\tr\bar{b}\bar{a}ab,\label{eq:trace-operators}
\end{equation}
 with coefficients scaling as $1/N$. Such a choice ensures the action
of the Hamiltonian to the trace states survives at the large $N$
limit. It then studied a special form of such a Hamiltonian 
\begin{equation}
H_{0}=\frac{2}{N}\tr\left[\left(\bar{a}^{2}-i\bar{b}^{2}\right)a^{2}-\left(\bar{b}^{2}-i\bar{a}^{2}\right)b^{2}+\left(\bar{a}\bar{b}+\bar{b}\bar{a}\right)ba+\left(\bar{a}\bar{b}-\bar{b}\bar{a}\right)ab\right],\label{eq:H_0}
\end{equation}
 which produces the Green-Schwarz Hamiltonian\cite{Green:1980zg,Green:1983hw}
at $N=\infty$. By the variational method, it shows that the ground
states of the Hamiltonian only survive at $N>\left(M-1\right)/2$.
Then a numerical study of the Hamiltonian at $M=3$ is performed.

In this paper, we will investigate more general forms of the supersymmetric
Hamiltonian and their energy spectrum at the large $N$ limit. We
will perform a numerical study of the Hamiltonian $H_{0}$ for $M\leq11$.
We will plot the energy levels as a function of $N$ at fixed values
of $M$ and show numerically that the would-be ground state disappears
at $N\leq\left(M-1\right)/2$ for odd $M\leq11$. Such a pattern is
spoiled when we add to $H_{0}$ an additional $\Delta H$ term, which
does not affect the large $N$ limit. For the Hamiltonians $\pm\left(H_{0}-\Delta H\right)$,
the disappearance of the ground state occurs at $N<\left(M-1\right)/2$,
which might suggest that the ground states can survive when $M$ is
large and $N$ is much smaller than $M$. We will also plot the ground
energy and excitation energy as a function of $M$ at fixed $N$ to
check whether the system manifests stringy behavior. For stringy behavior,
the ground energy should be a linear function of $M$ with negative
slope and the excitation energy proportional to $M^{-1}$ with positive
coefficient. It turns out that, for $N$ large enough, the ground
energies do drop almost linearly. For excitation energies, although
there are not enough data for an unquestioned pattern, it still shows
tendencies to go roughly as $M^{-1}$ when $N$ is large. 

The rest of this paper is organized as follows. In Sec. \ref{sec:Supersymmetry},
we discuss the general constraint on a supersymmetric Hamiltonian.
In Sec. \ref{sec:Large-N}, we investigate the energy spectrum of
the system in the large $N$ limit. In Sec. \ref{sec:Finite-N}, we
compute the energy spectrum at finite $N$ numerically and present
the plots from the numerical study. The Hamiltonian $H_{0}$ and its
variations will be studied in the section. The main text is closed
with a section of a summary and conclusion. Finally, we include seven
appendices covering technical details.

\section{\label{sec:Supersymmetry}Supersymmetric Hamiltonian}

In the toy model with $d=0,\, s=1$, while the spacetime supersymmetry
is explicitly broken, there still exists a form of supersymmetry between
bosonic and fermionic trace states. As the mathematical proof in Appendix
\ref{app:Counting-Trace-States} shows, the numbers of bosonic and
fermionic trace states are equal at any value of $M$. This is not
a coincidence. The physical interpretation is that the bit number
operator $M=\tr\left(\bar{a}a+\bar{b}b\right)$ commutes with the
supersymmetry operator 
\begin{equation}
Q=\exp\left(\frac{i\pi}{4}\right)\tr\bar{a}b+\exp\left(-\frac{i\pi}{4}\right)\tr\bar{b}a.\label{eq:Susy-susy-operator}
\end{equation}

Also we notice that $M=Q^{2}$. A Hamiltonian $H$ is supersymmetric
if $\left[H,Q\right]=0$. As we will show in the next section, a nice
feature of the supersymmetric Hamiltonian is that its excitation energy
vanishes at large $M$. 

Now, let us investigate possible forms of a supersymmetric Hamiltonian
and generalizations of $H_{0}$. The general form of a Hermitian Hamiltonian
built out of the trace operators in (\ref{eq:trace-operators}) reads
\begin{eqnarray}
H & = & \frac{1}{N}\Big[c_{1}\tr\bar{a}^{2}a^{2}+c_{2}\tr\bar{b}^{2}b^{2}+iz_{1}\tr\bar{a}^{2}b^{2}-iz_{1}^{*}\tr\bar{b}^{2}a^{2}\nonumber \\
 &  & +c_{3}\tr\bar{a}\bar{b}ba+c_{4}\tr\bar{b}\bar{a}ab+z_{2}\tr\bar{a}\bar{b}ab+z_{2}^{*}\tr\bar{b}\bar{a}ba\Big],\label{eq:Susy-general-H}
\end{eqnarray}
 where $c_{i}$ are real and $z_{i}$ are complex. Imposing the constraint
$\left[H,Q\right]=0$ yields%
\footnote{Appendix \ref{App:Calculation-of-HQ} details the calculation of $\left[H,Q\right]$.%
}
\begin{equation}
\begin{cases}
\Im z_{1} & =\Im z_{2}\\
c_{1}-c_{2} & =2\Re z_{2}\\
c_{3}-c_{4} & =2\Re z_{1}\\
c_{1}+c_{2} & =c_{3}+c_{4}
\end{cases},\label{eq:Susy-constraint}
\end{equation}
 which implies that a supersymmetric Hamiltonian can be written as
\begin{eqnarray}
H & = & H_{0}+\frac{2\xi}{N}\tr\left(\bar{a}\bar{b}ba+\bar{b}\bar{a}ab+\bar{a}^{2}a^{2}+\bar{b}^{2}b^{2}\right)\nonumber \\
 &  & +\frac{2\eta}{N}\tr\left(\bar{b}^{2}a^{2}+\bar{a}^{2}b^{2}+i\bar{a}\bar{b}ab-i\bar{b}\bar{a}ba\right)\nonumber \\
 &  & +\frac{2\zeta}{N}\tr\left(i\bar{b}^{2}a^{2}-i\bar{a}^{2}b^{2}-\bar{a}\bar{b}ba+\bar{b}\bar{a}ab\right),\label{eq:general-susy-H}
\end{eqnarray}
 where $\xi$, $\eta$, $\zeta$ are real parameters. Note that each
term in (\ref{eq:general-susy-H}) is Hermitian and supersymmetric. 

The Hamiltonian $H_{0}$ is the special case of (\ref{eq:general-susy-H})
when $\xi=\eta=\zeta=0$. But we can also obtain a generalization
of $H_{0}$ by keeping a twisted $\xi$ term. As noted in Ref. \cite{Sun:2014dga},
we are free to add the terms 
\begin{equation}
\Delta H^{\prime}=\frac{1}{N}\mathrm{Tr}\left[2\xi_{1}\bar{a}\bar{b}ba+2\xi_{2}\bar{b}\bar{a}ab+\left(\xi_{1}+\xi_{2}\right)\left(\bar{a}^{2}a^{2}+\bar{b}^{2}b^{2}-\tilde{M}\right)\right],\label{eq:Delta-H-Prime}
\end{equation}
 to a Hamiltonian without affecting the large $N$ limit. Here, $\tilde{M}$
is a supersymmetric term given by%
\footnote{Reference \cite{Sun:2014dga} uses the bit operator $M=\tr\left(\bar{a}a+\bar{b}b\right)$
instead of $\tilde{M}$ in $\Delta H^{\prime}$. Our calculation shows
that, in order for $\Delta H^{\prime}$ to vanish in the large $N$
limit, $M$ must be replaced by $\tilde{M}$.%
} 
\[
\tilde{M}=\tr\left(\bar{a}a+\bar{b}b\right)-\frac{1}{N}\left(\tr\bar{a}\tr a+\tr\bar{b}\tr b\right).
\]
 Setting $\xi_{1}-1=\xi_{2}+1=\xi$, we obtain a supersymmetric $\Delta H^{\prime}$
term which equals the $\xi$ term in (\ref{eq:general-susy-H}) minus
a $\tilde{M}$ term. Therefore, $H_{0}$ can be generalized to 
\begin{equation}
H=H_{0}+\xi\Delta H,\label{eq:H-general-1}
\end{equation}
 where 
\[
\Delta H=\frac{2}{N}\tr\left[\bar{a}\bar{b}ba+\bar{b}\bar{a}ab+\bar{a}^{2}a^{2}+\bar{b}^{2}b^{2}-\tilde{M}\right].
\]

In (\ref{eq:H-general-1}), $H_{0}$ makes a $\mathcal{O}\left(1\right)$
contribution, while $\Delta H$ makes only a $\mathcal{O}\left(\frac{1}{N}\right)$
contribution. The values of $\xi$ are constrained by the requirement
that a well-defined Hamiltonian should be stable for large $M$. The
$\tr\bar{a}^{2}a^{2}$ term can produce about $M^{2}$ terms by attacking
to the trace state $\tr\bar{a}^{M}\ket{0}$. This would cause a dangerous
instability if the coefficient of $\tr\bar{a}^{2}a^{2}$ is negative.
To maintain a positive $\tr\bar{a}^{2}a^{2}$ term, we must choose
$\xi\geq-1$. Therefore, we obtain a form of the well-defined Hamiltonian,
\begin{equation}
H=H_{0}+\xi\Delta H,\quad\xi\geq-1.\label{eq:Hamiltonian-3-1}
\end{equation}

In addition to (\ref{eq:Hamiltonian-3-1}), there exists another form
of the supersymmetric Hamiltonian. As suggested in Ref. \cite{Sun:2014dga},
we can replace $H_{0}$ with $-H_{0}$ and obtain 
\begin{equation}
H=-H_{0}+\xi\Delta H,\quad\xi\geq1,\label{eq:Hamiltonian-4-1}
\end{equation}
 where the constraint $\xi\geq1$ comes from the stability condition. 

One might wonder if there exist other supersymmetric operators that
are capable of stabilizing $-H_{0}$ and make only $\mathcal{O}\left(\frac{1}{N}\right)$
contributions. As suggested by Ref. \cite{Thorn:1991fv}, one possibility
is to use the $\tr\bar{a}a\bar{a}a$ operator, which also produces
about $M^{2}$ terms when acting on $\tr\bar{a}^{M}\ket{0}$. A combination
like 
\[
H^{\prime}=\frac{2}{N}\tr\left(\bar{a}a\bar{a}a+\bar{b}b\bar{a}a-\bar{a}b\bar{b}a\right)
\]
 meets such a requirement. However, as Appendix \ref{app:HPrime-Equals-DeltaH}
shows, $H^{\prime}$ equals $\Delta H$ for all trace states, i.e.,
\[
\left(H^{\prime}-\Delta H\right)\ket{\text{Any trace state}}=0.
\]
While we are not sure if there exist other variations of $H_{0}$,
for the time being, we leave the question for further research and
only study Hamiltonians as (\ref{eq:Hamiltonian-3-1}) and (\ref{eq:Hamiltonian-4-1})
in this paper.

\section{\label{sec:Large-N}Energy spectrum in large $N$ limit}

In this section, we will study the energy spectrum of our toy string
bit model in the large $N$ limit by both analytic and numerical methods.
We first show that the supersymmetry guarantees the excitation energy
to be vanishing at large $M$ and then present the energy spectrum
graphically.

\subsection{General $H$}

For convenience, we introduce a super creation operator using a Grassmann
anticommuting number $\theta$, 
\[
\psi(\theta)=\bar{a}+\bar{b}\theta,\quad\bar{b}=-\frac{d}{d\theta}\psi,\quad\bar{a}=\left(1-\theta\frac{d}{d\theta}\right)\psi.
\]
 We then choose 
\begin{equation}
\ket{\theta_{1}\theta_{2}\cdots\theta_{M}}=\tr\left[\psi\left(\theta_{1}\right)\psi\left(\theta_{2}\right)\cdots\psi\left(\theta_{M}\right)\right]\ket{0}\label{eq:LargeN-single-trace-basis}
\end{equation}
 to be a basis of $M$-bit single trace states. A general single trace
energy eigenstate at large $N$ reads 
\begin{equation}
\ket{E}=\int d^{M}\theta\,\Psi\left(\theta_{1}\cdots\theta_{M}\right)\ket{\theta_{1}\theta_{2}\cdots\theta_{M}},\label{eq:LargeN-energy-eigenstate}
\end{equation}
 where $\Psi\left(\theta_{1}\cdots\theta_{M}\right)$ is the wave
function in terms of $\theta_{i}$. Under the cyclic transformation,
$\theta_{i}\to\theta_{i+1}$, $\ket{\theta_{1}\cdots\theta_{M}}$
is invariant and the Jacobi $d^{M}\theta$ obtain a factor of $\left(-1\right)^{M-1}$.
It follows that we can constrain the wave function by a cyclic symmetry,
\begin{equation}
\Psi\left(\theta_{1}\theta_{2}\cdots\theta_{M}\right)=\left(-1\right)^{M-1}\Psi\left(\theta_{M}\theta_{1}\cdots\theta_{M-1}\right).\label{eq:LargeN-cyclic-symmetry}
\end{equation}

In the basis (\ref{eq:LargeN-single-trace-basis}), the leading term
of trace operators in (\ref{eq:trace-operators}) can be expressed
in terms of $\theta_{i}$ and $\frac{d}{d\theta_{i}}$, as shown in
Eqs. (9) to (16) of Ref. \cite{Sun:2014dga}, by which we rewrite
(\ref{eq:Susy-general-H}) in the large $N$ limit as
\[
H\ket{\theta_{1}\cdots\theta_{M}}=\hat{h}\ket{\theta_{1}\cdots\theta_{M}}+\mathcal{O}\left(\frac{1}{N}\right),
\]

\begin{eqnarray}
\hat{h} & = & \sum_{k=1}^{M}\Big[iz_{1}\theta_{k+1}\theta_{k}-iz_{1}^{\dagger}\frac{d}{d\theta_{k}}\frac{d}{d\theta_{k+1}}+z_{2}\theta_{k}\frac{d}{d\theta_{k+1}}\nonumber \\
 &  & +z_{2}^{\dagger}\theta_{k+1}\frac{d}{d\theta_{k}}+\left(-2c_{1}+c_{3}+c_{4}\right)\theta_{k}\frac{d}{d\theta_{k}}\nonumber \\
 &  & +\left(c_{1}+c_{2}-c_{3}-c_{4}\right)\theta_{k}\frac{d}{d\theta_{k}}\theta_{k+1}\frac{d}{d\theta_{k+1}}\Big]+c_{1}M.\label{eq:LargeN-H-in-theta}
\end{eqnarray}
Performing integration by parts as 
\[
\int d^{M}\theta\,\Psi\left(\theta_{1}\cdots\theta_{M}\right)\hat{h}\ket{\theta_{1}\theta_{2}\cdots\theta_{M}}=\int d^{M}\theta\, h\Psi\left(\theta_{1}\cdots\theta_{M}\right)\ket{\theta_{1}\theta_{2}\cdots\theta_{M}},
\]
 we obtain
\begin{eqnarray*}
h & = & \sum_{k=1}^{M}\Big[iz_{1}\theta_{k+1}\theta_{k}-iz_{1}^{\dagger}\frac{d}{d\theta_{k}}\frac{d}{d\theta_{k+1}}-z_{2}\theta_{k}\frac{d}{d\theta_{k+1}}\\
 &  & -z_{2}^{\dagger}\theta_{k+1}\frac{d}{d\theta_{k}}+\left(2c_{1}-c_{3}-c_{4}\right)\theta_{k}\frac{d}{d\theta_{k}}\Big]\\
 &  & +\left(c_{3}+c_{4}-c_{1}\right)M,
\end{eqnarray*}
 where for simplicity we drop the quartic term, which vanishes automatically
under the supersymmetry constraint (\ref{eq:Susy-constraint}). We
then introduce the Fourier transforms

\begin{eqnarray*}
\alpha_{n} & = & \frac{1}{\sqrt{M}}\sum_{k=1}^{M}\theta_{k}e^{2\pi ikn/M},\quad\beta_{n}=\frac{1}{\sqrt{M}}\sum_{k=1}^{M}\frac{d}{d\theta_{k}}e^{2\pi ikn/M},\quad n=0,\dots M-1,\\
\theta_{k} & = & \frac{1}{\sqrt{M}}\sum_{n=0}^{M-1}\alpha_{n}e^{-2\pi ikn/M},\quad\frac{d}{d\theta_{k}}=\frac{1}{\sqrt{M}}\sum_{n=0}^{M-1}\beta_{n}e^{-2\pi ikn/M},\quad k=1,\dots M,
\end{eqnarray*}
satisfying 
\[
\left\{ \alpha_{n},\beta_{m}\right\} =\delta_{m+n,M}+\delta_{m,0}\delta_{n,0}.
\]
A little algebra yields 
\begin{eqnarray*}
h & = & \sum_{n=1}^{M-1}\Big[\left(z_{1}\alpha_{n}\alpha_{M-n}+z_{1}^{\dagger}\beta_{n}\beta_{M-n}\right)\sin\frac{2n\pi}{M}\\
 &  & +2\left(c-\Re\left(z_{2}e^{2\pi in/M}\right)\right)\alpha_{n}\beta_{M-n}\Big]\\
 &  & +2\left(c-\Re z_{2}\right)\alpha_{0}\beta_{0}+\left(c_{1}-2c\right)M,
\end{eqnarray*}
where we have defined $c=c_{1}-\frac{1}{2}\left(c_{3}+c_{4}\right)$.
Note that we have $c=\Re z_{2}$ under the supersymmetry constraint
(\ref{eq:Susy-constraint}). 

We now find the ladder operators of $h$, which we denote as $L_{k}$.
We use the ansatz $L_{k}=a\alpha_{k}+b\beta_{k}$ and impose the constraint

\begin{equation}
\left[h,L_{k}\right]=\epsilon_{k}L_{k}.\label{eq:LargeN-ladder-constraint}
\end{equation}
 By direct calculation, we have 
\[
\left[h,a\alpha_{k}+ib\beta_{k}\right]=2\left(ad_{k}+bz_{1}\sin\frac{2k\pi}{M}\right)\alpha_{k}+2\left(az_{1}^{\dagger}\sin\frac{2k\pi}{M}-bd_{M-k}\right)\beta_{k},
\]
 where $d_{k}\equiv c-\Re\left(z_{2}e^{2\pi ik/M}\right)$. Constraint
(\ref{eq:LargeN-ladder-constraint}) yields
\begin{equation}
\begin{cases}
2\left(ad_{k}+bz_{1}\sin\frac{2k\pi}{M}\right) & =a\epsilon_{k}\\
2\left(az_{1}^{\dagger}\sin\frac{2k\pi}{M}-bd_{M-k}\right) & =b\epsilon_{k}
\end{cases}\label{eq:LargeN-ladder-equation}
\end{equation}

Let us first consider the $k=0$ case. If $d_{0}\equiv c-\Re z_{2}\neq0$,
there are two solutions:
\begin{eqnarray*}
\text{when }a\neq0,\, b & = & 0,\quad\epsilon_{0}=2\left(\Re z_{2}-c\right);\\
\text{when }a=0,\, b & \neq & 0,\quad\epsilon_{0}=-2\left(\Re z_{2}-c\right).
\end{eqnarray*}
 The corresponding ladder operators are $\alpha_{0}$ and $\beta_{0}$,
respectively. If $c-\Re z_{2}=0$, i.e., the supersymmetry case, then
$a,b$ can be any value, and $\epsilon_{0}=0$ , which implies there
is no ladder operator for $k=0$. In the supersymmetry case, the linear
combination $\exp\left(\frac{i\pi}{4}\right)\alpha_{0}+\exp\left(-\frac{i\pi}{4}\right)\beta_{0}$
is just the supersymmetry operator (\ref{eq:Susy-susy-operator}). 

For $k\neq0$, we solve for $\epsilon_{k}$,
\[
\epsilon_{k}^{\pm}=2\Im z_{2}\sin\frac{2k\pi}{M}\pm2\sqrt{\left(c-\Re z_{2}\cos\frac{2k\pi}{M}\right)^{2}+\left|z_{1}\right|^{2}\sin^{2}\frac{2k\pi}{M}}.
\]
 In general, $\epsilon_{k}$ is finite at large $M$, and the energy
levels are discrete. But under the supersymmetry constraint (\ref{eq:Susy-constraint}),
\begin{equation}
\epsilon_{k}^{\pm}=4\left(-\Im z_{1}\cos\frac{\pi k}{M}\pm\sqrt{\left(\Re z_{2}\right)^{2}\sin^{2}\frac{k\pi}{M}+|z_{1}|^{2}\cos^{2}\frac{k\pi}{M}}\right)\sin\frac{k\pi}{M},\label{eq:LargeN-En-susy}
\end{equation}
 which vanishes for finite $k$ at large $M$. Therefore, supersymmetry
ensures a continuous energy spectrum and stringy behavior.

\subsection{$H=H_{0}$}

In the case of $H=H_{0}$, we have $c_{1}=-c_{2}=c_{3}=-c_{4}=c=2$,
$z_{1}=z_{2}=2$, and 
\[
\epsilon_{k}^{\pm}=\pm8\sin\frac{k\pi}{M},\quad r_{k}^{\pm}\equiv\frac{a}{b}=\tan\frac{k\pi}{M}\pm\sec\frac{k\pi}{M},\quad k=1,\cdots,M-1.
\]
As $r_{M/2}^{+}=\infty$ and $r_{M/2}^{-}=0$, we choose the raising
and lowering operators to be 
\[
L_{k}^{+}=\alpha_{k}+\frac{1}{r_{k}^{+}}\beta_{k},\quad L_{k}^{-}=r_{k}^{-}\alpha_{k}+\beta_{k},\quad k=1,\cdots,M-1.
\]
Now, we can construct the ground function, which is annihilated by
all lowering operators. Observing that 
\begin{eqnarray*}
L_{k}^{-}\left(1+r_{k}^{-}\alpha_{k}\alpha_{M-k}\right) & = & L_{M-k}^{-}\left(1+r_{k}^{-}\alpha_{k}\alpha_{M-k}\right)=0,
\end{eqnarray*}
 and that $\alpha_{0}$ commutes with all $L_{k}^{-}$, we obtain
ground wave functions, 
\[
\Phi_{M}^{b}=\prod_{k=1}^{\left\lfloor M/2\right\rfloor }\left(1+r_{k}^{-}\alpha_{k}\alpha_{M-k}\right),\quad\Phi_{M}^{f}=\alpha_{0}\prod_{k=1}^{\left\lfloor M/2\right\rfloor }\left(1+r_{k}^{-}\alpha_{k}\alpha_{M-k}\right)
\]
 with $\left\lfloor M/2\right\rfloor $ the integral part of $M/2$.
Clearly $\Phi_{M}^{b}$ is bosonic and $\Phi_{M}^{f}$ is fermionic.
A direct calculation shows they have the same eigenvalue 
\begin{equation}
E_{G}=-4\sum_{k=1}^{M-1}\sin\frac{k\pi}{M}=-4\cot\frac{\pi}{2M}.\label{eq:LargeN-ground-energy}
\end{equation}
For each $k<M/2$, we have four different choices to attack the ground
functions, i.e., using 1, $L_{k}^{+}$, $L_{M-k}^{+}$, and $L_{k}^{+}L_{M-k}^{+}$,
which correspond to the energy level increasing by 0, $\epsilon_{k}^{+}$,
$\epsilon_{k}^{+}$, and $2\epsilon_{k}^{+}$. For $k=M/2$, there
are two choices to attack $\Phi_{M}$, by 1 and $L_{M/2}^{+}$, with
energy increments of 0 and $\epsilon_{M/2}^{+}$. Therefore, for each
choice of ground function, the energy levels can be written as 
\begin{eqnarray}
E\left(\left\{ \eta_{k}\right\} \right) & = & E_{G}+8\sum_{k=1}^{\left\lfloor M/2\right\rfloor }\sin\frac{k\pi}{M}+8\sum_{k=1}^{\left\lfloor M/2\right\rfloor }\eta_{k}\sin\frac{k\pi}{M}\nonumber \\
 &  & =8\sum_{k=1}^{\left\lfloor M/2\right\rfloor }\eta_{k}\sin\frac{k\pi}{M}+\begin{cases}
0 & \;\text{for odd }M\\
4 & \;\text{for even }M
\end{cases}\label{eq:LargeN-energy-levels}
\end{eqnarray}
\begin{equation}
\eta_{k}=-1,\,0,\,0,\,1,\;\text{for }k<M/2;\;\eta_{M/2}=-1,0.\label{eq:LargeN-eta}
\end{equation}
 Here, we reproduced Eqs. (94) and (95) of Ref. \cite{Sun:2014dga}
with a different approach.

Now, consider the cyclic constraint (\ref{eq:LargeN-cyclic-symmetry}).
The eigenfunctions should be changed by a factor of $\left(-1\right)^{M-1}$
under the transformation $\alpha_{k}\to\exp\left(2ik\pi/M\right)\alpha_{k}$
and $\beta_{k}\to\exp\left(2ik\pi/M\right)\beta_{k}$. Clearly the
ground eigenfunction $\Phi_{M}$ is invariant under the transformation,
and $L_{k}^{+}$ changes as $L_{k}^{+}\to\exp\left(2ik\pi/M\right)L_{k}^{+}$,
from which it follows that $\eta_{k}$ must satisfy
\begin{equation}
\sum_{\eta_{k}=0}^{M/2}k=\begin{cases}
nM, & \quad\text{for odd }M\\
\left(n+\frac{1}{2}\right)M, & \quad\text{for even }M
\end{cases},\quad n=0,1,2,\cdots.\label{eq:LargeN-cyclic-constraint}
\end{equation}
This constraint has several interesting consequences:
\begin{itemize}
\item For odd $M$, the lowest energy state of the $M$-bit system is comprised
of $M$-bit single trace states, which are generated by setting all
$\eta_{k}$ to $-1$, i.e.,
\begin{equation}
E_{\mathrm{min}}=E_{\mathrm{min}}^{\left(1\right)}=-4\cot\frac{\pi}{2M}=-\frac{8M}{\pi}+\frac{2\pi}{3M}+\mathcal{O}\left(M^{-3}\right),\label{eq:LargeN-E-min-odd}
\end{equation}
 where we use the superscript $\left(1\right)$ to denote single trace
states.
\item For even $M$, the lowest energy of single trace states, $E_{\mathrm{min}}^{\left(1\right)}$,
is achieved when $\eta_{M/2}=0$ and all other $\eta_{k}=-1$; while
the lowest energy state of the system is comprised of double trace
states with each trace of $M/2$ bits (if $M/2$ is even, the two
traces are of $M/2-1$ and $M/2+1$ bits). So we have 
\begin{eqnarray*}
E_{\mathrm{min}}^{\left(1\right)} & = & -\frac{8M}{\pi}+\frac{2\pi}{3M}+8+\mathcal{O}\left(M^{-3}\right),\\
E_{\mathrm{min}} & = & E_{\mathrm{min}}^{\left(2\right)}=-\frac{8M}{\pi}+\frac{4\pi}{3M}+\mathcal{O}\left(M^{-2}\right).
\end{eqnarray*}
When $M/2$ is even, the lowest energy states have extra degeneracy,
because the bosonic ground functions can be $\Phi_{M/2-1}^{b}\Phi_{M/2+1}^{b}$
and $\Phi_{M/2-1}^{f}\Phi_{M/2+1}^{f}$. 
\item For large $M$, the excitation energy is very small, and the discrete
energy levels become a continuous energy band. The difference of $E_{\mathrm{min}}^{\left(1\right)}$
between odd and even $M$ is much large than the excitation energy,
which implies only odd-bit chains participate in the low energy physics.
Particularly, it also means a low energy odd-bit chain cannot decay
into two chains.
\end{itemize}
Now, let us consider the first excitation energy of the odd $M$ system.
From the above analysis, there are no double trace states in the low
energy region, so we consider the triple trace states. From (\ref{eq:LargeN-E-min-odd}),
the lowest energy of triple trace states is achieved when each trace
has $M/3$ bits. Hence, we have 
\[
E_{1}=-\frac{8M}{\pi}+\frac{16\pi}{M}+\mathcal{O}\left(M^{-2}\right),
\]
from which it follows that the energy gap between the ground energy
(\ref{eq:LargeN-E-min-odd}) and first excitation energy is $\frac{16\pi}{3M}$.
If $M$ is divisible by 3, the first excitation energy has no extra
degeneracy. If $M=3n\pm1$, it has extra degeneracy: for $M=3n+1$,
the bosonic ground function can be $\Phi_{n-1}^{b}\Phi_{n+1}^{b}\Phi_{n+1}^{b}$
and $\Phi_{n-1}^{f}\Phi_{n+1}^{f}\Phi_{n+1}^{b}$; for $M=3n+1$,
the bosonic ground function can be $\Phi_{n-1}^{b}\Phi_{n-1}^{b}\Phi_{n+1}^{b}$
and $\Phi_{n-1}^{b}\Phi_{n-1}^{f}\Phi_{n+1}^{f}$.

Figure \ref{fig:LargeN-energy-levels} shows the energy spectrum at
$N=\infty$ for $M$ at $11$, 21, 51, and 101. In the plot, energy
states are represented by horizontal lines, with the red color for
single trace states and yellow color for triple trace states. The
vertical coordinate is $M\times\left(E-E_{\mathrm{min}}\right)$,
the product of $M$ with the difference between energy level and the
lowest energy. The threshold for triple trace states is a blue line. 

From the figure, it is clear that the energy gaps go smaller as $M$
increases and the energy levels become continuous at large $M$. The
energy of single trace states tends to distribute near multiples of
$\frac{16\pi}{M}$, and the first excitation energy appears near $\frac{16\pi}{3M}$.
The energy levels of triple trace states are even denser than single
trace states. At $M=101$, they almost filled the gap between consecutive
single trace energy levels. All these behaviors illustrate that the
chains behave as continuous strings at large $M$. 

\begin{center}
\begin{figure}
\begin{centering}
\includegraphics[bb=120bp 250bp 490bp 560bp,clip,width=1\columnwidth]{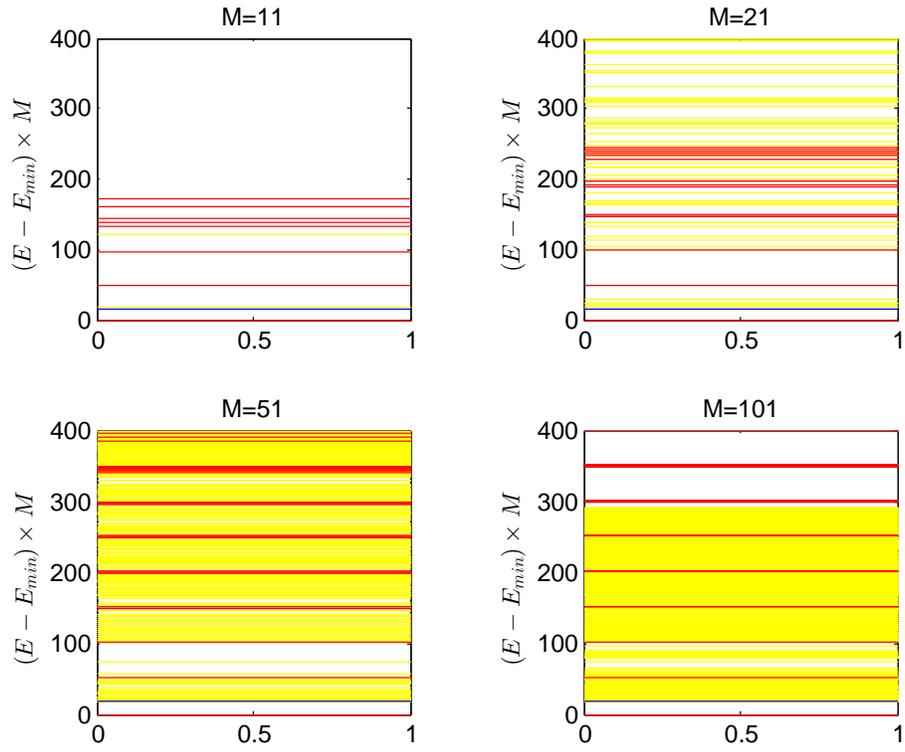}
\par\end{centering}

\centering{}\caption{\label{fig:LargeN-energy-levels}The energy levels of single trace
states (red lines) and triple trace states (yellow lines) at $M=11$,
M=21, M=51, and M=101 and the large $N$ limit. The blue dashed line
is the threshold for multitrace states energy when $M=\infty$.}
\end{figure}

\par\end{center}

\section{\label{sec:Finite-N}Energy spectrum at finite $N$}

In this section, we show numerically how the energy levels change
with respect to $N$ and the bit number $M$. We first introduce the
methods to calculate the energy states of the system. We then analyze
the result of the original Hamiltonian $H=H_{0}$, for which the $M=3$
case has been investigated in Ref. \cite{Sun:2014dga}. Next, we move
to the Hamiltonians of the form $H=H_{0}+\xi\Delta H$ and investigate
how the parameter $\xi$ affects the energy levels. Finally, we explore
the Hamiltonians of the form $H=-H_{0}+\xi\Delta H$. For each case,
we first analyze the change of energy levels with respect to $N$
when $M$ is fixed and then with respect to $M$ when $N$ is fixed.

\subsection{\label{subsec:Method} $\mathcal{H}$ matrices}

We have two methods to calculate the energy states of the system%
\footnote{In this subsection, we just state the properties of these two methods.
The relevant mathematical proofs are provided in Appendix \ref{app:Hamiltonian-Eigenvalue-Problem}.%
}. Both methods involve the $\mathcal{H}$ matrix defined as

\[
H\ket{i}=\sum_{j}\ket{j}\mathcal{H}_{ji},
\]
 where $\ket{i}$ and $\ket{j}$ are $M$-bit trace states. Note that,
since the trace state basis is not orthonormal, $\mathcal{H}$ is
not the Hamiltonian matrix and even not Hermitian. 

The first method, used in Ref. \cite{Sun:2014dga}, is to calculate
the eigenvalues of the $\mathcal{H}$ from the equation 
\begin{equation}
\mathcal{H}\ket{E}=E\ket{E}.\label{eq:eigenvalue-problem}
\end{equation}
The relation between eigenvalues of $\mathcal{H}$ and of the Hamiltonian
matrix is determined by the norm matrix, $G=\dprod{i}{j}$, as follows:
\begin{itemize}
\item If $G$ is positive definite, i.e., all its eigenvalues are positive,
there is a one-to-one correspondence between the eigenvalues of $\mathcal{H}$
and the Hamiltonian. In this case, all the eigenstates of $\mathcal{H}$
are physical and have positive norm, which is defined as 
\[
\dprod{E}{E}=\sum_{ij}v^{i*}\dprod{i}{j}v^{j}=v^{\dagger}Gv
\]
 for an eigenstate $\ket{E}=\sum_{i}\ket{i}v^{i}$. Our numerical
calculation shows that when $N\geq M$ the norm matrix $G$ is always
positive definite.
\item When $N$ is an integer and less than $M$, the norm matrix $G$ is
positive semidefinite; i.e., some eigenvalues are zero, and the others
are positive. In this case, only those eigenstates of $\mathcal{H}$
with positive norm correspond to energy states of the Hamiltonian,
while those eigenstates of $\mathcal{H}$ with zero norm are unphysical. 
\item When $N$ is a noninteger and less than $M$, the norm matrix $G$
is indefinite; i.e., $G$ has both positive and negative eigenvalues.
There is a subtlety in this case. The eigenstates of $\mathcal{H}$
can be of positive norm, of zero norm, and of negative norm. The negative
norm eigenstates of $\mathcal{H}$ stem from their coupling to ghost
states, the eigenstates of $G$ of which the eigenvalues are negative.
The zero and negative norm eigenstates are still unphysical. But positive
norm eigenstates cannot be simply taken as energy states anymore.
A positive norm eigenstate is a physical energy state if it is orthogonal
to every ghost state. 
\end{itemize}
From the above statements, we should treat positive norm eigenstates
of $\mathcal{H}$ physical when $N$ is large enough or a small integral.
Moreover, the eigenvalues of $\mathcal{H}$ can be nonreal. This occurs
for both positive-semidefinite and indefinite cases. For a nonreal
eigenvalue of $\mathcal{H}$, the norm of its eigenstate must be zero,
and its complex conjugate is also an eigenvalue of $\mathcal{H}$.

The second method is to solve a generalized eigenvalue problem, 
\begin{equation}
\left(G\mathcal{H}\right)\ket{E}=EG\ket{E}.\label{eq:generalised-eigenvalue-problem}
\end{equation}
This method is helpful for filtering unphysical states when $G$ is
positive semidefinite. If $G$ is a full-rank matrix, this is a regular
generalized eigenvalue problem. If $G$ is not a full-rank matrix,
to solve the equation, we need to remove some rows and columns from
$G$ and $G\mathcal{H}$. If the rank of $G$ is $r$, we can pick
$r$ independent rows and columns from $G$ and $\left(G\mathcal{H}\right)$
to form two $r\times r$ matrices as 
\begin{eqnarray*}
\tilde{G} & = & \left(\begin{array}{cccc}
G_{i_{1}i_{1}} & G_{i_{1}i_{2}} & \cdots & G_{i_{1}i_{r}}\\
G_{i_{2}i_{2}} & G_{i_{2}i_{2}} & \cdots & G_{i_{2}i_{r}}\\
\vdots & \vdots & \vdots & \vdots\\
G_{i_{r}i_{1}} & G_{i_{r}i_{2}} & \cdots & G_{i_{r}i_{r}}
\end{array}\right),\\
\widetilde{G\mathcal{H}} & = & \left(\begin{array}{cccc}
\left(G\mathcal{H}\right)_{i_{1}i_{1}} & \left(G\mathcal{H}\right)_{i_{1}i_{2}} & \cdots & \left(G\mathcal{H}\right)_{i_{1}i_{r}}\\
\left(G\mathcal{H}\right)_{i_{2}i_{2}} & \left(G\mathcal{H}\right)_{i_{2}i_{2}} & \cdots & \left(G\mathcal{H}\right)_{i_{2}i_{r}}\\
\vdots & \vdots & \vdots & \vdots\\
\left(G\mathcal{H}\right)_{i_{r}i_{1}} & \left(G\mathcal{H}\right)_{i_{r}i_{2}} & \cdots & \left(G\mathcal{H}\right)_{i_{r}i_{r}}
\end{array}\right).
\end{eqnarray*}
 Then, Eq. (\ref{eq:generalised-eigenvalue-problem}) becomes 
\[
\left(\widetilde{G\mathcal{H}}\right)\ket{E}=E\tilde{G}\ket{E},
\]
the eigenvalues and eigenstates of which are all physical. 

The first method is used to investigate the change of eigenstates,
including both physical and unphysical states, with respect to $N$
for fixed $M$, while the second one is for the change of physical
energy levels with respect to $M$ for fixed $N$. For different values
of $M$, we calculated the $\mathcal{H}$ and $G$ matrices, the entries
of which are expressed in terms of $N$. Then we solve Eq. (\ref{eq:eigenvalue-problem})
or (\ref{eq:generalised-eigenvalue-problem}) to find their eigenstates.
Since the number of trace states increases exponentially as $M$ increases,
it is only feasible to perform the calculation for small $M$. The
highest value of $M$ we study is 11, at which $\mathcal{H}$ and
$G$ are $1473\times1473$ matrices%
\footnote{The source code of the project can be found in \cite{Chen:2015GitHub}.%
}.

\subsection{$H=H_{0}$}

Let us first consider the case of odd $M$. Figures \ref{fig:H=00003DH0M=00003D3}
to \ref{fig:H=00003DH0M=00003D11} show the lowest five eigenvalues
of $\mathcal{H}$ as a function of $1/N$ for odd $M$ from 3 to 11.
We use different line styles for different norm types: solid, dotted,
and dash-dotted curves correspond to positive, negative, and zero
norm eigenstates, respectively. Dash-dotted curves are actually associated
with two complex eigenvalues which are conjugate to each other and
hence represent only the real part of the eigenvalues. For higher
$M$, the eigenvalues decline dramatically in higher $1/N$, which
would squeeze the lower $1/N$ part into a small vertical size. To
show more details in lower $1/N$, we split some plots into a lower
$1/N$ part and a higher $1/N$ part, between which curves of the
same color represent the same eigenstate. See Fig. \ref{fig:H=00003DH0M=00003D7}
as an example. 

\begin{center}
\begin{figure}
\begin{centering}
\includegraphics[bb=54bp 200bp 550bp 610bp,clip,width=1\columnwidth]{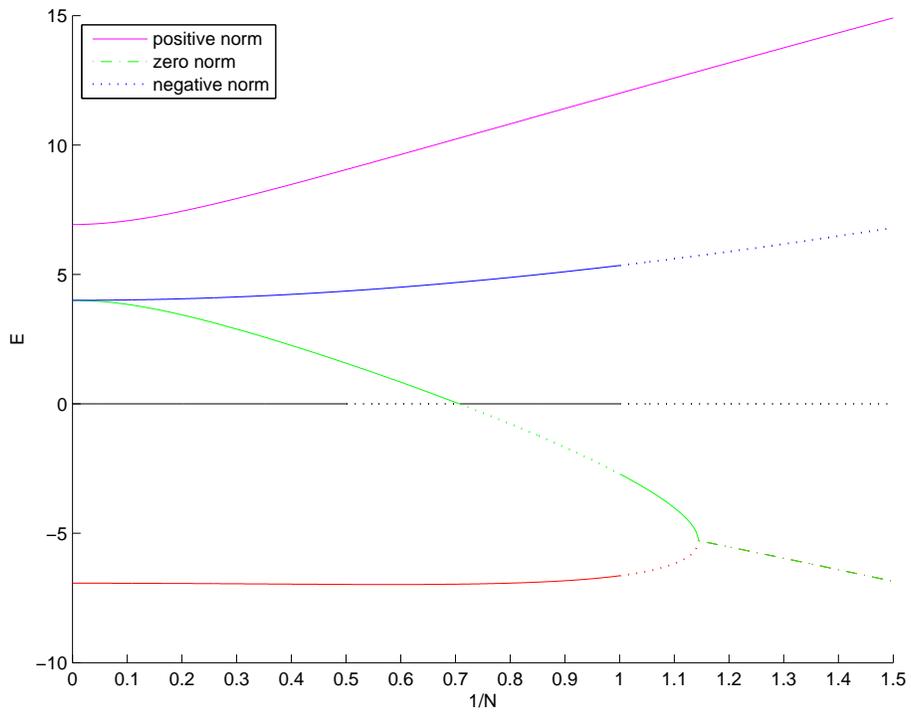}
\par\end{centering}

\caption{\label{fig:H=00003DH0M=00003D3}Lowest five energy states of the 3-bit
system with Hamiltonian $H=H_{0}$.}
\end{figure}

\par\end{center}

From these figures, we see several features of the eigenstates of
$\mathcal{H}$. At $N=\infty$, the ground states are nondegenerate,
while the first excited states are nondegenerate for $M$ divisible
by 3 and degenerate otherwise. This is consistent with the analytic
discussion of the previous section. As $1/N$ increases, degeneracies
are broken and the solid curves turn to dotted or dash-dotted curves,
which implies the disappearance of physical states. If a physical
state disappears at an integer value $N=n$, it also disappears at
$N=n-1,\, n-2$, and etc. For convenience, we denote as $N_{M}^{*}$
the maximum value of $N$ where the first disappearance of the ground
state occurs for bit number $M$. From the figures, we see that $N_{M}^{*}=\left(M-1\right)/2$
for $M\leq11$. If it is true for all $M$, it follows that, for ground
states surviving, $N$ must increase linearly as $M$ increases. The
eigenvalues drop dramatically at large $1/N$, as the right parts
of Figs. \ref{fig:H=00003DH0M=00003D7} to \ref{fig:H=00003DH0M=00003D11}
show. But it does not imply the decrease of energy levels, since all
these eigenstates are actually unphysical. 

\begin{center}
\begin{figure}
\begin{centering}
\includegraphics[bb=54bp 200bp 550bp 610bp,clip,width=1\columnwidth]{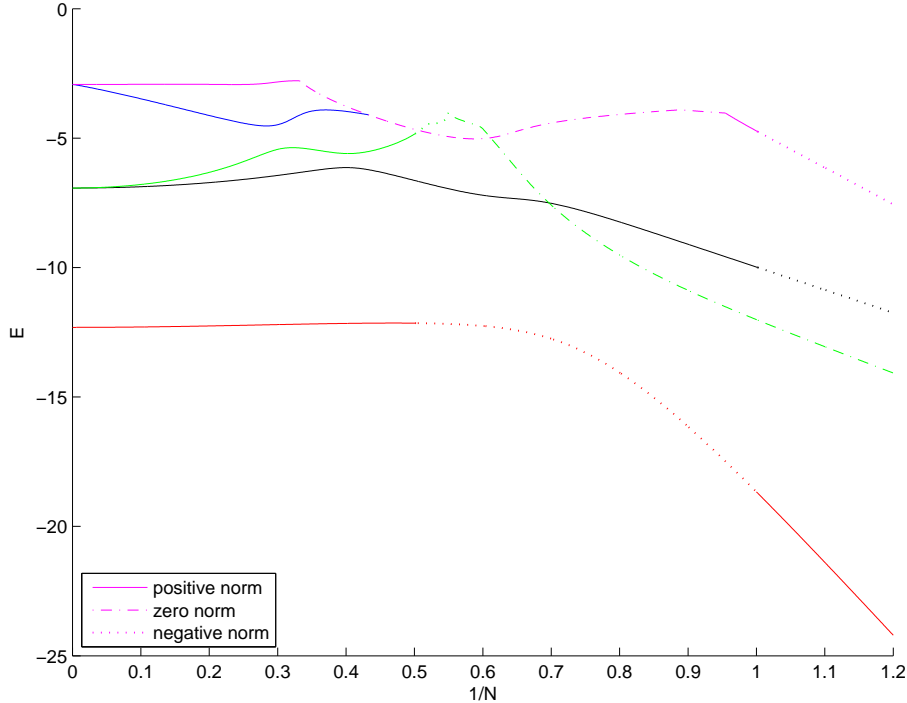}
\par\end{centering}

\caption{\label{fig:H=00003DH0M=00003D5}Lowest five eigenstates of $\mathcal{H}$
at $M=5$ for $H=H_{0}$.}
\end{figure}

\par\end{center}

\begin{center}
\begin{figure}
\begin{centering}
\includegraphics[bb=54bp 200bp 550bp 610bp,clip,width=1\columnwidth]{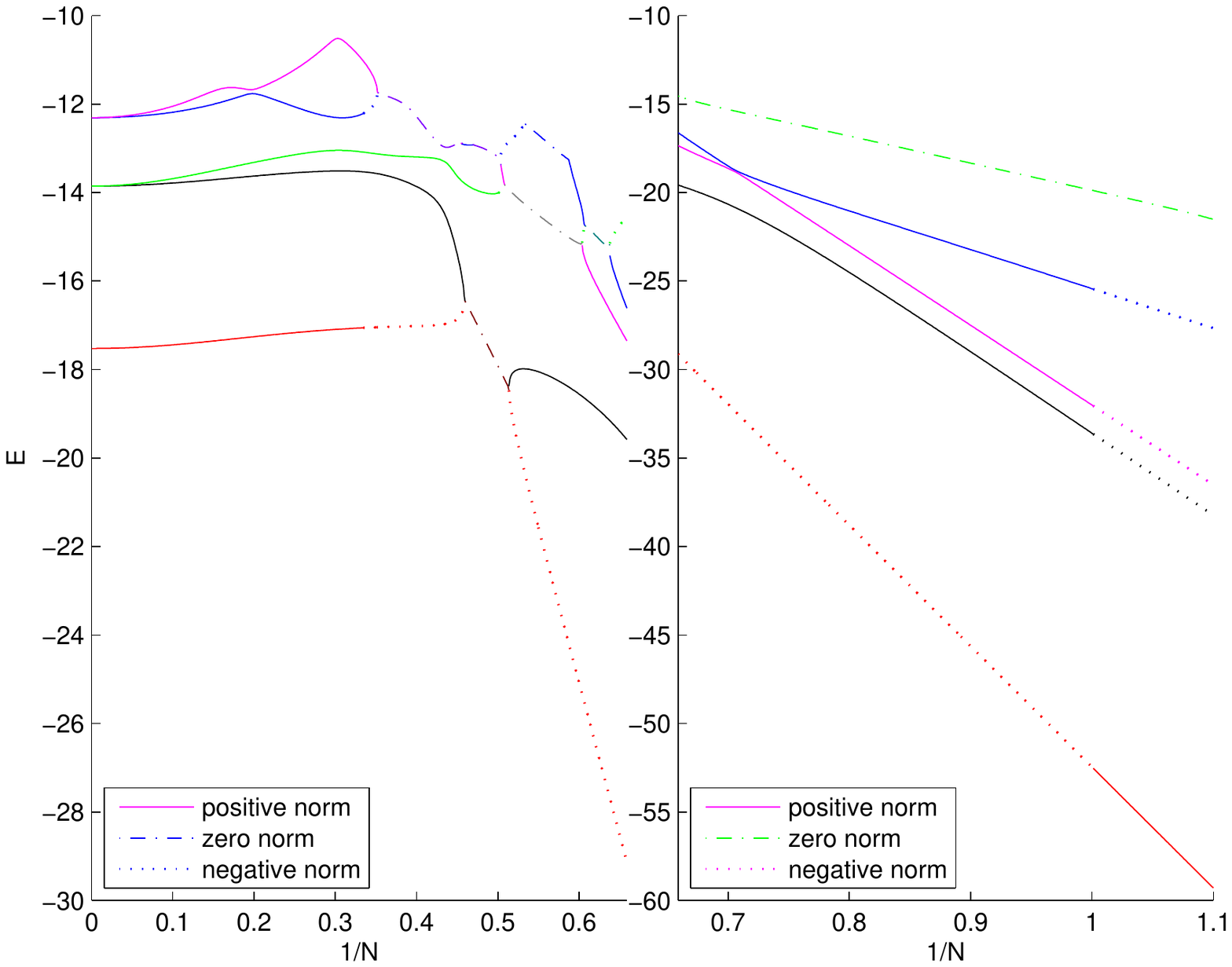}
\par\end{centering}

\caption{\label{fig:H=00003DH0M=00003D7}Lowest five eigenstates of $\mathcal{H}$
at $M=7$ for $H=H_{0}$.}
\end{figure}

\par\end{center}

\begin{center}
\begin{figure}
\begin{centering}
\includegraphics[bb=54bp 200bp 550bp 610bp,clip,width=1\columnwidth]{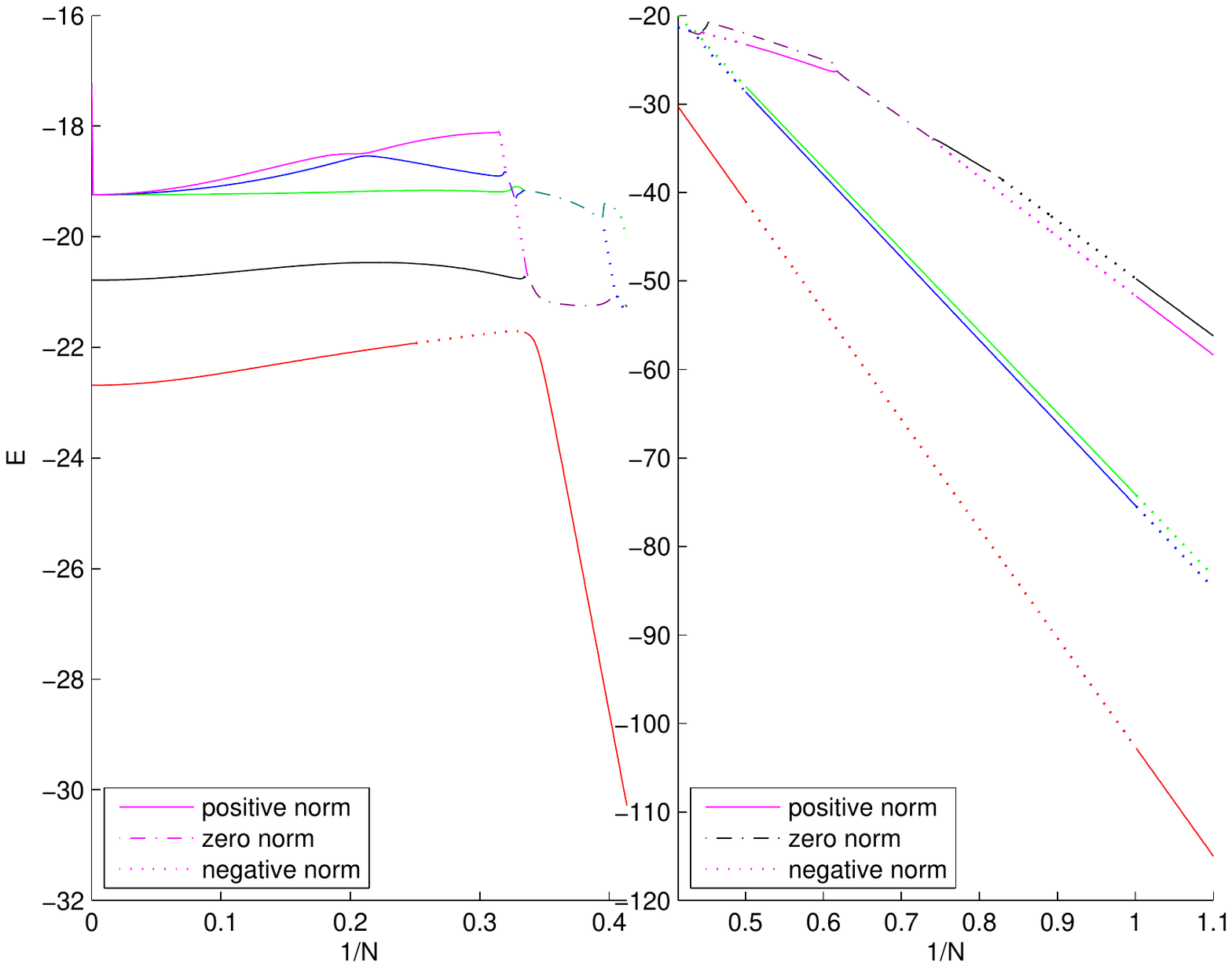}
\par\end{centering}

\caption{\label{fig:H=00003DH-M=00003D9}Lowest five eigenstates of $\mathcal{H}$
at $M=9$ for $H=H_{0}$.}
\end{figure}

\par\end{center}

\begin{center}
\begin{figure}
\begin{centering}
\includegraphics[bb=54bp 200bp 550bp 610bp,clip,width=1\columnwidth]{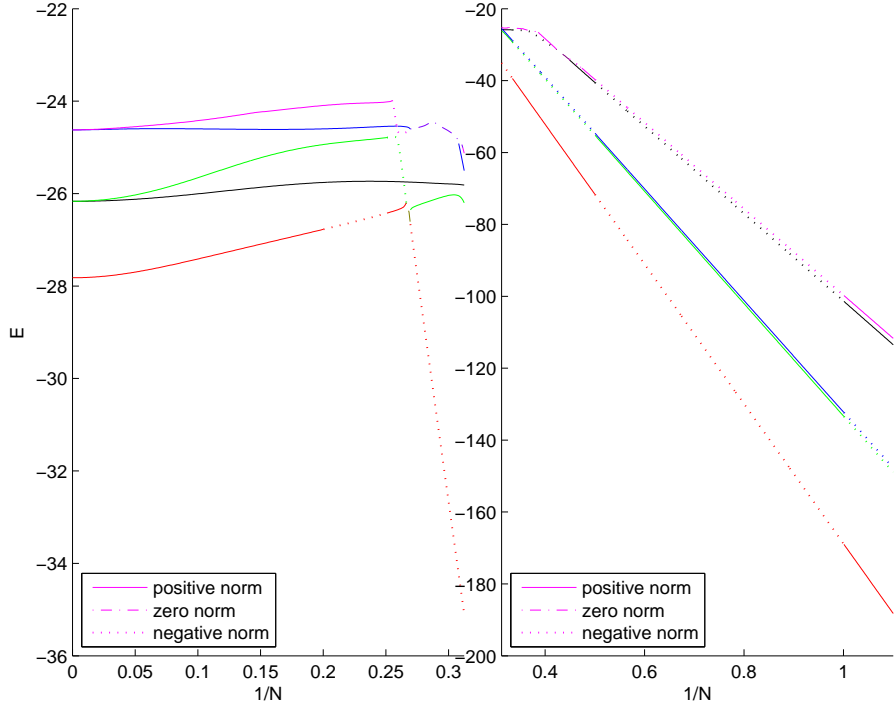}
\par\end{centering}

\caption{\label{fig:H=00003DH0M=00003D11}Lowest five eigenstates of $\mathcal{H}$
at $M=11$ for $H=H_{0}$.}
\end{figure}

\par\end{center}

For even $M$, we have similar plots as Fig. \ref{figH=00003Dh0M=00003Deven}.
At $N=\infty$, the lowest eigenstates are degenerate for $M=4$ and
$8$ and nondegenerate for $M=6$ and 10. It is again consistent with
our analysis in the previous section. The lowest states also disappear
when $N$ is small. But unlike the odd $M$ case, there is no simple
formula to determine $N_{M}^{*}$. The reason is that the lowest energy
of $E\left(\eta_{i}\right)$ in (\ref{eq:LargeN-energy-levels}) is
excluded by the cyclic constraint (\ref{eq:LargeN-cyclic-constraint}).

\begin{center}
\begin{figure}
\begin{centering}
\includegraphics[bb=54bp 200bp 550bp 610bp,clip,width=1\columnwidth]{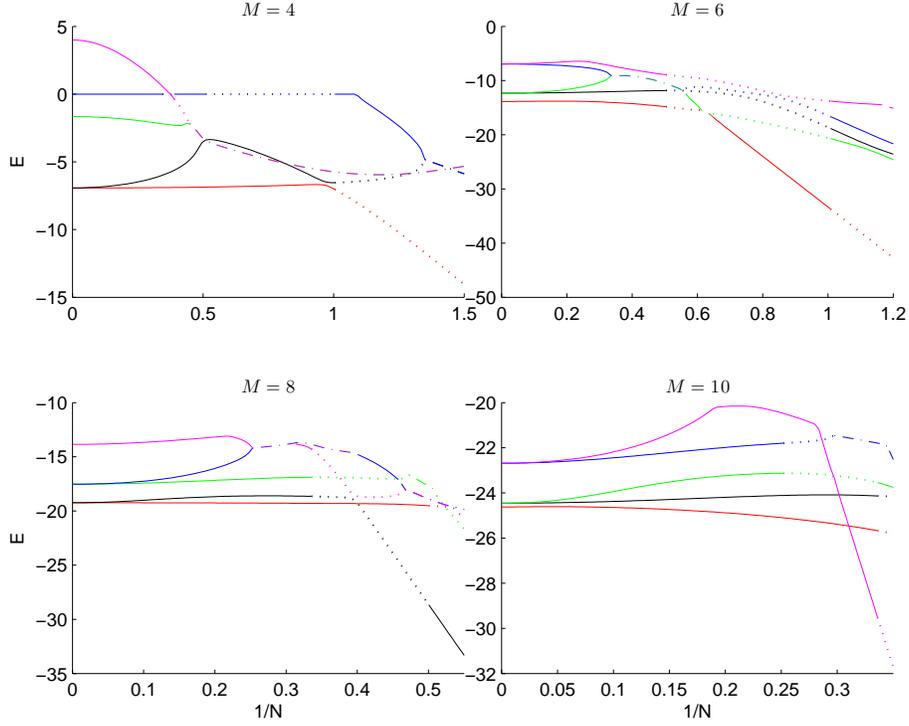}
\par\end{centering}

\caption{\label{figH=00003Dh0M=00003Deven}Lowest five eigenstates of $\mathcal{H}$
for $H=H_{0}$ at $M=4,6,8,10$}
\end{figure}

\par\end{center}

We now consider the physical ground energy as a function of $M$ when
$N$ is fixed, shown as Fig. \ref{fig:physical-state-H0}. The physical
ground states have different trends at different values of $N$. For
$N=1$, the physical ground state climbs significantly. This is consistent
with analytical calculation, which shows the ground state is a quadratic
function of $M$ when $N=1$. For $N=2$, the ground state only goes
up slightly. When $N\geq3$, it turns downward. For large $N$, the
physical ground energy drops almost linearly with respect to $M$
at rate $-8/\pi$, as predicted by Eq. (\ref{eq:LargeN-E-min-odd}).
This indicates the system becomes stringy when $N$ is large enough.

\begin{center}
\begin{figure}
\centering{}\includegraphics[bb=50bp 400bp 560bp 600bp,clip,width=1\columnwidth]{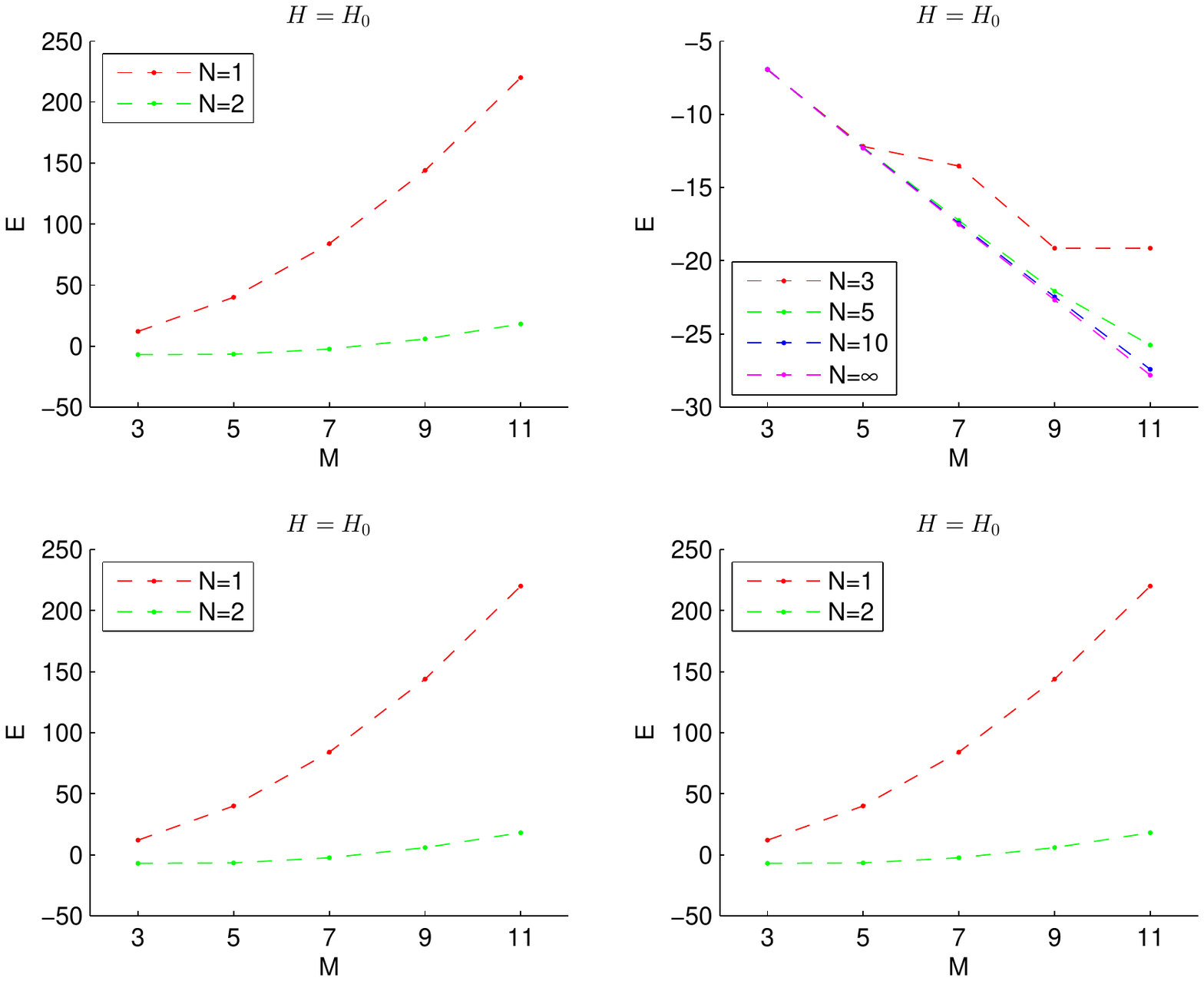}\caption{\label{fig:physical-state-H0}Change of physical ground states with
respect to $M$ for fixed $N$. Only the ground energies at odd $M$
are sampled. }
\end{figure}

\par\end{center}

Fig. \ref{fig:First-excitation-H0} shows how the excitation energy
changes with respect to $M$ for fixed $N$. The vertical axis of
Fig. \ref{fig:First-excitation-H0} is $M\times\left(E_{1}-E_{0}\right)$,
where $E_{1}-E_{0}$ is the gap between the first excited energy and
lowest energy. For stringy behavior, $M\times\left(E_{1}-E_{0}\right)$
should be constant for large $M$. Though we only calculate up to
$M=11$, we still see the trend that, for $N$ large enough, $M\times\left(E_{1}-E_{0}\right)$
is almost a constant between 15 and 20. As a reference, the analytic
prediction of the gap at $N=\infty$ is $16\pi/3M$. That being said,
there is no inconsistency between the numerical results and stringy
behavior. 

\begin{center}
\begin{figure}
\begin{centering}
\includegraphics[bb=54bp 200bp 550bp 610bp,clip,width=1\columnwidth]{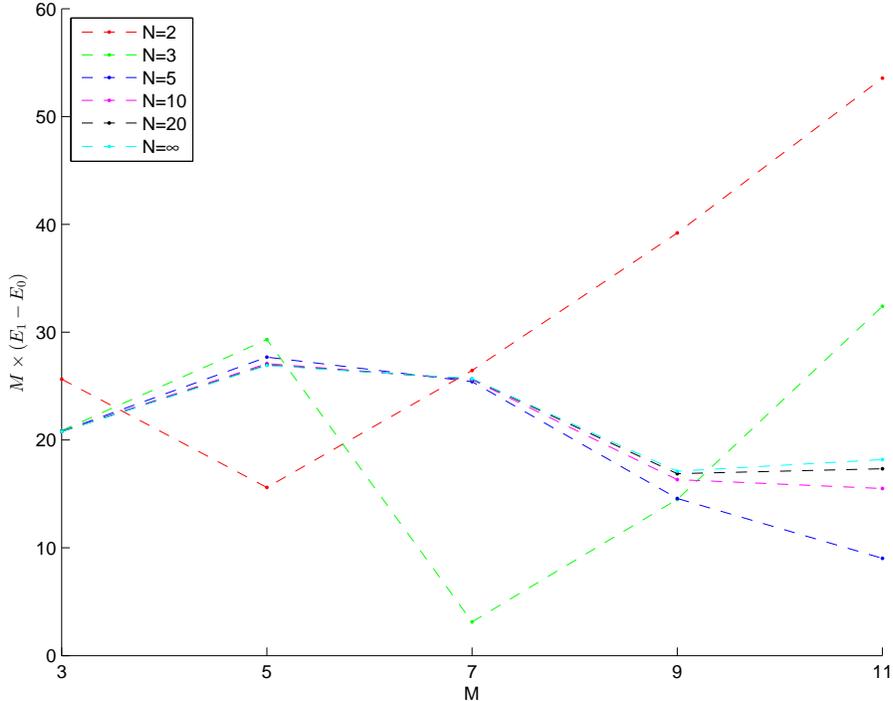}
\par\end{centering}

\caption{\label{fig:First-excitation-H0} $\left(E_{1}-E_{0}\right)\times M$
as a function of $M$}
\end{figure}

\par\end{center}

\subsection{Variations of $H$}

In this subsection, we will analyze the energy levels of two variations
of the Hamiltonian, $H=H_{0}+\xi\Delta H$ and $H=-H_{0}+\xi\Delta H$. 

Figure \ref{fig:xi=00003DmanyM=00003D3} shows the eigenvalues of
$\mathcal{H}$ as a function of $1/N$ when $M=3$ and the Hamiltonian
is of the form $H=H_{0}+\xi\Delta H$. As $\xi$ increases, the disappearance
point of the highest eigenstate moves in the small $N$ direction:
for $\xi=-1$, it is at $N=2$; for $\xi=-0.6$, it is at $N=1$;
when $\xi\geq-0.1$, the disappearance point occurs after $N<1/2$.
The disappearance point of the ground state, $N_{3}^{*}$, moves in
the opposite direction: for $-1\leq\xi\leq-0.1$, $N_{3}^{*}=1$;
for $\xi=0.5$, $1<N_{3}^{*}<2$; for $\xi=3$, $N_{3}^{*}=2$. 

\begin{center}
\begin{figure}
\centering{}\includegraphics[bb=55bp 210bp 550bp 720bp,clip,width=1\columnwidth]{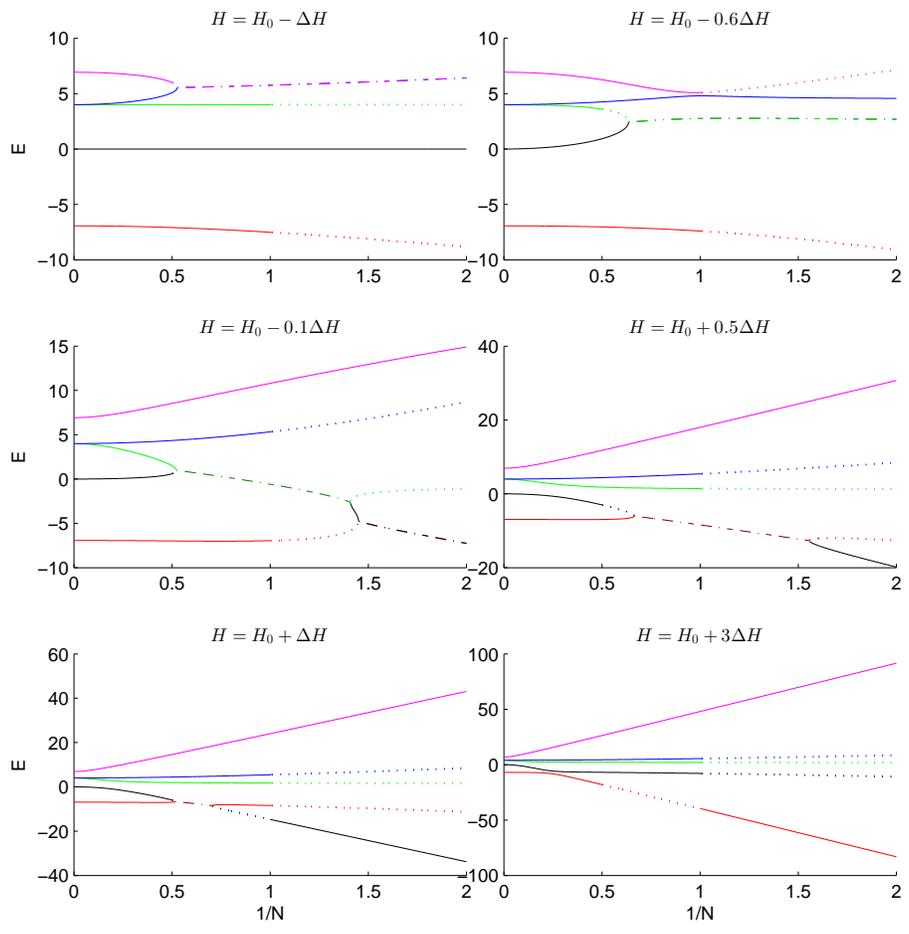}\caption{\label{fig:xi=00003DmanyM=00003D3}Eigenvalues of $\mathcal{H}$ matrices
at $M=3$ for Hamiltonian $H=H_{0}+\xi\Delta H$, with $\xi=-1$,
$-0.6$, $-0.1$, $0.5$, $1$, and $3$.}
\end{figure}

\par\end{center}

Since all eigenstates of $\mathcal{H}$ are physical when $N\geq M$,
the largest value of $N_{M}^{*}$ is $M-1$. Particularly, for $M\leq11$,
we find $N_{M}^{*}=M-1$ can be achieved when $\xi\geq2$. $N_{M}^{*}$
is minimal when $\xi=-1$, the lower bound of $\xi$ under the stabilization
constraint. The $\xi=-1$ case is shown in Fig. \ref{fig:xi=00003D-1M=00003Dodd}.
While $N_{M}^{*}=\left(M-1\right)/2$ still holds for $M=5$ and $7$,
$N_{9}^{*}=2$ and $N_{11}^{*}=3$ spoil the pattern. We do not have
results for $M>11$, but it seems that $M/N_{M}^{*}$ could be large
for large $M$. If it is true, it means that the ground eigenstates
could survive when $M$ is large and $N\ll M$. 

\begin{center}
\begin{figure}
\begin{centering}
\includegraphics[bb=54bp 200bp 550bp 610bp,clip,width=1\columnwidth]{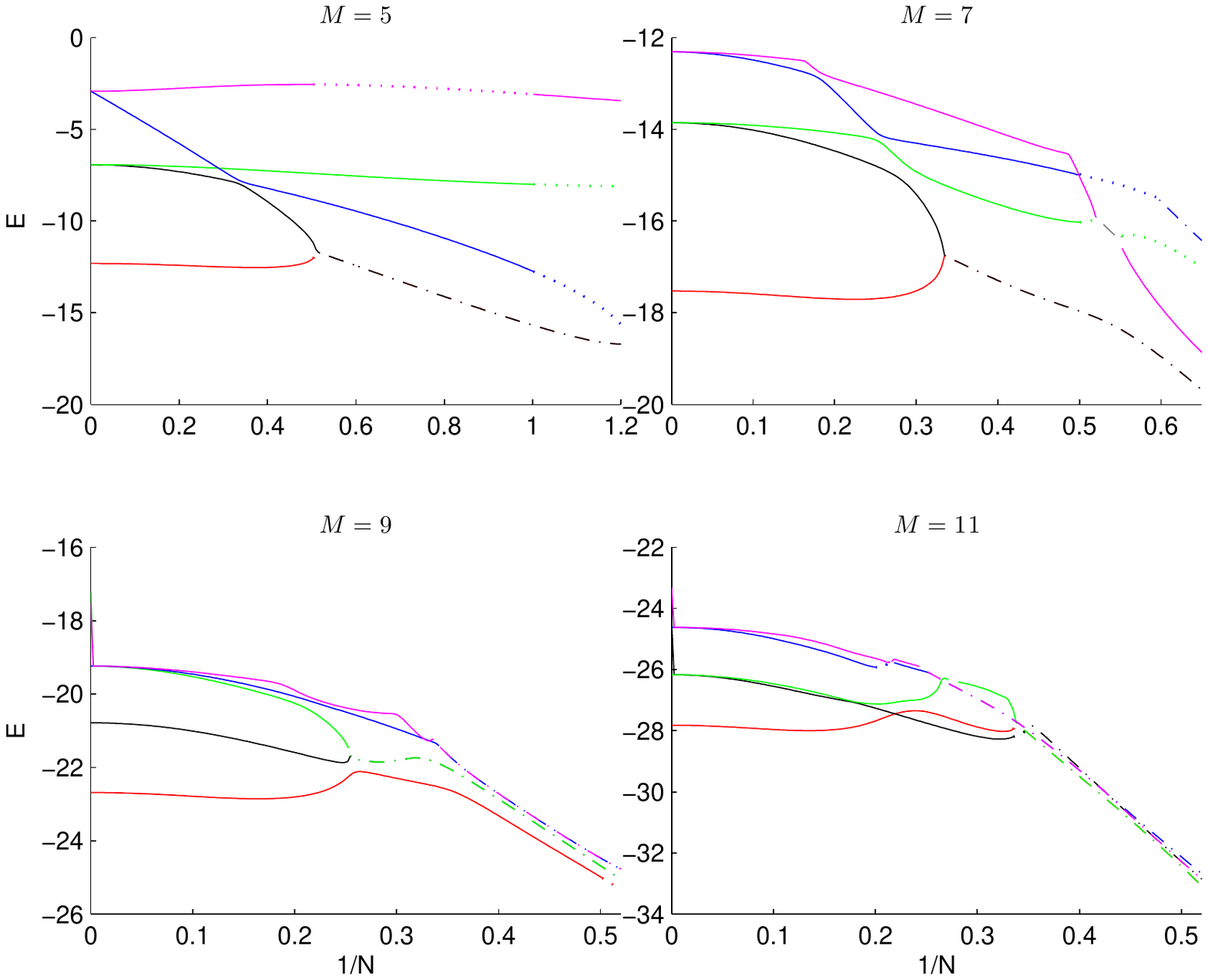}
\par\end{centering}

\caption{\label{fig:xi=00003D-1M=00003Dodd}Lowest five eigenstates of $\mathcal{H}$
for $H=H_{0}-\Delta H$ at $M=5,7,9,11$}

\end{figure}

\par\end{center}

Figure \ref{fig:EvsM-N>=00003D3} shows the change of physical ground
energy with respect to $M$ for a fixed value of $N$. Note that only
ground energies at odd $M$ are evaluated. The ground energies have
different trends for $\xi<-1$, $\xi=-1$, and $\xi>-1$: when $\xi=-1$,
the ground energies decrease almost linearly for all $N$; when $\xi<-1$,
the ground energies decline faster than linearly, which implies the
system is not stable; when $\xi>-1$, the ground energy first declines
and then increases for small $N$, and it declines linearly for large
$N$. It follows that the system has stringy behavior if $\xi\geq-1$
and $N$ is not too small.

\begin{center}
\begin{figure}
\centering{}\includegraphics[bb=54bp 200bp 550bp 610bp,clip,width=1\columnwidth]{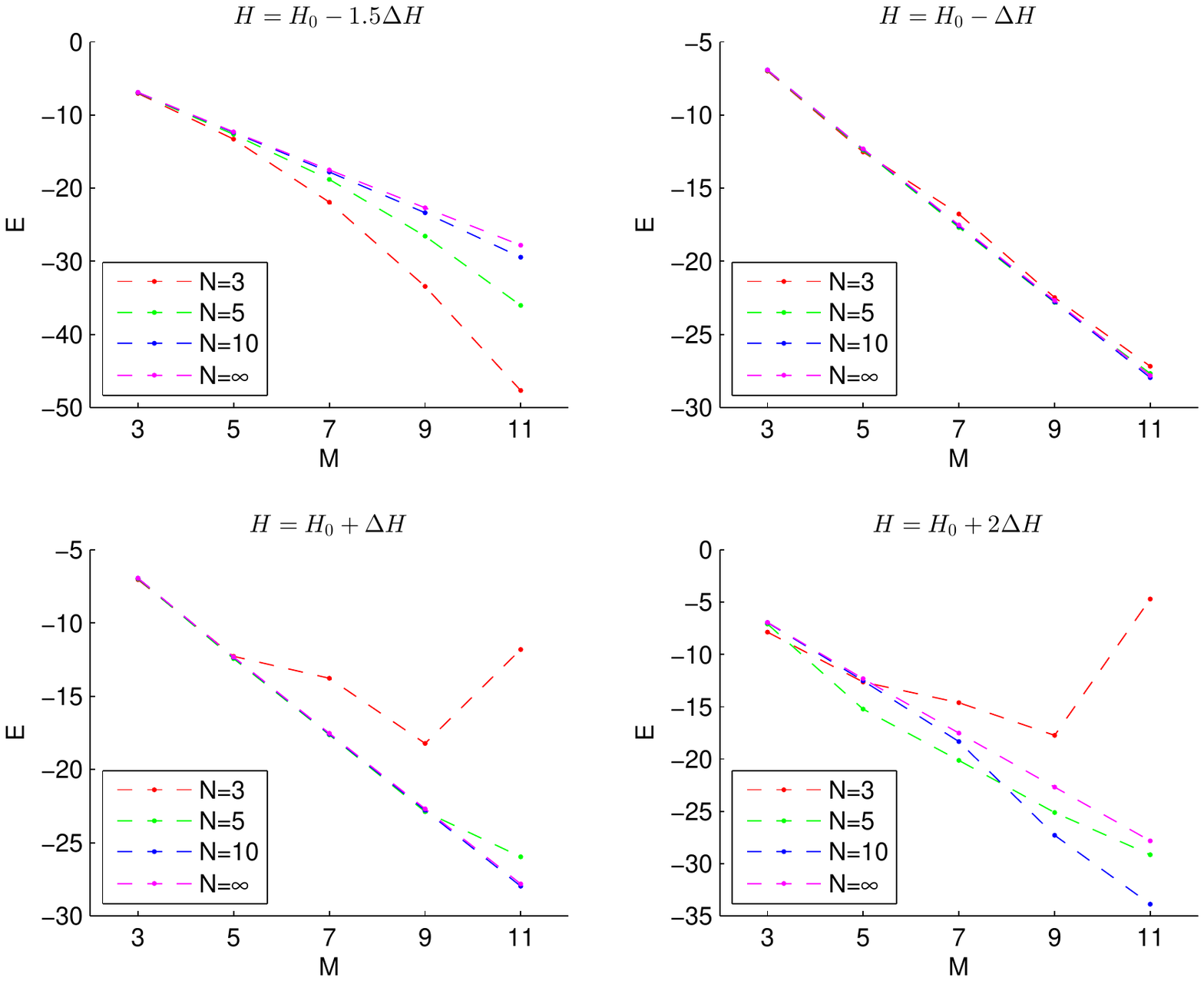}\caption{\label{fig:EvsM-N>=00003D3}Change of physical ground energy with
respect to $M$ at fixed $N$}
\end{figure}

\par\end{center}

For $H=-H_{0}+\xi\Delta H$, in the large $N$ limit, the maximum
value of $E\left(\eta_{i}\right)$ in (\ref{eq:LargeN-energy-levels})
is allowed for both odd and even $M$. Consequently, the ground eigenstates
are nondegenerate for all $M$, as shown in Fig. \ref{fig:nxi=00003D1.5M=00003Dmany}
for $H=-H_{0}+1.5\Delta H$. From the figure, we see that $N_{M}^{*}=M-1$. 

\begin{center}
\begin{figure}
\begin{centering}
\includegraphics[bb=54bp 200bp 550bp 710bp,clip,width=1\columnwidth]{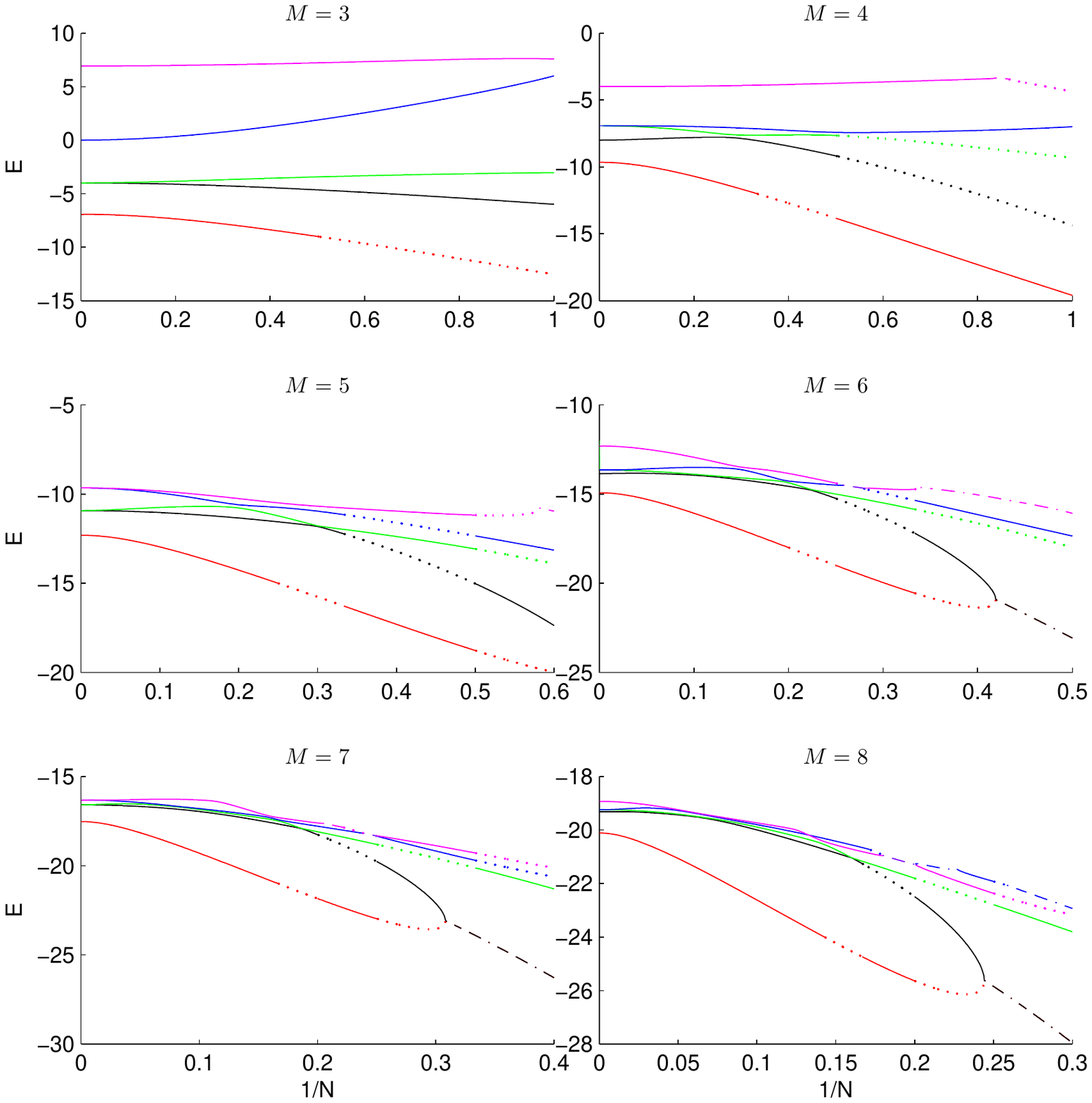}
\par\end{centering}

\caption{\label{fig:nxi=00003D1.5M=00003Dmany}Eigenvalues of $\mathcal{H}$
for $H=-H_{0}+1.5\Delta H$. For each $M$, the ground state disappears
at $N=M-1$.}
\end{figure}

\par\end{center}

$\xi$ has a similar impact on $N_{M}^{*}$ as the $H=H_{0}+\xi\Delta H$
case. Figure \ref{fig:nxi=00003D1M=00003Dmany} plots the eigenstates
of $\mathcal{H}$ for $\xi=1$, when $N_{M}^{*}$ is minimal. There
is no simple pattern for $N_{M}^{*}$: for odd $M$, $N_{3}^{*}=2$,
$N_{5}^{*}=2$, $N_{7}^{*}=3$, and $N_{9}^{*}=3$; for even $M$,
$N_{4}^{*}=3$, $N_{6}^{*}=2$, $N_{8}=2$, and $N_{10}^{*}=3$. It
seems to suggest that the ground eigenstate could survive when $M$
is large and $N\ll M$. 

\begin{center}
\begin{figure}
\begin{centering}
\includegraphics[bb=55bp 90bp 550bp 595bp,clip,width=1\columnwidth]{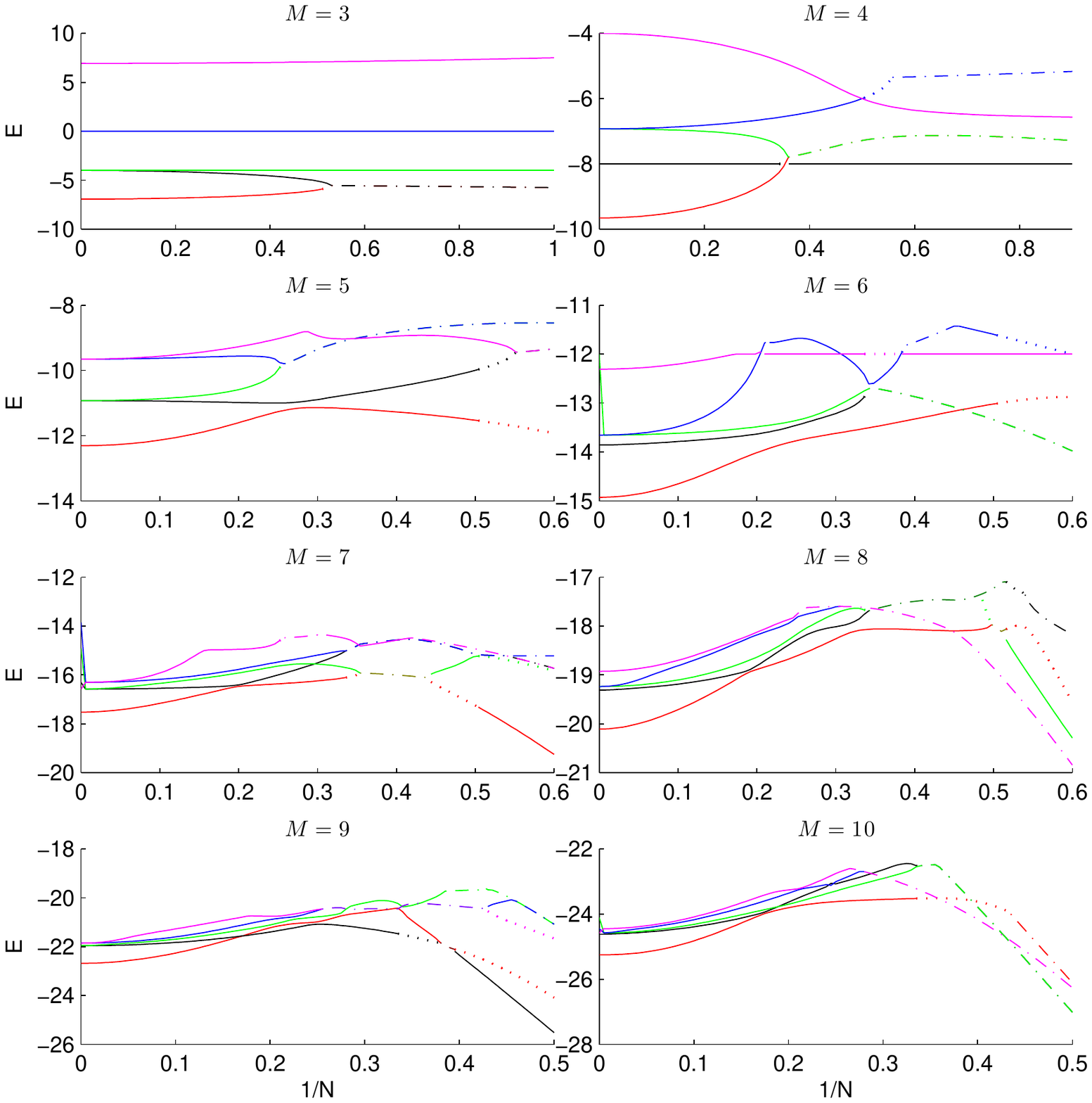}
\par\end{centering}

\caption{\label{fig:nxi=00003D1M=00003Dmany}Eigenvalues of $\mathcal{H}$
for $H=-H_{0}+\Delta H$ and $3\leq M\leq10$.}

\end{figure}

\par\end{center}

Figure \ref{fig:nH-Physical-ground-energy} shows the change of physical
ground energy with respect to $M$ at fixed $N$ for $H=-H_{0}+\xi\Delta H$.
It is similar to the $H=H_{0}+\xi\Delta H$ case. When $\xi=0.5$,
the system is not stable at finite $N$ as the curves decline faster
than linearly. $\xi=1$ is the marginal case, in which all the physical
ground energies drop almost linearly. When $\xi=1.5$ or $\xi=3$,
the curves for small $N$ are zig-zag, and particularly, when $\xi=3$
and $N=3$, the trend is  slightly upward. It implies that the system
is stable for large $M$. 

\begin{center}
\begin{figure}
\centering{}\includegraphics[bb=54bp 200bp 550bp 610bp,clip,width=1\columnwidth]{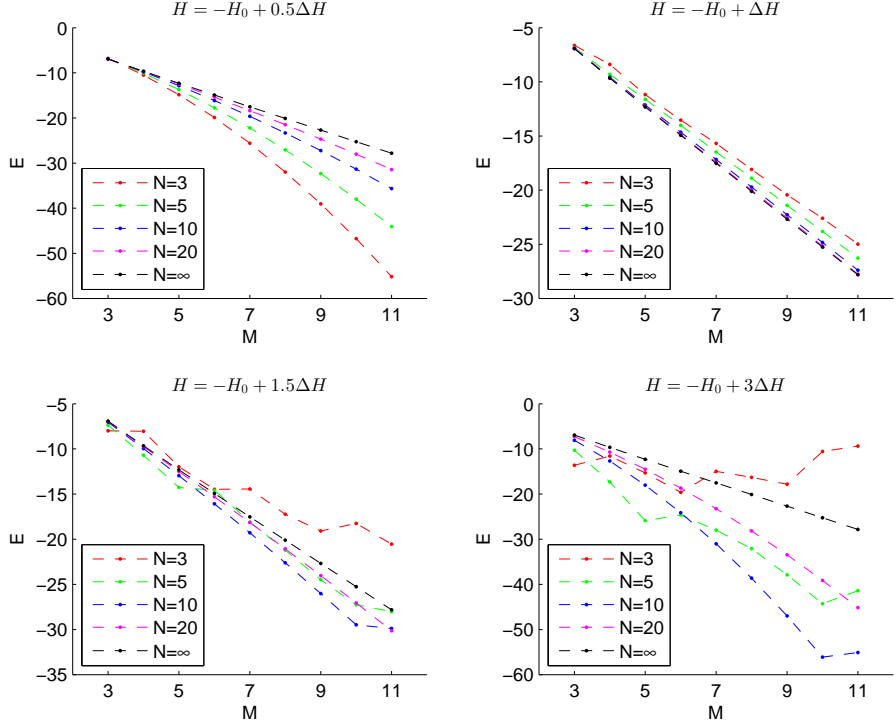}\caption{\label{fig:nH-Physical-ground-energy}Physical ground energy of $-H_{0}+\xi\Delta H$
at $\xi=0.5,\,1,\,1.5,\,3$ and $N=3,\,5,\,10,\,20,\,\infty$. }
\end{figure}

\par\end{center}

\section{Summary and conclusion}

In this paper we have studied the string bit model with $s=1$, $d=0$.
We studied possible forms of the supersymmetric Hamiltonian and their
excitation energies in the large $N$ limit. We also performed a numerical
study of energy levels at finite $N$ for Hamiltonians $H=\pm H_{0}+\xi\Delta H$,
where, at $N=\infty$, $\Delta H$ vanishes and $H_{0}$ produces
the Green-Schwarz Hamiltonian. 

We showed that the supersymmetry plays a crucial role in the model.
The general Hamiltonian is chosen to be a linear combination of eight
single trace operators, which contain two consecutive creation operators
followed by two annihilation operators. With the supersymmetry constraint,
we reduce the number of parameters in the Hamiltonian to 3. Another
interesting consequence of supersymmetry is that, after imposing the
supersymmetry constraint on the Hamiltonian, the excitation energy
becomes of order $M^{-1}$, which implies the energy spectrum of the
model is continuous when $M$ is large.

In finite $N$, we numerically studied the energy spectrum of the
model up to $M\leq11$. There exists a maximal integer $N_{M}^{*}$
that when $N\leq N_{M}^{*}$ the would-be ground energy eigenstate
of the $M$-bit system is unphysical. For $H=H_{0}$ and odd $M\leq11$,
the numerical computation shows $N_{M}^{*}=\left(M-1\right)/2$. If
such a simple relation holds for all odd $M$, then, at large $M$,
the surviving of ground state requires $N$ to be large as well. For
$H=\pm H_{0}+\xi\Delta H$, $N_{M}^{*}$ increases (decreases) as
$\xi$ increases (decreases). The maximum value of $N_{M}^{*}$ is
$\left(M-1\right)$. The minimum of $N_{M}^{*}$ is achieved when
$H=\pm H_{0}\mp\Delta H$ because of the stabilization constraint
$\xi\geq\mp1$. In the minimum cases, one find that $N_{M}^{*}$ is
less than $\left(M-1\right)/2$ when $7<M\leq11$ . If such a trend
continues for $M>11$, it means that the ground energy state might
be able to survive at very large $M$ and $M\gg N$. 

For fixed finite $N$ and $H=\pm H_{0}+\xi\Delta H$, the system is
stable only when $\xi\geq\mp1$. The ground energy drops almost linearly
with respect to $M$ when $\xi\geq\mp1$ and faster than linearly
when $\xi<\mp1$. The numerical computation also reveals the excitation
energy is roughly proportional to $M^{-1}$. While we do not have
data for $M>11$, the trend is still evident. These properties indicate
that the model has stringy behavior when $\xi\geq\mp1$.

The numerical computation is performed up to $M=11$. The bottleneck
is the calculation of norm matrices. Our algorithm has $\mathcal{O}\left(M!\right)$
time complexity for computing each entry of the matrix. It needs significant
improvement for numerical computation of higher $M$. This is one
of the issues we need to address in future research.

We can also extend our work in other directions. Our strategy can
be applied to the model with $s>1,d=0$ or $d>0$ cases. We can also
analytically calculate $1/N$ expansion of the model, in which some
progress has been made by Ref. \cite{Thorn:2015wli}.

\section{Acknowledgments}

We thank Charles Thorn for his guidance on this work. This research
was supported in part by the Department of Energy under Grant No.
DE-SC0010296.

\newpage{}

\appendix
\numberwithin{equation}{section}

\section{\label{app:Trace-States-List}Bosonic trace states}

\subsection*{1. 1 bit}

One bosonic state: 
\begin{align*}
\left|1\right\rangle  & =\mathrm{Tr}\bar{a}\left|0\right\rangle 
\end{align*}

\subsection*{2. 2 bits}

Two bosonic states: 
\begin{align*}
\left|1\right\rangle  & =\mathrm{Tr}\bar{a}\bar{a}\left|0\right\rangle  & \left|2\right\rangle  & =\mathrm{Tr}\bar{a}\mathrm{Tr}\bar{a}\left|0\right\rangle 
\end{align*}

\subsection*{3. 3 bits}

Five bosonic states: 
\begin{align*}
\left|1\right\rangle  & =\mathrm{Tr}\bar{a}\bar{a}\bar{a}\left|0\right\rangle  & \left|2\right\rangle  & =\mathrm{Tr}\bar{a}\mathrm{Tr}\bar{a}\bar{a}\left|0\right\rangle  & \left|3\right\rangle  & =\mathrm{Tr}\bar{a}\mathrm{Tr}\bar{a}\mathrm{Tr}\bar{a}\left|0\right\rangle \\
\left|4\right\rangle  & =\mathrm{Tr}\bar{a}\bar{b}\bar{b}\left|0\right\rangle  & \left|5\right\rangle  & =\mathrm{Tr}\bar{b}\mathrm{Tr}\bar{a}\bar{b}\left|0\right\rangle 
\end{align*}

\subsection*{4. 4 bits}

Ten bosonic states: 
\begin{align*}
\left|1\right\rangle  & =\mathrm{Tr}\bar{a}\bar{a}\bar{a}\bar{a}\left|0\right\rangle  & \left|2\right\rangle  & =\mathrm{Tr}\bar{a}\mathrm{Tr}\bar{a}\bar{a}\bar{a}\left|0\right\rangle  & \left|3\right\rangle  & =\mathrm{Tr}\bar{a}\bar{a}\mathrm{Tr}\bar{a}\bar{a}\left|0\right\rangle \\
\left|4\right\rangle  & =\mathrm{Tr}\bar{a}\mathrm{Tr}\bar{a}\mathrm{Tr}\bar{a}\bar{a}\left|0\right\rangle  & \left|5\right\rangle  & =\mathrm{Tr}\bar{a}\mathrm{Tr}\bar{a}\mathrm{Tr}\bar{a}\mathrm{Tr}\bar{a}\left|0\right\rangle  & \left|6\right\rangle  & =\mathrm{Tr}\bar{a}\bar{a}\bar{b}\bar{b}\left|0\right\rangle \\
\left|7\right\rangle  & =\mathrm{Tr}\bar{a}\mathrm{Tr}\bar{a}\bar{b}\bar{b}\left|0\right\rangle  & \left|8\right\rangle  & =\mathrm{Tr}\bar{b}\mathrm{Tr}\bar{a}\bar{a}\bar{b}\left|0\right\rangle  & \left|9\right\rangle  & =\mathrm{Tr}\bar{a}\mathrm{Tr}\bar{b}\mathrm{Tr}\bar{a}\bar{b}\left|0\right\rangle \\
\left|10\right\rangle  & =\mathrm{Tr}\bar{b}\mathrm{Tr}\bar{b}\bar{b}\bar{b}\left|0\right\rangle 
\end{align*}

\subsection*{5. 5 bits}

Twenty-one bosonic states: 
\begin{align*}
\left|1\right\rangle  & =\mathrm{Tr}\bar{a}\bar{a}\bar{a}\bar{a}\bar{a}\left|0\right\rangle  & \left|2\right\rangle  & =\mathrm{Tr}\bar{a}\mathrm{Tr}\bar{a}\bar{a}\bar{a}\bar{a}\left|0\right\rangle  & \left|3\right\rangle  & =\mathrm{Tr}\bar{a}\bar{a}\mathrm{Tr}\bar{a}\bar{a}\bar{a}\left|0\right\rangle \\
\left|4\right\rangle  & =\mathrm{Tr}\bar{a}\mathrm{Tr}\bar{a}\mathrm{Tr}\bar{a}\bar{a}\bar{a}\left|0\right\rangle  & \left|5\right\rangle  & =\mathrm{Tr}\bar{a}\mathrm{Tr}\bar{a}\bar{a}\mathrm{Tr}\bar{a}\bar{a}\left|0\right\rangle  & \left|6\right\rangle  & =\mathrm{Tr}\bar{a}\mathrm{Tr}\bar{a}\mathrm{Tr}\bar{a}\mathrm{Tr}\bar{a}\bar{a}\left|0\right\rangle \\
\left|7\right\rangle  & =\mathrm{Tr}\bar{a}\mathrm{Tr}\bar{a}\mathrm{Tr}\bar{a}\mathrm{Tr}\bar{a}\mathrm{Tr}\bar{a}\left|0\right\rangle  & \left|8\right\rangle  & =\mathrm{Tr}\bar{a}\bar{a}\bar{a}\bar{b}\bar{b}\left|0\right\rangle  & \left|9\right\rangle  & =\mathrm{Tr}\bar{a}\bar{a}\bar{b}\bar{a}\bar{b}\left|0\right\rangle \\
\left|10\right\rangle  & =\mathrm{Tr}\bar{a}\mathrm{Tr}\bar{a}\bar{a}\bar{b}\bar{b}\left|0\right\rangle  & \left|11\right\rangle  & =\mathrm{Tr}\bar{b}\mathrm{Tr}\bar{a}\bar{a}\bar{a}\bar{b}\left|0\right\rangle  & \left|12\right\rangle  & =\mathrm{Tr}\bar{a}\bar{a}\mathrm{Tr}\bar{a}\bar{b}\bar{b}\left|0\right\rangle \\
\left|13\right\rangle  & =\mathrm{Tr}\bar{a}\bar{b}\mathrm{Tr}\bar{a}\bar{a}\bar{b}\left|0\right\rangle  & \left|14\right\rangle  & =\mathrm{Tr}\bar{a}\mathrm{Tr}\bar{a}\mathrm{Tr}\bar{a}\bar{b}\bar{b}\left|0\right\rangle  & \left|15\right\rangle  & =\mathrm{Tr}\bar{a}\mathrm{Tr}\bar{b}\mathrm{Tr}\bar{a}\bar{a}\bar{b}\left|0\right\rangle \\
\left|16\right\rangle  & =\mathrm{Tr}\bar{b}\mathrm{Tr}\bar{a}\bar{a}\mathrm{Tr}\bar{a}\bar{b}\left|0\right\rangle  & \left|17\right\rangle  & =\mathrm{Tr}\bar{a}\mathrm{Tr}\bar{a}\mathrm{Tr}\bar{b}\mathrm{Tr}\bar{a}\bar{b}\left|0\right\rangle  & \left|18\right\rangle  & =\mathrm{Tr}\bar{a}\bar{b}\bar{b}\bar{b}\bar{b}\left|0\right\rangle \\
\left|19\right\rangle  & =\mathrm{Tr}\bar{b}\mathrm{Tr}\bar{a}\bar{b}\bar{b}\bar{b}\left|0\right\rangle  & \left|20\right\rangle  & =\mathrm{Tr}\bar{a}\bar{b}\mathrm{Tr}\bar{b}\bar{b}\bar{b}\left|0\right\rangle  & \left|21\right\rangle  & =\mathrm{Tr}\bar{a}\mathrm{Tr}\bar{b}\mathrm{Tr}\bar{b}\bar{b}\bar{b}\left|0\right\rangle 
\end{align*}

\subsection*{6. 6 bits}

Forty-four bosonic states: 
\begin{align*}
\left|1\right\rangle  & =\mathrm{Tr}\bar{a}\bar{a}\bar{a}\bar{a}\bar{a}\bar{a}\left|0\right\rangle  & \left|2\right\rangle  & =\mathrm{Tr}\bar{a}\mathrm{Tr}\bar{a}\bar{a}\bar{a}\bar{a}\bar{a}\left|0\right\rangle  & \left|3\right\rangle  & =\mathrm{Tr}\bar{a}\bar{a}\mathrm{Tr}\bar{a}\bar{a}\bar{a}\bar{a}\left|0\right\rangle \\
\left|4\right\rangle  & =\mathrm{Tr}\bar{a}\bar{a}\bar{a}\mathrm{Tr}\bar{a}\bar{a}\bar{a}\left|0\right\rangle  & \left|5\right\rangle  & =\mathrm{Tr}\bar{a}\mathrm{Tr}\bar{a}\mathrm{Tr}\bar{a}\bar{a}\bar{a}\bar{a}\left|0\right\rangle  & \left|6\right\rangle  & =\mathrm{Tr}\bar{a}\mathrm{Tr}\bar{a}\bar{a}\mathrm{Tr}\bar{a}\bar{a}\bar{a}\left|0\right\rangle \\
\left|7\right\rangle  & =\mathrm{Tr}\bar{a}\bar{a}\mathrm{Tr}\bar{a}\bar{a}\mathrm{Tr}\bar{a}\bar{a}\left|0\right\rangle  & \left|8\right\rangle  & =\mathrm{Tr}\bar{a}\mathrm{Tr}\bar{a}\mathrm{Tr}\bar{a}\mathrm{Tr}\bar{a}\bar{a}\bar{a}\left|0\right\rangle  & \left|9\right\rangle  & =\mathrm{Tr}\bar{a}\mathrm{Tr}\bar{a}\mathrm{Tr}\bar{a}\bar{a}\mathrm{Tr}\bar{a}\bar{a}\left|0\right\rangle \\
\left|10\right\rangle  & =\mathrm{Tr}\bar{a}\mathrm{Tr}\bar{a}\mathrm{Tr}\bar{a}\mathrm{Tr}\bar{a}\mathrm{Tr}\bar{a}\bar{a}\left|0\right\rangle  & \left|11\right\rangle  & =\mathrm{Tr}\bar{a}\mathrm{Tr}\bar{a}\mathrm{Tr}\bar{a}\mathrm{Tr}\bar{a}\mathrm{Tr}\bar{a}\mathrm{Tr}\bar{a}\left|0\right\rangle  & \left|12\right\rangle  & =\mathrm{Tr}\bar{a}\bar{a}\bar{a}\bar{a}\bar{b}\bar{b}\left|0\right\rangle \\
\left|13\right\rangle  & =\mathrm{Tr}\bar{a}\bar{a}\bar{a}\bar{b}\bar{a}\bar{b}\left|0\right\rangle  & \left|14\right\rangle  & =\mathrm{Tr}\bar{a}\mathrm{Tr}\bar{a}\bar{a}\bar{a}\bar{b}\bar{b}\left|0\right\rangle  & \left|15\right\rangle  & =\mathrm{Tr}\bar{a}\mathrm{Tr}\bar{a}\bar{a}\bar{b}\bar{a}\bar{b}\left|0\right\rangle \\
\left|16\right\rangle  & =\mathrm{Tr}\bar{b}\mathrm{Tr}\bar{a}\bar{a}\bar{a}\bar{a}\bar{b}\left|0\right\rangle  & \left|17\right\rangle  & =\mathrm{Tr}\bar{a}\bar{a}\mathrm{Tr}\bar{a}\bar{a}\bar{b}\bar{b}\left|0\right\rangle  & \left|18\right\rangle  & =\mathrm{Tr}\bar{a}\bar{b}\mathrm{Tr}\bar{a}\bar{a}\bar{a}\bar{b}\left|0\right\rangle \\
\left|19\right\rangle  & =\mathrm{Tr}\bar{a}\bar{a}\bar{a}\mathrm{Tr}\bar{a}\bar{b}\bar{b}\left|0\right\rangle  & \left|20\right\rangle  & =\mathrm{Tr}\bar{a}\mathrm{Tr}\bar{a}\mathrm{Tr}\bar{a}\bar{a}\bar{b}\bar{b}\left|0\right\rangle  & \left|21\right\rangle  & =\mathrm{Tr}\bar{a}\mathrm{Tr}\bar{b}\mathrm{Tr}\bar{a}\bar{a}\bar{a}\bar{b}\left|0\right\rangle \\
\left|22\right\rangle  & =\mathrm{Tr}\bar{a}\mathrm{Tr}\bar{a}\bar{a}\mathrm{Tr}\bar{a}\bar{b}\bar{b}\left|0\right\rangle  & \left|23\right\rangle  & =\mathrm{Tr}\bar{a}\mathrm{Tr}\bar{a}\bar{b}\mathrm{Tr}\bar{a}\bar{a}\bar{b}\left|0\right\rangle  & \left|24\right\rangle  & =\mathrm{Tr}\bar{b}\mathrm{Tr}\bar{a}\bar{a}\mathrm{Tr}\bar{a}\bar{a}\bar{b}\left|0\right\rangle \\
\left|25\right\rangle  & =\mathrm{Tr}\bar{b}\mathrm{Tr}\bar{a}\bar{b}\mathrm{Tr}\bar{a}\bar{a}\bar{a}\left|0\right\rangle  & \left|26\right\rangle  & =\mathrm{Tr}\bar{a}\mathrm{Tr}\bar{a}\mathrm{Tr}\bar{a}\mathrm{Tr}\bar{a}\bar{b}\bar{b}\left|0\right\rangle  & \left|27\right\rangle  & =\mathrm{Tr}\bar{a}\mathrm{Tr}\bar{a}\mathrm{Tr}\bar{b}\mathrm{Tr}\bar{a}\bar{a}\bar{b}\left|0\right\rangle \\
\left|28\right\rangle  & =\mathrm{Tr}\bar{a}\mathrm{Tr}\bar{b}\mathrm{Tr}\bar{a}\bar{a}\mathrm{Tr}\bar{a}\bar{b}\left|0\right\rangle  & \left|29\right\rangle  & =\mathrm{Tr}\bar{a}\mathrm{Tr}\bar{a}\mathrm{Tr}\bar{a}\mathrm{Tr}\bar{b}\mathrm{Tr}\bar{a}\bar{b}\left|0\right\rangle  & \left|30\right\rangle  & =\mathrm{Tr}\bar{a}\bar{a}\bar{b}\bar{b}\bar{b}\bar{b}\left|0\right\rangle \\
\left|31\right\rangle  & =\mathrm{Tr}\bar{a}\bar{b}\bar{a}\bar{b}\bar{b}\bar{b}\left|0\right\rangle  & \left|32\right\rangle  & =\mathrm{Tr}\bar{a}\bar{b}\bar{b}\bar{a}\bar{b}\bar{b}\left|0\right\rangle  & \left|33\right\rangle  & =\mathrm{Tr}\bar{a}\mathrm{Tr}\bar{a}\bar{b}\bar{b}\bar{b}\bar{b}\left|0\right\rangle \\
\left|34\right\rangle  & =\mathrm{Tr}\bar{b}\mathrm{Tr}\bar{a}\bar{a}\bar{b}\bar{b}\bar{b}\left|0\right\rangle  & \left|35\right\rangle  & =\mathrm{Tr}\bar{b}\mathrm{Tr}\bar{a}\bar{b}\bar{a}\bar{b}\bar{b}\left|0\right\rangle  & \left|36\right\rangle  & =\mathrm{Tr}\bar{a}\bar{b}\mathrm{Tr}\bar{a}\bar{b}\bar{b}\bar{b}\left|0\right\rangle \\
\left|37\right\rangle  & =\mathrm{Tr}\bar{a}\bar{a}\bar{b}\mathrm{Tr}\bar{b}\bar{b}\bar{b}\left|0\right\rangle  & \left|38\right\rangle  & =\mathrm{Tr}\bar{a}\bar{b}\bar{b}\mathrm{Tr}\bar{a}\bar{b}\bar{b}\left|0\right\rangle  & \left|39\right\rangle  & =\mathrm{Tr}\bar{a}\mathrm{Tr}\bar{b}\mathrm{Tr}\bar{a}\bar{b}\bar{b}\bar{b}\left|0\right\rangle \\
\left|40\right\rangle  & =\mathrm{Tr}\bar{a}\mathrm{Tr}\bar{a}\bar{b}\mathrm{Tr}\bar{b}\bar{b}\bar{b}\left|0\right\rangle  & \left|41\right\rangle  & =\mathrm{Tr}\bar{b}\mathrm{Tr}\bar{a}\bar{a}\mathrm{Tr}\bar{b}\bar{b}\bar{b}\left|0\right\rangle  & \left|42\right\rangle  & =\mathrm{Tr}\bar{b}\mathrm{Tr}\bar{a}\bar{b}\mathrm{Tr}\bar{a}\bar{b}\bar{b}\left|0\right\rangle \\
\left|43\right\rangle  & =\mathrm{Tr}\bar{a}\mathrm{Tr}\bar{a}\mathrm{Tr}\bar{b}\mathrm{Tr}\bar{b}\bar{b}\bar{b}\left|0\right\rangle  & \left|44\right\rangle  & =\mathrm{Tr}\bar{b}\mathrm{Tr}\bar{b}\bar{b}\bar{b}\bar{b}\bar{b}\left|0\right\rangle 
\end{align*}

\subsection*{7. 7 bits}

Eighty-nine bosonic states:

\begin{align*}
\left|1\right\rangle  & =\mathrm{Tr}\bar{a}\bar{a}\bar{a}\bar{a}\bar{a}\bar{a}\bar{a}\left|0\right\rangle  & \left|2\right\rangle  & =\mathrm{Tr}\bar{a}\mathrm{Tr}\bar{a}\bar{a}\bar{a}\bar{a}\bar{a}\bar{a}\left|0\right\rangle  & \left|3\right\rangle  & =\mathrm{Tr}\bar{a}\bar{a}\mathrm{Tr}\bar{a}\bar{a}\bar{a}\bar{a}\bar{a}\left|0\right\rangle \\
\left|4\right\rangle  & =\mathrm{Tr}\bar{a}\bar{a}\bar{a}\mathrm{Tr}\bar{a}\bar{a}\bar{a}\bar{a}\left|0\right\rangle  & \left|5\right\rangle  & =\mathrm{Tr}\bar{a}\mathrm{Tr}\bar{a}\mathrm{Tr}\bar{a}\bar{a}\bar{a}\bar{a}\bar{a}\left|0\right\rangle  & \left|6\right\rangle  & =\mathrm{Tr}\bar{a}\mathrm{Tr}\bar{a}\bar{a}\mathrm{Tr}\bar{a}\bar{a}\bar{a}\bar{a}\left|0\right\rangle \\
\left|7\right\rangle  & =\mathrm{Tr}\bar{a}\mathrm{Tr}\bar{a}\bar{a}\bar{a}\mathrm{Tr}\bar{a}\bar{a}\bar{a}\left|0\right\rangle  & \left|8\right\rangle  & =\mathrm{Tr}\bar{a}\bar{a}\mathrm{Tr}\bar{a}\bar{a}\mathrm{Tr}\bar{a}\bar{a}\bar{a}\left|0\right\rangle  & \left|9\right\rangle  & =\mathrm{Tr}\bar{a}\mathrm{Tr}\bar{a}\mathrm{Tr}\bar{a}\mathrm{Tr}\bar{a}\bar{a}\bar{a}\bar{a}\left|0\right\rangle \\
\left|10\right\rangle  & =\mathrm{Tr}\bar{a}\mathrm{Tr}\bar{a}\mathrm{Tr}\bar{a}\bar{a}\mathrm{Tr}\bar{a}\bar{a}\bar{a}\left|0\right\rangle  & \left|11\right\rangle  & =\mathrm{Tr}\bar{a}\mathrm{Tr}\bar{a}\bar{a}\mathrm{Tr}\bar{a}\bar{a}\mathrm{Tr}\bar{a}\bar{a}\left|0\right\rangle  & \left|12\right\rangle  & =\mathrm{Tr}\bar{a}\mathrm{Tr}\bar{a}\mathrm{Tr}\bar{a}\mathrm{Tr}\bar{a}\mathrm{Tr}\bar{a}\bar{a}\bar{a}\left|0\right\rangle \\
\left|13\right\rangle  & =\mathrm{Tr}\bar{a}\mathrm{Tr}\bar{a}\mathrm{Tr}\bar{a}\mathrm{Tr}\bar{a}\bar{a}\mathrm{Tr}\bar{a}\bar{a}\left|0\right\rangle  & \left|14\right\rangle  & =\mathrm{Tr}\bar{a}\mathrm{Tr}\bar{a}\mathrm{Tr}\bar{a}\mathrm{Tr}\bar{a}\mathrm{Tr}\bar{a}\mathrm{Tr}\bar{a}\bar{a}\left|0\right\rangle  & \left|15\right\rangle  & =\mathrm{Tr}\bar{a}\mathrm{Tr}\bar{a}\mathrm{Tr}\bar{a}\mathrm{Tr}\bar{a}\mathrm{Tr}\bar{a}\mathrm{Tr}\bar{a}\mathrm{Tr}\bar{a}\left|0\right\rangle \\
\left|16\right\rangle  & =\mathrm{Tr}\bar{a}\bar{a}\bar{a}\bar{a}\bar{a}\bar{b}\bar{b}\left|0\right\rangle  & \left|17\right\rangle  & =\mathrm{Tr}\bar{a}\bar{a}\bar{a}\bar{a}\bar{b}\bar{a}\bar{b}\left|0\right\rangle  & \left|18\right\rangle  & =\mathrm{Tr}\bar{a}\bar{a}\bar{a}\bar{b}\bar{a}\bar{a}\bar{b}\left|0\right\rangle \\
\left|19\right\rangle  & =\mathrm{Tr}\bar{a}\mathrm{Tr}\bar{a}\bar{a}\bar{a}\bar{a}\bar{b}\bar{b}\left|0\right\rangle  & \left|20\right\rangle  & =\mathrm{Tr}\bar{a}\mathrm{Tr}\bar{a}\bar{a}\bar{a}\bar{b}\bar{a}\bar{b}\left|0\right\rangle  & \left|21\right\rangle  & =\mathrm{Tr}\bar{b}\mathrm{Tr}\bar{a}\bar{a}\bar{a}\bar{a}\bar{a}\bar{b}\left|0\right\rangle \\
\left|22\right\rangle  & =\mathrm{Tr}\bar{a}\bar{a}\mathrm{Tr}\bar{a}\bar{a}\bar{a}\bar{b}\bar{b}\left|0\right\rangle  & \left|23\right\rangle  & =\mathrm{Tr}\bar{a}\bar{a}\mathrm{Tr}\bar{a}\bar{a}\bar{b}\bar{a}\bar{b}\left|0\right\rangle  & \left|24\right\rangle  & =\mathrm{Tr}\bar{a}\bar{b}\mathrm{Tr}\bar{a}\bar{a}\bar{a}\bar{a}\bar{b}\left|0\right\rangle \\
\left|25\right\rangle  & =\mathrm{Tr}\bar{a}\bar{a}\bar{a}\mathrm{Tr}\bar{a}\bar{a}\bar{b}\bar{b}\left|0\right\rangle  & \left|26\right\rangle  & =\mathrm{Tr}\bar{a}\bar{a}\bar{b}\mathrm{Tr}\bar{a}\bar{a}\bar{a}\bar{b}\left|0\right\rangle  & \left|27\right\rangle  & =\mathrm{Tr}\bar{a}\bar{b}\bar{b}\mathrm{Tr}\bar{a}\bar{a}\bar{a}\bar{a}\left|0\right\rangle \\
\left|28\right\rangle  & =\mathrm{Tr}\bar{a}\mathrm{Tr}\bar{a}\mathrm{Tr}\bar{a}\bar{a}\bar{a}\bar{b}\bar{b}\left|0\right\rangle  & \left|29\right\rangle  & =\mathrm{Tr}\bar{a}\mathrm{Tr}\bar{a}\mathrm{Tr}\bar{a}\bar{a}\bar{b}\bar{a}\bar{b}\left|0\right\rangle  & \left|30\right\rangle  & =\mathrm{Tr}\bar{a}\mathrm{Tr}\bar{b}\mathrm{Tr}\bar{a}\bar{a}\bar{a}\bar{a}\bar{b}\left|0\right\rangle \\
\left|31\right\rangle  & =\mathrm{Tr}\bar{a}\mathrm{Tr}\bar{a}\bar{a}\mathrm{Tr}\bar{a}\bar{a}\bar{b}\bar{b}\left|0\right\rangle  & \left|32\right\rangle  & =\mathrm{Tr}\bar{a}\mathrm{Tr}\bar{a}\bar{b}\mathrm{Tr}\bar{a}\bar{a}\bar{a}\bar{b}\left|0\right\rangle  & \left|33\right\rangle  & =\mathrm{Tr}\bar{a}\mathrm{Tr}\bar{a}\bar{a}\bar{a}\mathrm{Tr}\bar{a}\bar{b}\bar{b}\left|0\right\rangle \\
\left|34\right\rangle  & =\mathrm{Tr}\bar{b}\mathrm{Tr}\bar{a}\bar{a}\mathrm{Tr}\bar{a}\bar{a}\bar{a}\bar{b}\left|0\right\rangle  & \left|35\right\rangle  & =\mathrm{Tr}\bar{b}\mathrm{Tr}\bar{a}\bar{b}\mathrm{Tr}\bar{a}\bar{a}\bar{a}\bar{a}\left|0\right\rangle  & \left|36\right\rangle  & =\mathrm{Tr}\bar{b}\mathrm{Tr}\bar{a}\bar{a}\bar{a}\mathrm{Tr}\bar{a}\bar{a}\bar{b}\left|0\right\rangle \\
\left|37\right\rangle  & =\mathrm{Tr}\bar{a}\bar{a}\mathrm{Tr}\bar{a}\bar{a}\mathrm{Tr}\bar{a}\bar{b}\bar{b}\left|0\right\rangle  & \left|38\right\rangle  & =\mathrm{Tr}\bar{a}\bar{a}\mathrm{Tr}\bar{a}\bar{b}\mathrm{Tr}\bar{a}\bar{a}\bar{b}\left|0\right\rangle  & \left|39\right\rangle  & =\mathrm{Tr}\bar{a}\mathrm{Tr}\bar{a}\mathrm{Tr}\bar{a}\mathrm{Tr}\bar{a}\bar{a}\bar{b}\bar{b}\left|0\right\rangle \\
\left|40\right\rangle  & =\mathrm{Tr}\bar{a}\mathrm{Tr}\bar{a}\mathrm{Tr}\bar{b}\mathrm{Tr}\bar{a}\bar{a}\bar{a}\bar{b}\left|0\right\rangle  & \left|41\right\rangle  & =\mathrm{Tr}\bar{a}\mathrm{Tr}\bar{a}\mathrm{Tr}\bar{a}\bar{a}\mathrm{Tr}\bar{a}\bar{b}\bar{b}\left|0\right\rangle  & \left|42\right\rangle  & =\mathrm{Tr}\bar{a}\mathrm{Tr}\bar{a}\mathrm{Tr}\bar{a}\bar{b}\mathrm{Tr}\bar{a}\bar{a}\bar{b}\left|0\right\rangle \\
\left|43\right\rangle  & =\mathrm{Tr}\bar{a}\mathrm{Tr}\bar{b}\mathrm{Tr}\bar{a}\bar{a}\mathrm{Tr}\bar{a}\bar{a}\bar{b}\left|0\right\rangle  & \left|44\right\rangle  & =\mathrm{Tr}\bar{a}\mathrm{Tr}\bar{b}\mathrm{Tr}\bar{a}\bar{b}\mathrm{Tr}\bar{a}\bar{a}\bar{a}\left|0\right\rangle  & \left|45\right\rangle  & =\mathrm{Tr}\bar{b}\mathrm{Tr}\bar{a}\bar{a}\mathrm{Tr}\bar{a}\bar{a}\mathrm{Tr}\bar{a}\bar{b}\left|0\right\rangle \\
\left|46\right\rangle  & =\mathrm{Tr}\bar{a}\mathrm{Tr}\bar{a}\mathrm{Tr}\bar{a}\mathrm{Tr}\bar{a}\mathrm{Tr}\bar{a}\bar{b}\bar{b}\left|0\right\rangle  & \left|47\right\rangle  & =\mathrm{Tr}\bar{a}\mathrm{Tr}\bar{a}\mathrm{Tr}\bar{a}\mathrm{Tr}\bar{b}\mathrm{Tr}\bar{a}\bar{a}\bar{b}\left|0\right\rangle  & \left|48\right\rangle  & =\mathrm{Tr}\bar{a}\mathrm{Tr}\bar{a}\mathrm{Tr}\bar{b}\mathrm{Tr}\bar{a}\bar{a}\mathrm{Tr}\bar{a}\bar{b}\left|0\right\rangle \\
\left|49\right\rangle  & =\mathrm{Tr}\bar{a}\mathrm{Tr}\bar{a}\mathrm{Tr}\bar{a}\mathrm{Tr}\bar{a}\mathrm{Tr}\bar{b}\mathrm{Tr}\bar{a}\bar{b}\left|0\right\rangle  & \left|50\right\rangle  & =\mathrm{Tr}\bar{a}\bar{a}\bar{a}\bar{b}\bar{b}\bar{b}\bar{b}\left|0\right\rangle  & \left|51\right\rangle  & =\mathrm{Tr}\bar{a}\bar{a}\bar{b}\bar{a}\bar{b}\bar{b}\bar{b}\left|0\right\rangle \\
\left|52\right\rangle  & =\mathrm{Tr}\bar{a}\bar{a}\bar{b}\bar{b}\bar{a}\bar{b}\bar{b}\left|0\right\rangle  & \left|53\right\rangle  & =\mathrm{Tr}\bar{a}\bar{a}\bar{b}\bar{b}\bar{b}\bar{a}\bar{b}\left|0\right\rangle  & \left|54\right\rangle  & =\mathrm{Tr}\bar{a}\bar{b}\bar{a}\bar{b}\bar{a}\bar{b}\bar{b}\left|0\right\rangle \\
\left|55\right\rangle  & =\mathrm{Tr}\bar{a}\mathrm{Tr}\bar{a}\bar{a}\bar{b}\bar{b}\bar{b}\bar{b}\left|0\right\rangle  & \left|56\right\rangle  & =\mathrm{Tr}\bar{a}\mathrm{Tr}\bar{a}\bar{b}\bar{a}\bar{b}\bar{b}\bar{b}\left|0\right\rangle  & \left|57\right\rangle  & =\mathrm{Tr}\bar{a}\mathrm{Tr}\bar{a}\bar{b}\bar{b}\bar{a}\bar{b}\bar{b}\left|0\right\rangle \\
\left|58\right\rangle  & =\mathrm{Tr}\bar{b}\mathrm{Tr}\bar{a}\bar{a}\bar{a}\bar{b}\bar{b}\bar{b}\left|0\right\rangle  & \left|59\right\rangle  & =\mathrm{Tr}\bar{b}\mathrm{Tr}\bar{a}\bar{a}\bar{b}\bar{a}\bar{b}\bar{b}\left|0\right\rangle  & \left|60\right\rangle  & =\mathrm{Tr}\bar{b}\mathrm{Tr}\bar{a}\bar{a}\bar{b}\bar{b}\bar{a}\bar{b}\left|0\right\rangle \\
\left|61\right\rangle  & =\mathrm{Tr}\bar{b}\mathrm{Tr}\bar{a}\bar{b}\bar{a}\bar{b}\bar{a}\bar{b}\left|0\right\rangle  & \left|62\right\rangle  & =\mathrm{Tr}\bar{a}\bar{a}\mathrm{Tr}\bar{a}\bar{b}\bar{b}\bar{b}\bar{b}\left|0\right\rangle  & \left|63\right\rangle  & =\mathrm{Tr}\bar{a}\bar{b}\mathrm{Tr}\bar{a}\bar{a}\bar{b}\bar{b}\bar{b}\left|0\right\rangle \\
\left|64\right\rangle  & =\mathrm{Tr}\bar{a}\bar{b}\mathrm{Tr}\bar{a}\bar{b}\bar{a}\bar{b}\bar{b}\left|0\right\rangle  & \left|65\right\rangle  & =\mathrm{Tr}\bar{a}\bar{a}\bar{b}\mathrm{Tr}\bar{a}\bar{b}\bar{b}\bar{b}\left|0\right\rangle  & \left|66\right\rangle  & =\mathrm{Tr}\bar{a}\bar{b}\bar{b}\mathrm{Tr}\bar{a}\bar{a}\bar{b}\bar{b}\left|0\right\rangle \\
\left|67\right\rangle  & =\mathrm{Tr}\bar{b}\bar{b}\bar{b}\mathrm{Tr}\bar{a}\bar{a}\bar{a}\bar{b}\left|0\right\rangle  & \left|68\right\rangle  & =\mathrm{Tr}\bar{a}\mathrm{Tr}\bar{a}\mathrm{Tr}\bar{a}\bar{b}\bar{b}\bar{b}\bar{b}\left|0\right\rangle  & \left|69\right\rangle  & =\mathrm{Tr}\bar{a}\mathrm{Tr}\bar{b}\mathrm{Tr}\bar{a}\bar{a}\bar{b}\bar{b}\bar{b}\left|0\right\rangle \\
\left|70\right\rangle  & =\mathrm{Tr}\bar{a}\mathrm{Tr}\bar{b}\mathrm{Tr}\bar{a}\bar{b}\bar{a}\bar{b}\bar{b}\left|0\right\rangle  & \left|71\right\rangle  & =\mathrm{Tr}\bar{a}\mathrm{Tr}\bar{a}\bar{b}\mathrm{Tr}\bar{a}\bar{b}\bar{b}\bar{b}\left|0\right\rangle  & \left|72\right\rangle  & =\mathrm{Tr}\bar{a}\mathrm{Tr}\bar{a}\bar{a}\bar{b}\mathrm{Tr}\bar{b}\bar{b}\bar{b}\left|0\right\rangle \\
\left|73\right\rangle  & =\mathrm{Tr}\bar{a}\mathrm{Tr}\bar{a}\bar{b}\bar{b}\mathrm{Tr}\bar{a}\bar{b}\bar{b}\left|0\right\rangle  & \left|74\right\rangle  & =\mathrm{Tr}\bar{b}\mathrm{Tr}\bar{a}\bar{a}\mathrm{Tr}\bar{a}\bar{b}\bar{b}\bar{b}\left|0\right\rangle  & \left|75\right\rangle  & =\mathrm{Tr}\bar{b}\mathrm{Tr}\bar{a}\bar{b}\mathrm{Tr}\bar{a}\bar{a}\bar{b}\bar{b}\left|0\right\rangle \\
\left|76\right\rangle  & =\mathrm{Tr}\bar{b}\mathrm{Tr}\bar{a}\bar{a}\bar{a}\mathrm{Tr}\bar{b}\bar{b}\bar{b}\left|0\right\rangle  & \left|77\right\rangle  & =\mathrm{Tr}\bar{b}\mathrm{Tr}\bar{a}\bar{a}\bar{b}\mathrm{Tr}\bar{a}\bar{b}\bar{b}\left|0\right\rangle  & \left|78\right\rangle  & =\mathrm{Tr}\bar{a}\bar{a}\mathrm{Tr}\bar{a}\bar{b}\mathrm{Tr}\bar{b}\bar{b}\bar{b}\left|0\right\rangle \\
\left|79\right\rangle  & =\mathrm{Tr}\bar{a}\mathrm{Tr}\bar{a}\mathrm{Tr}\bar{b}\mathrm{Tr}\bar{a}\bar{b}\bar{b}\bar{b}\left|0\right\rangle  & \left|80\right\rangle  & =\mathrm{Tr}\bar{a}\mathrm{Tr}\bar{a}\mathrm{Tr}\bar{a}\bar{b}\mathrm{Tr}\bar{b}\bar{b}\bar{b}\left|0\right\rangle  & \left|81\right\rangle  & =\mathrm{Tr}\bar{a}\mathrm{Tr}\bar{b}\mathrm{Tr}\bar{a}\bar{a}\mathrm{Tr}\bar{b}\bar{b}\bar{b}\left|0\right\rangle \\
\left|82\right\rangle  & =\mathrm{Tr}\bar{a}\mathrm{Tr}\bar{b}\mathrm{Tr}\bar{a}\bar{b}\mathrm{Tr}\bar{a}\bar{b}\bar{b}\left|0\right\rangle  & \left|83\right\rangle  & =\mathrm{Tr}\bar{a}\mathrm{Tr}\bar{a}\mathrm{Tr}\bar{a}\mathrm{Tr}\bar{b}\mathrm{Tr}\bar{b}\bar{b}\bar{b}\left|0\right\rangle  & \left|84\right\rangle  & =\mathrm{Tr}\bar{a}\bar{b}\bar{b}\bar{b}\bar{b}\bar{b}\bar{b}\left|0\right\rangle \\
\left|85\right\rangle  & =\mathrm{Tr}\bar{b}\mathrm{Tr}\bar{a}\bar{b}\bar{b}\bar{b}\bar{b}\bar{b}\left|0\right\rangle  & \left|86\right\rangle  & =\mathrm{Tr}\bar{a}\bar{b}\mathrm{Tr}\bar{b}\bar{b}\bar{b}\bar{b}\bar{b}\left|0\right\rangle  & \left|87\right\rangle  & =\mathrm{Tr}\bar{b}\bar{b}\bar{b}\mathrm{Tr}\bar{a}\bar{b}\bar{b}\bar{b}\left|0\right\rangle \\
\left|88\right\rangle  & =\mathrm{Tr}\bar{a}\mathrm{Tr}\bar{b}\mathrm{Tr}\bar{b}\bar{b}\bar{b}\bar{b}\bar{b}\left|0\right\rangle  & \left|89\right\rangle  & =\mathrm{Tr}\bar{b}\mathrm{Tr}\bar{a}\bar{b}\bar{b}\mathrm{Tr}\bar{b}\bar{b}\bar{b}\left|0\right\rangle 
\end{align*}

\section{\label{app:Counting-Trace-States}Counting problems on trace states}

How many trace states are there for a fixed bit number $M$? In this
Appendix, we will first count the single trace states and then the
trace states which includes both single and multiple trace states.

\subsection*{1. Counting single trace states}

There are $2^{M}$ combinations of an $M$-bit string consisting of
$\ba,\bb$. By the property of trace, a trace state is equivalent
to its cyclic permutations. For example, $\tr\bar{b}\bar{a}$ and
$\tr\ba\bb$ are equivalent states, and so are $\tr\bar{b}\bar{a}\bar{a}\bar{b}$
and $\tr\ba\ba\bb\bb$. Actually, the latter case differs by a negative
sign,
\[
\tr\ba\ba\bb\bb=-\tr\bb\ba\ba\bb.
\]
The rule is that each swap of two $\bar{b}$ introduces a minus sign.
It follows that some trace states are vanishing, for example, $\tr\bar{b}\bar{b}=-\tr\bar{b}\bar{b}=0.$

To count the single trace states, we need the following definition
and theorem \cite{rotman1999introduction}. 
\begin{defn*}
Given a group $G$ acting on a set $X$, the orbit of $x\in X$ is
the set $Gx=\left\{ g\cdot x|g\in G\right\} $. The set of orbits
is denoted by $X/G$. 
\end{defn*}
In our case, the cyclic group $C_{M}$ is the group $G$. $X$ is
the $2^{M}$ combinations of $M$-bit operators, and $x$ corresponds
to one particular combination. $X/G$ is the set of different combinations
under the action of the cyclic group. 
\begin{thm*}
(Burnside\textquoteright{}s counting theorem). \textemdash{}If $G$
is a finite group acting on a finite set $X$, then 
\[
\left|X/G\right|=\frac{1}{\left|G\right|}\sum_{g\in G}\left|\mathrm{Fix}\left(g\right)\right|,
\]
 where $\mathrm{Fix}\left(g\right)$ is the set of $x$ that is invariant
under action of $g$, i.e., 
\[
\mathrm{Fix}\left(g\right)=\left\{ x\in X|g\cdot x=x\right\} .
\]
 
\end{thm*}
To find the number of states, we need to find $\left|\mathrm{Fix}\left(g\right)\right|$
for each group member.

We first consider the odd $M$ case. Let $c_{k}\in C_{M},\, k=1,2,\cdots M$,
be the group member that shifts $k$ operators from the tail of the
trace to the beginning. The identity of the group is $e=c_{M}$. Let
$\left(M,k\right)$ denote the greatest common divisor of $M$ and
$k$. For group member $c_{k}$, we equally partition the $M$ bits
into $M/\left(M,k\right)$ consecutive parts: the first part starts
from bit 1 to bit $\left(M,k\right)$, the second part starts from
bit $\left(M,k\right)+1$ to bit $2\left(M,k\right)$, etc. Under
the action of $c_{k}$, the $i$th part transfers as 
\[
i\mathrm{th}\text{ part}\to\left(i+\frac{k}{\left(M,k\right)}\right)\mathrm{\text{th part}}.
\]
The trace is invariant under $c_{k}$ if and only if all the parts
are identical to each other. For bosonic trace states, each part need
to bosonic, from which it follows that 
\begin{equation}
\left|\mathrm{Fix}\left(c_{k}\right)\right|=\sum_{\text{even }i}\binom{\left(M,k\right)}{i}=\frac{1}{2}2^{\left(M,k\right)}.\label{eq:A-Fix-1}
\end{equation}
 Similarly, for fermionic single trace states, each part needs to
be fermionic, 
\begin{equation}
\left|\mathrm{Fix}\left(c_{k}\right)\right|=\sum_{\text{odd }i}\binom{\left(M,k\right)}{i}=\frac{1}{2}2^{\left(M,k\right)},\label{eq:A-Fix-2}
\end{equation}
which implies there is the same number of bosonic and fermionic single
trace states for odd $M$. By Burnside's theorem, this number is given
by 
\begin{equation}
S_{M}=\frac{1}{2M}\sum_{k=1}^{M}2^{\left(M,k\right)}.\label{eq:A-SM-Odd}
\end{equation}

For even $M$, let us first consider the fermionic states. For a group
member $c_{k}$, $\left|\mathrm{Fix}\left(c_{k}\right)\right|=0$
if $M/\left(M,k\right)$ is even. The reason is that an odd number
of $\bar{b}$ cannot be equally partitioned into even parts. Therefore,
only odd $M/\left(M,k\right)$ contributes to $\left|\mathrm{Fix}\left(c_{k}\right)\right|$,
which is still given by (\ref{eq:A-Fix-2}). And Eq. (\ref{eq:A-SM-Odd})
becomes
\begin{equation}
S_{M}=\frac{1}{2M}\sum_{M/\left(M,k\right)\text{ is odd}}2^{\left(M,k\right)}.\label{eq:A-SM-Even}
\end{equation}
Let $i=M/\left(M,k\right)$; Eq. (\ref{eq:A-SM-Even}) can be written
as 
\begin{equation}
S_{M}=\frac{1}{2M}\sum_{\mathrm{odd}\, i,i|M}\varphi\left(i\right)2^{\frac{M}{i}},\label{eq:A-SM-Euler}
\end{equation}
where $\varphi\left(i\right)$ is the Euler totient function and $i|M$
means $M$ is divisible by $i$. We see that Eq. (\ref{eq:A-SM-Odd})
can also be written as Eqs. (\ref{eq:A-SM-Even}) and (\ref{eq:A-SM-Euler}). 

For bosonic states, because there exist vanishing states, like $\tr\ba\bb\ba\bb=-\tr\ba\bb\ba\bb=0$,
the number of bosonic states equals the number of even-$\bar{b}$
state minus the number of vanishing states. Consider the number of
even-$\bar{b}$-states, which is denoted as $B_{M}$ for convenience.
For a group member $c_{k}$, we partition $M$ bits equally into $M/\left(M,k\right)$
consecutive parts with each part $\left(M,k\right)$ bits: if $M/\left(M,k\right)$
is odd, we need even number of $\bar{b}$ in each part; if $M/\left(M,k\right)$
is even, there can be any number of $\bar{b}$ in each part, from
which it follows that
\begin{eqnarray}
B_{M} & = & \frac{1}{M}\left(\sum_{M/\left(M,k\right)\text{ is odd}}2^{\left(M,k\right)-1}+\sum_{M/\left(M,k\right)\text{ is even}}2^{\left(M,k\right)}\right)\nonumber \\
 & = & \frac{1}{2M}\left(\sum_{\mathrm{odd}\, i,i|M}\varphi\left(i\right)2^{\frac{M}{i}}+2\sum_{\mathrm{even}\, i,i|M}\varphi\left(i\right)2^{\frac{M}{i}}\right).\label{eq:A-Even-b-States}
\end{eqnarray}
Now, consider the number of vanishing states, which is denoted as
$V_{M}$. For each $c_{k}$, we again partition $M$ bits into $M/\left(M,k\right)$
consecutive parts. If $M/\left(M,k\right)$ is even and all parts
are identical with an odd number of $\bar{b}$, then it is a vanishing
state. But this does not cover all the possibilities. If $\left(M,k\right)$
is even, we can perform finer partition: divide $M$-bits into $2M/\left(M,k\right)$
parts with each part of $\left(M,k\right)/2$ bits. If all the $2M/\left(M,k\right)$
parts are the same and contain an odd number of $\bar{b}$, it is
a vanishing state. We can continue to perform the finer partition
$i$ times until $\left(M,k\right)/2^{i}$ is odd. There is a difference
between odd $M/\left(M,k\right)$ and even $M/\left(M,k\right)$:
it needs to perform at least one finer partition for odd $M/\left(M,k\right)$,
while for even $M/\left(M,k\right)$ it does not. Therefore, the number
of vanishing states reads 
\begin{eqnarray}
V_{M} & = & \frac{1}{M}\sum_{\text{odd }M/\left(M,k\right)}\left(\sum_{i\geq1\text{ and }2^{i}|\left(M,k\right)}2^{\frac{\left(M,k\right)}{2^{i}}-1}\right)\nonumber \\
 &  & +\frac{1}{M}\sum_{\text{even }M/\left(M,k\right)}\left(\sum_{i\geq0\text{ and }2^{i}|\left(M,k\right)}2^{\frac{\left(M,k\right)}{2^{i}}-1}\right)\nonumber \\
 & = & \frac{1}{2M}\left[\sum_{k}\left(\sum_{i\geq1,2^{i}|\left(M,k\right)}2^{\frac{\left(M,k\right)}{2^{i}}}\right)+\sum_{\text{even }M/\left(M,k\right)}2^{\left(M,k\right)}\right]\nonumber \\
 & = & \frac{1}{2M}\left[\sum_{k}\left(\sum_{i\geq1,2^{i}|\left(M,k\right)}2^{\frac{\left(M,k\right)}{2^{i}}}\right)+\sum_{\mathrm{even}\, i,i|M}\varphi\left(i\right)2^{\frac{M}{i}}\right]\label{eq:A-Vanish-States-1}
\end{eqnarray}
 Let $\left(M,k\right)/2^{i}=\frac{M}{j}$; then we have $2^{i}|j$
and $\left(M,k\right)=2^{i}M/j$. The number of $k$ satisfying $\left(M,k\right)=2^{i}M/j$
is equal to 
\[
\varphi\left(\frac{M}{2^{i}M/j}\right)=\varphi\left(\frac{j}{2^{i}}\right).
\]
Now, the first term inside the parentheses of Eq. (\ref{eq:A-Vanish-States-1})
can be written as
\begin{equation}
\sum_{k}\left(\sum_{i\geq1,2^{i}|\left(M,k\right)}2^{\frac{\left(M,k\right)}{2^{i}}}\right)=\sum_{\text{ even }j,j|M}\left[\sum_{i\geq1,2^{i}|j}\varphi\left(\frac{j}{2^{i}}\right)2^{\frac{M}{j}}\right].\label{eq:A-Vanish-States-2}
\end{equation}
 With the following property of the function $\varphi$, 
\[
\varphi\left(2m\right)=\begin{cases}
2\varphi\left(m\right) & \quad\text{if }m\text{ is even}\\
\varphi\left(m\right) & \quad\text{if }m\text{ is odd}
\end{cases},
\]
we see that 
\[
\sum_{i\geq1,2^{i}|j}\varphi\left(\frac{j}{2^{i}}\right)=\varphi\left(j\right),\quad\text{if }j\text{ is even},
\]
Now Eq.(\ref{eq:A-Vanish-States-2}) becomes 
\[
\sum_{k}\left(\sum_{i\geq1,2^{i}|\left(M,k\right)}2^{\frac{\left(M,k\right)}{2^{i}}}\right)=\sum_{\text{even }j,j|M}\varphi\left(j\right)2^{\frac{M}{j}},
\]
 from which it follows that 
\[
V_{M}=\frac{1}{M}\sum_{\mathrm{even}\, i,i|M}\varphi\left(i\right)2^{\frac{M}{i}}.
\]

The difference of Eqs. (\ref{eq:A-Even-b-States}) and (\ref{eq:A-Vanish-States-1})
is 
\[
S_{M}=B_{M}-V_{M}=\frac{1}{2M}\sum_{\mathrm{odd}\, i,i|M}\varphi\left(i\right)2^{\frac{M}{i}},
\]
 which is the same as the formula for fermionic states. 

In summation, we conclude that there is an equal number of bosonic
and fermionic states for a given bit number $M$ and both can be written
as 
\begin{equation}
S_{M}=\frac{1}{2M}\sum_{\mathrm{odd}\, n,n|M}\varphi\left(n\right)2^{\frac{M}{n}}.\label{eq:Single-Trace-Number}
\end{equation}

\subsection*{2. Counting trace states}

Now, consider the general trace states, including single and multiple
trace states. Let $T_{m,r}^{\left(0\right)}$ be the number of $r$-bit
bosonic trace states built out of single trace states of bits less
than or equal to $m$. $T_{m,r}^{\left(1\right)}$ is defined similarly
for fermionic trace states. We can build the recursive relation of
$T_{m,r}^{\left(b\right)}$ as follows. Out of $r$ string bits, we
can assign $i\times m$ bits to $i$ bosonic $m$-bit single trace
states and $j\times m$ bits to $j$ fermionic $m$-bit single trace
states provided $\left(i+j\right)m\leq r$. There are $\binom{S_{m}}{i}$
ways to $ $pick $i$ fermionic $m$-bit single trace states and $\binom{S_{m}+j-1}{j}$
ways to pick $j$ bosonic $m$-bit single trace states. The remaining
$r-\left(i+j\right)m$ bits need to be built out of single trace states
of bits less than $m$. Summation over all non-negative $i$, $j$
yields 
\begin{equation}
T_{m,r}^{\left(b\right)}=\sum_{\left(i+j\right)m\leq r}\binom{S_{m}}{i}\binom{S_{m}+j-1}{j}T_{m-1,r-\left(i+j\right)m}^{\left(\left(b+i\right)\mod2\right)}.\label{eq:Multiple-Trace-States}
\end{equation}

We can actually drop the superscript of $T$ because $T_{m,r}^{\left(0\right)}$
equals $T_{m,r}^{\left(0\right)}$ for all $m,r$. It can be proved
by mathematical induction that for $m=1$ the only $r$-bit bosonic
state is $\left(\tr\bar{a}\right)^{r}\ket{0}$ and the only $r$-bit
fermionic state is $\left(\tr\bar{a}\right)^{r-1}\tr\bar{b}\ket{0}$,
which implies $T_{1,r}^{\left(0\right)}=T_{1,r}^{\left(1\right)}$.
If $T_{m-1,r}^{\left(0\right)}=T_{m-1,r}^{\left(1\right)}$ holds
for all $r$, then Eq. (\ref{eq:Multiple-Trace-States}) gives the
same result for $T_{m,r}^{\left(0\right)}$ and $T_{m,r}^{\left(0\right)}$,
from which it follows that $T_{m,r}^{\left(0\right)}=T_{m,r}^{\left(0\right)}$
holds for all values of $m,r$. Therefore, we can simply write (\ref{eq:Multiple-Trace-States})
as 
\begin{equation}
T_{m,r}=\sum_{\left(i+j\right)m\leq r}\binom{S_{m}}{i}\binom{S_{m}+j-1}{j}T_{m-1,r}.\label{eq:Multiple-trace-states-2}
\end{equation}
The number of $M$-bit bosonic or fermionic trace states is simply
\begin{equation}
T_{M}=T_{M,M}.\label{eq:Multiple-trace-states-3}
\end{equation}

We use a computer program to calculate the values of $S_{M}$ and
$T_{M}$, as shown in Table \ref{tab:Number-of-trace-states}. The
results reveal that when $M$ is large 
\[
S_{M}\to\frac{2^{M-1}}{M},\qquad T_{M}\to\left(0.7261768212\cdots\right)\times2^{M}.
\]
The limit of $S_{M}$ shows that almost all the single trace states
have $M$ different cyclic permutations when $M$ is large. This is
not surprising: the density of the single trace with certain cyclic
symmetry goes down as $M$ increases. $T_{M}$ increases as $2^{M}$
with a magic prefactor we do not understand, which could be an interesting
mathematical problem to explore.

\begin{center}
\begin{table}
\begin{centering}
\begin{tabular}{|c|c|c|c|c|}
\hline 
M & $S_{M}$ & $T_{M}$ & $S_{M}\times\nicefrac{M}{2^{M}}$ & $\nicefrac{T_{M}}{2^{M}}$\tabularnewline
\hline 
\hline 
1 & 1 & 1 & 0.500000000000  & 0.500000000000 \tabularnewline
\hline 
2 & 1 & 2 & 0.500000000000  & 0.500000000000 \tabularnewline
\hline 
3 & 2 & 5 & 0.750000000000  & 0.625000000000 \tabularnewline
\hline 
4 & 2 & 10 & 0.500000000000  & 0.625000000000 \tabularnewline
\hline 
5 & 4 & 21 & 0.625000000000  & 0.656250000000 \tabularnewline
\hline 
6 & 6 & 44 & 0.562500000000  & 0.687500000000 \tabularnewline
\hline 
7 & 10 & 89 & 0.546875000000  & 0.695312500000 \tabularnewline
\hline 
8 & 16 & 180 & 0.500000000000  & 0.703125000000 \tabularnewline
\hline 
9 & 30 & 365 & 0.527343750000  & 0.712890625000 \tabularnewline
\hline 
10 & 52 & 734 & 0.507812500000  & 0.716796875000 \tabularnewline
\hline 
11 & 94 & 1473 & 0.504882812500  & 0.719238281250 \tabularnewline
\hline 
20 & 26216 & 761282  & 0.500030517578  & 0.726015090942 \tabularnewline
\hline 
30 & 17895736  & 779724424  & 0.500001087785  & 0.726174958050 \tabularnewline
\hline 
40 & 13743895360  & 798439834644  & 0.500000000466  & 0.726176799293 \tabularnewline
\hline 
50 & 11258999068468  & 817602415099946  & 0.500000000001  & 0.726176820986 \tabularnewline
\hline 
60 & 9607679205074672  & 837224873334502342  & 0.500000000001  & 0.726176821223 \tabularnewline
\hline 
\end{tabular}
\par\end{centering}

\caption{\label{tab:Number-of-trace-states}Number of trace states}
\end{table}

\par\end{center}

\section{Rank of norm matrix}

The rank of norm matrix $G_{ij}=\dprod{i}{j}$ is the dimension of
the trace state space and also the number of energy levels of the
system. In this section, we show some interesting patterns of the
rank of norm matrix. We only focus on the norm matrix of $M$-bit
bosonic trace states, which is a $T_{M}\times T_{M}$ real symmetric
matrix. By supersymmetry, the norm matrix of $M$-bit fermionic trace
state space has the same rank as the one of $M$-bit bosonic trace
state space.

We generate the norm matrices for $M\leq11$ and calculate their ranks
numerically. We find that when $N\geq M$ $G$ has full rank and when
$N<M$ it is rank deficient. As $N$ changes from $M$ to $1$, the
rank of $G$ changes from $T_{M}$ to $1$. We arrange the ranks of
norm matrices for $M\leq11$ and $N\leq M$ as a number triangle as
below:

\begin{center}
\begin{gather*}
1 \\
1 \quad 2 \\   
1 \quad 4 \quad 5 \\
1\quad 6 \quad 9 \quad 10 \\    
1\quad 8\quad 17\quad 20\quad 21 \\   
1\quad 10 \quad 31 \quad 40 \quad 43\quad 44 \\
1\quad\, 12 \quad 49 \quad\, 76 \quad\, 85\quad 88\quad 89 \\
1\quad 14 \quad 75 \quad 140 \quad 167\quad 176 \quad 179 \quad 180\\
1\quad 16 \quad 109 \quad 252 \quad 325\quad 352 \quad 361 \quad 364 \quad 365\\
1\quad 18 \quad 147 \quad 436 \quad 621\quad 694 \quad 721 \quad 730 \quad 733 \quad 734\\
1\quad 20 \quad 193 \quad 724 \;\; 1165\;\; 1360\;\; 1433 \;\; 1460 \;\; 1469 \;\; 1472 \;\; 1473
\end{gather*} 
\par\end{center}

The number at the $i$ th row and $j$ th column is the rank of $G$
for $M=i$ and $N=j$. For convenience, we denote it as $R_{i,j}$.
We immediately see several patterns: $R_{M,M}=T_{M}$, $R_{M,M-1}=T_{M}-1$,
$R_{M,1}=1$, and for $M$ greater than 1, $R_{M,2}=2M-2$. If we
define $R_{i,0}=0$, then we can define new variables $D_{i,j}=R_{i,j}-R_{i,j-1}$,
which represent the change of $G$'s rank when $M=i$ and $N$ change
from $j$ to $j-1$. We arrange $D_{ij}$ as another number triangle
as below:

\begin{center}
\begin{gather*}
1 \\   
1 \quad 1 \\   
1 \quad 3 \quad 1 \\    
1\quad 5 \quad 3 \quad 1 \\    
1\quad 7\quad 9\quad 3\quad 1 \\   
1\quad 9 \quad 21 \quad 9 \quad 3\quad 1 \\
1\quad 11 \quad 37 \quad 27 \quad 9\quad 3\quad 1 \\
1\quad 13 \quad 61 \quad 65 \quad 27\quad 9 \quad 3 \quad 1\\
1\quad 15 \quad 93 \quad 143 \quad 73\quad 27 \quad 9 \quad 3 \quad 1\\
1\quad 17 \quad 129 \quad 289 \quad 185\quad 73 \quad 27 \quad 9 \quad 3 \quad 1\\
1\quad 19 \quad 173 \quad 531 \quad 441\quad 195 \quad 73 \quad 27 \quad 9 \quad 3 \quad 1
\end{gather*} 
\par\end{center}

Going through each row from right to left, we find the following sequence:
\[
1,\quad3,\quad9,\quad27,\quad73,\quad195,\cdots.
\]
 For odd $M$, the sequence starts from $N=M$ and ends at $N=\left(M+1\right)/2$;
for even $M$, the sequence starts from $N=M$ and ends at $N=M/2$.
This means that, no matter what the value $M$ is, the changes of
$G$'s rank from $N$ to $N-1$ for $N\geq M$ are the same. 

Since we only obtain the norm matrices for $M\leq11$, we do not know
the next number of the sequence. Finding the pattern of the sequence
is an interesting problem for future research.

\section{\label{App:Calculation-of-HQ}Calculation of $\left[H,Q\right]$}

In this section, let us find the constraint of the supersymmetric
Hamiltonian, i.e., the condition for $\left[H,Q\right]=0$, where
\[
Q=\exp\left(\frac{i\pi}{4}\right)\tr\bar{a}b+\exp\left(-\frac{i\pi}{4}\right)\tr\bar{b}a.
\]

We first calculate the commutation between $Q$ and each trace operator
in (\ref{eq:trace-operators}). We have 
\[
\left[\mathrm{Tr}\bar{a}^{2}b^{2},\mathrm{Tr}\bar{a}b\right]=\mathrm{Tr}\bar{a}^{2}b^{2}\mathrm{Tr}\bar{a}b-\mathrm{Tr}\bar{a}b\mathrm{Tr}\bar{a}^{2}b^{2}=0,
\]

\begin{eqnarray*}
\left[\mathrm{Tr}\bar{a}^{2}b^{2},\tr\bar{b}a\right] & = & \mathrm{Tr}\bar{a}^{2}b^{2}\mathrm{Tr}\bar{b}a-\mathrm{Tr}\bar{b}a\mathrm{Tr}\bar{a}^{2}b^{2}\\
 & = & \mathrm{Tr}\bar{a}^{2}ba-\mathrm{Tr}b\bar{a}^{2}a+\normord{\mathrm{Tr}\bar{a}^{2}b^{2}\mathrm{Tr}\bar{b}a}\\
 &  & -\left(\mathrm{Tr}\bar{b}\bar{a}b^{2}+\tr\bar{b}b^{2}\bar{a}+\normord{\mathrm{Tr}\bar{a}^{2}b^{2}\mathrm{Tr}\bar{b}a}\right)\\
 & = & \tr\bar{a}^{2}\left(ba-ab\right)-\tr\left(\bar{b}\bar{a}+\bar{a}\bar{b}\right)b^{2},
\end{eqnarray*}
 where $\normord{\mathrm{Tr}\bar{a}^{2}b^{2}\mathrm{Tr}\bar{b}a}$
denotes the normal ordering of $\mathrm{Tr}\bar{a}^{2}b^{2}\mathrm{Tr}\bar{b}a$.
As we see, the normal ordering terms cancel out. This occurs for all
the trace operators. So in the following calculation, we simply drop
the normal ordering terms in most cases. From above two results, it
follows that 
\begin{eqnarray*}
\left[\mathrm{Tr}\bar{a}^{2}b^{2},Q\right] & = & \exp\left(-i\frac{\pi}{4}\right)\left[\mathrm{Tr}\bar{a}^{2}b^{2},\tr\bar{b}a\right]\\
 & = & \exp\left(-i\frac{\pi}{4}\right)\left[\tr\bar{a}^{2}\left(ba-ab\right)-\tr\left(\bar{b}\bar{a}+\bar{a}\bar{b}\right)b^{2}\right].
\end{eqnarray*}
 We repeat the calculation for the other trace operators as follows:

\begin{eqnarray*}
\left[\tr\bar{b}^{2}a^{2},\tr\bar{a}b\right] & = & \tr\bar{b}^{2}a^{2}\tr\bar{a}b-\tr\bar{a}b\tr\bar{b}^{2}a^{2}\\
 & = & \tr\bar{b}^{2}ab+\tr a\bar{b}^{2}b-\tr\bar{a}\bar{b}a^{2}+\tr\bar{a}a^{2}\bar{b}\\
 & = & \tr\bar{b}^{2}\left(ab+ba\right)+\tr\left(\bar{b}\bar{a}-\bar{a}\bar{b}\right)a^{2},
\end{eqnarray*}
\[
\left[\tr\bar{b}^{2}a^{2},\tr\bar{b}a\right]=0,
\]
 from which it follows that
\begin{eqnarray}
\left[\tr\bar{b}^{2}a^{2},Q\right] & = & \exp\left(i\frac{\pi}{4}\right)\left[\tr\bar{b}^{2}a^{2},\tr\bar{a}b\right]\nonumber \\
 & = & \exp\left(i\frac{\pi}{4}\right)\left[\tr\bar{b}^{2}\left(ab+ba\right)+\tr\left(\bar{b}\bar{a}-\bar{a}\bar{b}\right)a^{2}\right].
\end{eqnarray}
 
\begin{eqnarray*}
\left[\tr\bar{a}^{2}a^{2},\tr\bar{a}b\right] & = & \tr\bar{a}^{2}a^{2}\tr\bar{a}b-\tr\bar{a}b\tr\bar{a}^{2}a^{2}\\
 & = & \tr\bar{a}^{2}ab+\tr\bar{a}^{2}ba+\normord{\tr\bar{a}^{2}a^{2}\tr\bar{a}b}-\tr\bar{a}b\tr\bar{a}^{2}a^{2}\\
 & = & \tr\bar{a}^{2}\left(ab+ba\right),
\end{eqnarray*}
\begin{eqnarray*}
\left[\tr\bar{a}^{2}a^{2},\tr\bar{b}a\right] & = & \tr\bar{a}^{2}a^{2}\tr\bar{b}a-\tr\bar{b}a\tr\bar{a}^{2}a^{2}\\
 & = & \tr\bar{a}^{2}a^{2}\tr\bar{b}a-\tr\bar{b}\bar{a}a^{2}-\tr\bar{a}\bar{b}a^{2}-\normord{\tr\bar{b}a\tr\bar{a}^{2}a^{2}}\\
 & = & -\tr\left(\bar{a}\bar{b}+\bar{b}\bar{a}\right)a^{2},
\end{eqnarray*}
from which it follows that
\begin{eqnarray}
\left[\tr\bar{a}^{2}a^{2},Q\right] & = & \exp\left(i\frac{\pi}{4}\right)\left[\tr\bar{a}^{2}a^{2},\tr\bar{a}b\right]+\exp\left(-i\frac{\pi}{4}\right)\left[\tr\bar{a}^{2}a^{2},\tr\bar{b}a\right]\nonumber \\
 & = & \exp\left(i\frac{\pi}{4}\right)\tr\bar{a}^{2}\left(ab+ba\right)-\exp\left(-i\frac{\pi}{4}\right)\tr\left(\bar{a}\bar{b}+\bar{b}\bar{a}\right)a^{2}.
\end{eqnarray}
\begin{eqnarray*}
\left[\tr\bar{b}^{2}b^{2},\tr\bar{a}b\right] & = & \tr\bar{b}^{2}b^{2}\tr\bar{a}b-\tr\bar{a}b\tr\bar{b}^{2}b^{2}\\
 & = & \tr\bar{b}^{2}b^{2}\tr\bar{a}b-\tr\bar{a}\bar{b}b^{2}+\tr\bar{b}\bar{a}b^{2}-\normord{\tr\bar{a}b\tr\bar{b}^{2}b^{2}}\\
 & = & \tr\left(\bar{b}\bar{a}-\bar{a}\bar{b}\right)b^{2},
\end{eqnarray*}
\begin{eqnarray*}
\left[\tr\bar{b}^{2}b^{2},\tr\bar{b}a\right] & = & \tr\bar{b}^{2}b^{2}\tr\bar{b}a-\tr\bar{b}a\tr\bar{b}^{2}b^{2}\\
 & = & \tr\bar{b}^{2}ba-\tr\bar{b}^{2}ab+\normord{\tr\bar{b}^{2}b^{2}\tr\bar{b}a}-\tr\bar{b}a\tr\bar{b}^{2}b^{2}\\
 & = & \tr\bar{b}^{2}\left(ba-ab\right),
\end{eqnarray*}
 from which it follows that
\begin{eqnarray}
\left[\tr\bar{b}^{2}b^{2},Q\right] & = & \exp\left(i\frac{\pi}{4}\right)\left[\tr\bar{b}^{2}b^{2},\tr\bar{a}b\right]+\exp\left(-i\frac{\pi}{4}\right)\left[\tr\bar{b}^{2}b^{2},\tr\bar{b}a\right]\nonumber \\
 & = & \exp\left(i\frac{\pi}{4}\right)\tr\left(\bar{b}\bar{a}-\bar{a}\bar{b}\right)b^{2}+\exp\left(-i\frac{\pi}{4}\right)\tr\bar{b}^{2}\left(ba-ab\right).
\end{eqnarray}
\begin{eqnarray*}
\left[\tr\bar{b}\bar{a}ba,\tr\bar{a}b\right] & = & \tr\bar{b}\bar{a}ba\tr\bar{a}b-\tr\bar{a}b\tr\bar{b}\bar{a}ba\\
 & = & \tr\bar{b}\bar{a}bb+\normord{\tr\bar{b}\bar{a}ba\tr\bar{a}b}-\tr\bar{a}\bar{a}ba+\normord{\tr\bar{a}b\tr\bar{b}\bar{a}ba}\\
 & = & \tr\bar{b}\bar{a}bb+\normord{\tr\bar{b}\bar{a}ba\tr\bar{a}b}-\tr\bar{a}\bar{a}ba-\normord{\tr\bar{b}\bar{a}ba\tr\bar{a}b}\\
 & = & \tr\bar{b}\bar{a}bb-\tr\bar{a}\bar{a}ba,
\end{eqnarray*}
\begin{eqnarray*}
\left[\tr\bar{b}\bar{a}ba,\tr\bar{b}a\right] & = & \tr\bar{b}\bar{a}ba\tr\bar{b}a-\tr\bar{b}a\tr\bar{b}\bar{a}ba\\
 & = & \tr\bar{b}\bar{a}a^{2}-\normord{\tr\bar{b}\bar{a}ba\tr\bar{b}a}+\tr\bar{b}\bar{b}ba-\normord{\tr\bar{b}a\tr\bar{b}\bar{a}ba}\\
 & = & \tr\bar{b}\bar{a}a^{2}-\normord{\tr\bar{b}\bar{a}ba\tr\bar{b}a}+\tr\bar{b}\bar{b}ba+\normord{\tr\bar{b}\bar{a}ba\tr\bar{b}a}\\
 & = & \tr\bar{b}\bar{a}a^{2}+\tr\bar{b}\bar{b}ba,
\end{eqnarray*}
 from which it follows that
\begin{eqnarray}
\left[\tr\bar{b}\bar{a}ba,Q\right] & = & \exp\left(i\frac{\pi}{4}\right)\left[\tr\bar{b}\bar{a}ba,\tr\bar{a}b\right]+\exp\left(-i\frac{\pi}{4}\right)\left[\tr\bar{b}\bar{a}ba,\tr\bar{b}a\right]\nonumber \\
 & = & \exp\left(i\frac{\pi}{4}\right)\left[\tr\bar{b}\bar{a}bb-\tr\bar{a}\bar{a}ba\right]\nonumber \\
 &  & +\exp\left(-i\frac{\pi}{4}\right)\left[\tr\bar{b}\bar{a}a^{2}+\tr\bar{b}\bar{b}ba\right].
\end{eqnarray}
\begin{eqnarray*}
\left[\tr\bar{a}\bar{b}ab,\tr\bar{a}b\right] & = & \tr\bar{a}\bar{b}ab\tr\bar{a}b-\tr\bar{a}b\tr\bar{a}\bar{b}ab\\
 & = & -\tr\bar{a}\bar{b}b^{2}+\normord{\tr\bar{a}\bar{b}ab\tr\bar{a}b}-\tr\bar{a}^{2}ab+\normord{\tr\bar{a}b\tr\bar{a}\bar{b}ab}\\
 & = & -\tr\bar{a}\bar{b}b^{2}-\tr\bar{a}^{2}ab,
\end{eqnarray*}
\begin{eqnarray*}
\left[\tr\bar{a}\bar{b}ab,\tr\bar{b}a\right] & = & \tr\bar{a}\bar{b}ab\tr\bar{b}a-\tr\bar{b}a\tr\bar{a}\bar{b}ab\\
 & = & \tr\bar{a}\bar{b}a^{2}-\normord{\tr\bar{a}\bar{b}ab\tr\bar{b}a}-\tr\bar{b}^{2}ab-\normord{\tr\bar{a}\bar{b}ab\tr\bar{b}a}\\
 & = & \tr\bar{a}\bar{b}a^{2}-\tr\bar{b}^{2}ab,
\end{eqnarray*}
 which follows
\begin{eqnarray}
\left[\tr\bar{a}\bar{b}ab,Q\right] & = & \exp\left(i\frac{\pi}{4}\right)\left[\tr\bar{a}\bar{b}ab,\tr\bar{b}a\right]+\exp\left(-i\frac{\pi}{4}\right)\left[\tr\bar{a}\bar{b}ab,\tr\bar{a}b\right]\nonumber \\
 & = & \exp\left(i\frac{\pi}{4}\right)\left[-\tr\bar{a}\bar{b}b^{2}-\tr\bar{a}^{2}ab\right]\nonumber \\
 &  & +\exp\left(-i\frac{\pi}{4}\right)\left[\tr\bar{a}\bar{b}a^{2}-\tr\bar{b}^{2}ab\right].
\end{eqnarray}

\begin{eqnarray*}
\left[\tr\bar{a}\bar{b}ba,\tr\bar{a}b\right] & = & \tr\bar{a}\bar{b}ba\tr\bar{a}b-\tr\bar{a}b\tr\bar{a}\bar{b}ba\\
 & = & \tr\bar{a}\bar{b}bb+\normord{\tr\bar{a}\bar{b}ba\tr\bar{a}b}-\tr\bar{a}^{2}ba+\normord{\tr\bar{a}b\tr\bar{a}\bar{b}ba}\\
 & = & \tr\bar{a}\bar{b}bb-\tr\bar{a}^{2}ba,
\end{eqnarray*}
\begin{eqnarray*}
\left[\tr\bar{a}\bar{b}ba,\tr\bar{b}a\right] & = & \tr\bar{a}\bar{b}ba\tr\bar{b}a-\tr\bar{b}a\tr\bar{a}\bar{b}ba\\
 & = & \tr\bar{a}\bar{b}aa-\normord{\tr\bar{a}\bar{b}ba\tr\bar{b}a}-\tr\bar{b}\bar{b}ba-\normord{\tr\bar{a}\bar{b}ba\tr\bar{b}a}\\
 & = & \tr\bar{a}\bar{b}aa-\tr\bar{b}^{2}ba,
\end{eqnarray*}
 from which it follows that
\begin{eqnarray}
\left[\tr\bar{a}\bar{b}ba,Q\right] & = & \exp\left(i\frac{\pi}{4}\right)\left[\tr\bar{a}\bar{b}ba,\tr\bar{a}b\right]+\exp\left(-i\frac{\pi}{4}\right)\left[\tr\bar{a}\bar{b}ba,\tr\bar{b}a\right]\nonumber \\
 & = & \exp\left(i\frac{\pi}{4}\right)\left[\tr\bar{a}\bar{b}bb-\tr\bar{a}^{2}ba\right]\nonumber \\
 &  & +\exp\left(-i\frac{\pi}{4}\right)\left[\tr\bar{a}\bar{b}aa-\tr\bar{b}^{2}ba\right].
\end{eqnarray}
 
\begin{eqnarray*}
\left[\tr\bar{b}\bar{a}ab,\tr\bar{a}b\right] & = & \tr\bar{b}\bar{a}ab\tr\bar{a}b-\tr\bar{a}b\tr\bar{b}\bar{a}ab\\
 & = & -\tr\bar{b}\bar{a}b^{2}-\tr\bar{a}^{2}ab,
\end{eqnarray*}
\begin{eqnarray*}
\left[\tr\bar{b}\bar{a}ab,\tr\bar{b}a\right] & = & \tr\bar{b}\bar{a}ab\tr\bar{b}a-\tr\bar{b}a\tr\bar{b}\bar{a}ab\\
 & = & \tr\bar{b}\bar{a}a^{2}+\tr\bar{b}^{2}ab,
\end{eqnarray*}
 from which it follows that
\begin{eqnarray}
\left[\tr\bar{b}\bar{a}ab,Q\right] & = & \exp\left(i\frac{\pi}{4}\right)\left[\tr\bar{b}\bar{a}ab,\tr\bar{a}b\right]+\exp\left(-i\frac{\pi}{4}\right)\left[\tr\bar{b}\bar{a}ab,\tr\bar{b}a\right]\nonumber \\
 & = & \exp\left(i\frac{\pi}{4}\right)\left[-\tr\bar{b}\bar{a}b^{2}-\tr\bar{a}^{2}ab\right]\nonumber \\
 &  & +\exp\left(-i\frac{\pi}{4}\right)\left[\tr\bar{b}\bar{a}a^{2}+\tr\bar{b}^{2}ab\right].
\end{eqnarray}

As mentioned in the main text, the general form of Hermitian Hamiltonian
is 
\begin{eqnarray*}
H & = & \frac{1}{N}\Big[c_{1}\tr\bar{a}^{2}a^{2}+c_{2}\tr\bar{b}^{2}b^{2}+iz_{1}\tr\bar{a}^{2}b^{2}-iz_{1}^{*}\tr\bar{b}^{2}a^{2}\\
 &  & +c_{3}\tr\bar{a}\bar{b}ba+c_{4}\tr\bar{b}\bar{a}ab+z_{2}\tr\bar{a}\bar{b}ab+z_{2}^{*}\tr\bar{b}\bar{a}ba\Big].
\end{eqnarray*}
 With the above calculation, we have 
\begin{eqnarray*}
N\exp\left(\frac{i\pi}{4}\right)\left[H,Q\right] & = & c_{1}\left[i\tr\bar{a}^{2}\left(ab+ba\right)-\tr\left(\bar{a}\bar{b}+\bar{b}\bar{a}\right)a^{2}\right]\\
 &  & +c_{2}\left[i\tr\left(\bar{b}\bar{a}-\bar{a}\bar{b}\right)b^{2}+\tr\bar{b}^{2}\left(ba-ab\right)\right]\\
 &  & +iz_{1}\left[\tr\bar{a}^{2}\left(ba-ab\right)-\tr\left(\bar{b}\bar{a}+\bar{a}\bar{b}\right)b^{2}\right]\\
 &  & +z_{1}^{*}\left[\tr\bar{b}^{2}\left(ab+ba\right)+\tr\left(\bar{b}\bar{a}-\bar{a}\bar{b}\right)a^{2}\right]\\
 &  & +c_{3}\left[i\left(\tr\bar{a}\bar{b}bb-\tr\bar{a}^{2}ba\right)+\tr\bar{a}\bar{b}aa-\tr\bar{b}^{2}ba\right]\\
 &  & +c_{4}\left[i\left(-\tr\bar{b}\bar{a}b^{2}-\tr\bar{a}^{2}ab\right)+\tr\bar{b}\bar{a}a^{2}+\tr\bar{b}^{2}ab\right]\\
 &  & +z_{2}\left[i\left(-\tr\bar{a}\bar{b}b^{2}-\tr\bar{a}^{2}ab\right)+\tr\bar{a}\bar{b}a^{2}-\tr\bar{b}^{2}ab\right]\\
 &  & +z_{2}^{*}\left[i\left(\tr\bar{b}\bar{a}bb-\tr\bar{a}^{2}ba\right)+\tr\bar{b}\bar{a}a^{2}+\tr\bar{b}^{2}ba\right]\\
 & = & i\left(c_{1}-z_{1}-c_{4}-z_{2}\right)\tr\bar{a}^{2}ab-\left(c_{1}-z_{1}^{*}-c_{4}-z_{2}^{*}\right)\tr\bar{b}\bar{a}a^{2}\\
 &  & +i\left(c_{1}+z_{1}-c_{3}-z_{2}^{*}\right)\tr\bar{a}^{2}ba-\left(c_{1}+z_{1}^{*}-c_{3}-z_{2}\right)\tr\bar{a}\bar{b}a^{2}\\
 &  & -\left(c_{2}-z_{1}^{*}-c_{4}+z_{2}\right)\tr\bar{b}^{2}ab+i\left(c_{2}-z_{1}-c_{4}+z_{2}^{*}\right)\tr\bar{a}\bar{b}b^{2}\\
 &  & +\left(c_{2}+z_{1}^{*}-c_{3}+z_{2}^{*}\right)\tr\bar{b}^{2}ba-i\left(c_{2}+z_{1}-c_{3}+z_{2}\right)\tr\bar{a}\bar{b}b^{2}.
\end{eqnarray*}
 Then, $\left[H,Q\right]=0$ yields 
\[
\begin{cases}
c_{1}-z_{1}-c_{4}-z_{2} & =0\\
c_{1}+z_{1}-c_{3}-z_{2}^{*} & =0\\
c_{2}-z_{1}^{*}-c_{4}+z_{2} & =0\\
c_{2}+z_{1}^{*}-c_{3}+z_{2}^{*} & =0
\end{cases},
\]
from which it follows (\ref{eq:Susy-constraint}).

\section{\label{app:HPrime-Equals-DeltaH}Proof of $\left(H^{\prime}-\Delta H\right)\ket{\text{any trace state}}=0$}

$\Delta H$ and $H^{\prime}$ are defined as
\begin{eqnarray*}
\Delta H & = & \frac{2}{N}\tr\left[\bar{a}\bar{b}ba+\bar{b}\bar{a}ab+\bar{a}^{2}a^{2}+\bar{b}^{2}b^{2}-\tilde{M}\right],\\
H^{\prime} & = & \frac{2}{N}\tr\left(\bar{a}a\bar{a}a+\bar{b}b\bar{a}a-\bar{a}b\bar{b}a\right),
\end{eqnarray*}
 where 
\[
\tilde{M}=\tr\left(\bar{a}a+\bar{b}b\right)-\frac{1}{N}\left(\tr\bar{a}\tr a+\tr\bar{b}\tr b\right).
\]
We first prove that 
\begin{equation}
N\left(H^{\prime}-\Delta H\right)=\tr G^{2},\label{eq:HPrime-DeltaH-G2-1}
\end{equation}
 where the color operator $G_{\alpha}^{\beta}$ is defined as 

\[
G_{\alpha}^{\beta}=\left(\bar{a}a-\normord{a\bar{a}}+\bar{b}b-\normord{b\bar{b}}\right)_{\alpha}^{\beta},
\]
 then it is sufficient to prove that 
\begin{equation}
G_{\alpha}^{\beta}\ket{\text{Any trace state}}=0.\label{eq:HPrime-DeltaH-G2-2}
\end{equation}
 Expanding $\tr G^{2}$ yields
\[
\tr G^{2}=\tr\left(\bar{a}a-\normord{a\bar{a}}\right)^{2}+\tr\left(\bar{b}b-\normord{b\bar{b}}\right)^{2}+2\tr\left(\bar{b}b-\normord{b\bar{b}}\right)\left(\bar{a}a-\normord{a\bar{a}}\right).
\]
Expanding each term of the right-hand side, we obtain
\begin{eqnarray*}
\tr\left(\bar{a}a-\normord{a\bar{a}}\right)^{2} & = & \tr\bar{a}a\bar{a}a+\tr\left(\normord{a\bar{a}}\normord{a\bar{a}}\right)-\tr\left(\bar{a}a\normord{a\bar{a}}+\normord{a\bar{a}}\bar{a}a\right)\\
 & = & 2\tr\left(\normord{\bar{a}a\bar{a}a}\right)+2N\tr\bar{a}a-\left(2\tr\bar{a}^{2}a^{2}+\tr\bar{a}\tr a\right),
\end{eqnarray*}
\begin{eqnarray*}
\tr\left(\bar{b}b-\normord{b\bar{b}}\right)^{2} & = & \tr\left(\bar{b}b\bar{b}b+\normord{b\bar{b}}\normord{b\bar{b}}\right)-\tr\left(\bar{b}b\normord{b\bar{b}}+\normord{b\bar{b}}\bar{b}b\right)\\
 & = & 2N\tr\bar{b}b-2\left(\tr\bar{b}^{2}b^{2}+\tr\bar{b}\tr b\right),
\end{eqnarray*}
\[
\tr\left(\bar{b}b-\normord{b\bar{b}}\right)\left(\bar{a}a-\normord{a\bar{a}}\right)=\tr\bar{b}b\bar{a}a+\tr\left(\normord{\bar{a}b\bar{b}a}\right)-\tr\left(\bar{a}\bar{b}ba+\bar{b}\bar{a}ab\right).
\]
 It follows that
\begin{eqnarray*}
\tr G^{2} & = & 2\tr\left(\normord{\bar{a}a\bar{a}a}+\bar{b}b\bar{a}a+\normord{\bar{a}b\bar{b}a}\right)\\
 &  & -2\tr\left(\bar{a}\bar{b}ba+\bar{b}\bar{a}ab+\bar{a}^{2}a^{2}+\bar{b}^{2}b^{2}\right)\\
 &  & +2N\tr\left(\bar{a}a+\bar{b}b\right)-2\tr\bar{a}\tr a-2\tr\bar{b}\tr b\\
 & = & 2\tr\left(\bar{a}a\bar{a}a+\bar{b}b\bar{a}a-\bar{a}b\bar{b}a\right)\\
 &  & -2\tr\left(\bar{a}\bar{b}ba+\bar{b}\bar{a}ab+\bar{a}^{2}a^{2}+\bar{b}^{2}b^{2}-\tilde{M}\right)\\
 & = & N\left(H^{\prime}-\Delta H\right).
\end{eqnarray*}

Now let us prove (\ref{eq:HPrime-DeltaH-G2-2}). It is easy to check
that

\begin{eqnarray*}
\left[\bar{a}_{\alpha}^{\beta},G_{\gamma}^{\delta}\right] & = & \bar{a}_{\gamma}^{\beta}\delta_{\alpha}^{\delta}-\delta_{\gamma}^{\beta}\bar{a}_{\alpha}^{\delta},\\
\left[\bar{b}_{\alpha}^{\beta},G_{\gamma}^{\delta}\right] & = & \bar{b}_{\gamma}^{\beta}\delta_{\alpha}^{\delta}-\delta_{\gamma}^{\beta}\bar{b}_{\alpha}^{\delta}.
\end{eqnarray*}
Let $X$ be an $M$-bit chain 
\[
X_{\alpha}^{\beta}=\left(\bar{x}_{1}\bar{x}_{2}\cdots\bar{x}_{M}\right)_{\alpha}^{\beta},\quad\bar{x}_{i}=\bar{a}\,\text{or}\,\bar{b},
\]
 then
\begin{eqnarray*}
\left[X_{\alpha}^{\beta},G_{\gamma}^{\delta}\right] & = & \sum_{i=1}^{M}\left(\bar{x}_{1}\cdots\bar{x}_{i-1}\right)_{\sigma}^{\beta}\left[\bar{x}_{\rho}^{\sigma},G_{\gamma}^{\delta}\right]\left(\bar{x}_{i+1}\cdots\bar{x}_{M}\right)_{\alpha}^{\rho}\\
 & = & \sum_{i=1}^{M}\left(\bar{x}_{1}\cdots\bar{x}_{i-1}\right)_{\sigma}^{\beta}\left(\left(\bar{x}_{i}\right)_{\gamma}^{\sigma}\delta_{\rho}^{\delta}-\delta_{\gamma}^{\sigma}\left(\bar{x}_{i}\right)_{\rho}^{\delta}\right)\left(\bar{x}_{i+1}\cdots\bar{x}_{M}\right)_{\alpha}^{\rho}\\
 & = & \sum_{i=1}^{M-1}\left(\bar{x}_{1}\cdots\bar{x}_{i}\right)_{\gamma}^{\beta}\left(\bar{x}_{i+1}\cdots\bar{x}_{M}\right)_{\alpha}^{\delta}+\left(\bar{x}_{1}\cdots\bar{x}_{M}\right)_{\gamma}^{\beta}\delta_{\alpha}^{\delta}\\
 &  & -\left(\sum_{i=2}^{M}\left(\bar{x}_{1}\cdots\bar{x}_{i-1}\right)_{\gamma}^{\beta}\left(\bar{x}_{i}\cdots\bar{x}_{M}\right)_{\alpha}^{\delta}+\delta_{\gamma}^{\beta}\left(\bar{x}_{1}\cdots\bar{x}_{M}\right)_{\alpha}^{\delta}\right)\\
 & = & \left(\bar{x}_{1}\cdots\bar{x}_{M}\right)_{\gamma}^{\beta}\delta_{\alpha}^{\delta}-\delta_{\gamma}^{\beta}\left(\bar{x}_{1}\cdots\bar{x}_{M}\right)_{\alpha}^{\delta}.
\end{eqnarray*}
 On the other hand, 
\[
\left[X_{\alpha}^{\beta},G_{\gamma}^{\delta}\right]\ket{0}=X_{\alpha}^{\beta}G_{\gamma}^{\delta}\ket{0}-G_{\gamma}^{\delta}X_{\alpha}^{\beta}\ket{0}=-G_{\gamma}^{\delta}X_{\alpha}^{\beta}\ket{0},
\]
 from which it follows that
\[
G_{\gamma}^{\delta}X_{\alpha}^{\beta}\ket{0}=\left(\delta_{\gamma}^{\beta}\left(\bar{x}_{1}\cdots\bar{x}_{M}\right)_{\alpha}^{\delta}-\left(\bar{x}_{1}\cdots\bar{x}_{M}\right)_{\gamma}^{\beta}\delta_{\alpha}^{\delta}\right)\ket{0}.
\]
 Taking the trace on the indices of $X$ yields 
\[
G_{\gamma}^{\delta}\tr X\ket{0}=0.
\]
Therefore, we proved (\ref{eq:HPrime-DeltaH-G2-2}).

\section{\label{app:Hamiltonian-Eigenvalue-Problem}Hamiltonian eigenvalue
problem}

This section proves several claims on the eigenvalue problems of $\mathcal{H}$,
\begin{equation}
\left(\mathcal{H}-E\right)V=0,\label{eq:C-mH-Equation}
\end{equation}
where $V$ is a vector and $\mathcal{H}$ is given by 
\begin{equation}
H\ket{i}=\sum_{j}\ket{j}\mathcal{H}_{ji}.\label{eq:C-H-on-trace}
\end{equation}
First, let us prove the following two claims:
\begin{itemize}
\item If $E$ is an eigenvalue of $\mathcal{H}$, its complex conjugate
$E^{*}$ is also an eigenvalue of $\mathcal{H}$. 
\item If $E$ is not real, it must have $V^{\dagger}GV=0$, where $G$ is
the norm matrix $G_{ij}=\dprod{i}{j}$.\end{itemize}
\begin{proof}
\textemdash{}Using (\ref{eq:C-H-on-trace}), we have 
\[
\bra{i}H\ket{j}=\sum_{k}\dprod{i}{k}\mathcal{H}_{kj}=\left(G\mathcal{H}\right)_{ij}.
\]
 Since $H$ is Hermitian, we also have 
\[
\bra{i}H\ket{j}=\sum_{k}\mathcal{H}_{ik}^{\dagger}\dprod{k}{j}=\left(\mathcal{H}^{\dagger}G\right)_{ij},
\]
 which implies 
\begin{equation}
G\mathcal{H}=\mathcal{H}^{\dagger}G.\label{eq:GH-HG}
\end{equation}
Left multiplying Eq. (\ref{eq:C-mH-Equation}) by $G$ and taking
the complex conjugate yields 
\begin{equation}
V^{\dagger}\left(\mathcal{H}^{\dagger}G-E^{*}G\right)=0.\label{eq:E-Complex-Eq1}
\end{equation}
 Using Eq. (\ref{eq:GH-HG}) and taking the transpose of Eq. (\ref{eq:E-Complex-Eq1}),
we obtain 
\[
\left(\mathcal{H}^{T}-E^{*}\right)GV^{*}=0.
\]
 Since $\mathcal{H}$ has the same eigenvalues as $\mathcal{H}^{T}$,
$E^{*}$ is an eigenvalue of $\mathcal{H}$.

Using (\ref{eq:C-mH-Equation}), we have 
\begin{eqnarray*}
EV^{\dagger}GV & = & V^{\dagger}G\left(EV\right)=V^{\dagger}G\mathcal{H}V,\\
E^{*}V^{\dagger}GV & = & \left(E^{*}V^{\dagger}\right)GV=V^{\dagger}\mathcal{H}^{\dagger}GV,
\end{eqnarray*}
 from which it follows that
\[
\left(E-E^{*}\right)V^{\dagger}GV=V^{\dagger}\left(G\mathcal{H}-\mathcal{H}^{\dagger}G\right)V=0.
\]
Therefore, if $E$ is not real, it must have $V^{\dagger}GV=0$. 
\end{proof}
The remaining claims are related to whether or not $G$ is positive
semidefinite. Let us discuss them case by case.

\subsection*{1. Positive-semidefinite $G$ matrix}

If $G$ is a positive-semidefinite matrix, all its eigenvalues are
non-negative. There exists a set of orthonormal bases spanning the
trace state space. Suppose there are $r$ trace states $\ket{1},\cdots,\ket{r}$,
$ $with dimension $p\leq r$. We can build orthonormal bases $\ketp{i}$
using a $p\times r$ matrix $S$, 
\begin{equation}
\brap{i}=\sum_{j}S_{ij}\bra{j},\quad1\leq i\leq p,\,1\leq j\leq r,\label{eq:ortho-basis}
\end{equation}
 where the basis and the matrix $S$ satisfy 
\begin{eqnarray*}
\dprodp{i}{j} & = & \sum_{k,l}S\dprod{k}{l}S=\left(SGS\right)_{ij}=\delta_{ij}.
\end{eqnarray*}
 In this basis, the $p\times p$ Hamiltonian matrix $\mat{H}$ is
given by
\begin{eqnarray}
\mathbf{H}_{ij} & \equiv & \brap{i}H\ketp{j}\nonumber \\
 & = & \sum_{k,l}S_{ik}\bra{k}H\ket{l}S_{lj}^{\dagger}\nonumber \\
 & = & \sum_{k,l,m}S_{ik}\dprod{k}{m}\mathcal{H}_{ml}S_{lj}^{\dagger}\nonumber \\
 & = & \left(SG\mathcal{H}S^{\dagger}\right)_{ij}.\label{eq:Ham-definition}
\end{eqnarray}
 The eigenvalues of the Hamiltonian are given by the equation
\begin{equation}
\left(\mathbf{H}-E\right)W=0,\label{eq:Ham-Eigenvalue-Eq}
\end{equation}
 where $W$ is a $p$-dimensional vector. We claim:
\begin{itemize}
\item Every eigenvalue of $\mathbf{H}$ is an eigenvalue of $\mathcal{H}$. 
\item An eigenvalue E of $\mathcal{H}$ with an eigenvector $V$ is also
an eigenvalue of $\mathbf{H}$ if and only if $V^{\dagger}GV>0$. \end{itemize}
\begin{proof}
\textemdash{}We extend the $p$ basis vectors $\ketp{i}$ to $r$
vectors $\ketp{i}^{\prime}$ so that 
\[
\dprodp{i}{j}^{\prime}=\begin{cases}
\delta_{ij} & ,\quad\text{if }i,j\leq p\\
0 & ,\quad\text{if }i>p\,\text{ or }j>p
\end{cases}.
\]
This can be done by extending the $p\times r$ matrix $S$ to an $r\times r$
invertible matrix $R$. The matrix $R$ can be constructed as follows.
We pick any invertible $r\times r$ matrix which contains $S$ as
the first $p$ rows. For the $\left(p+1\right)$ th row vector, $R_{p+1}$,
we calculate $R_{p+1}^{\dagger}GR_{i}$ for each $i\leq p$. If $R_{p+1}^{\dagger}GR_{i}\neq0$,
we replace $R_{p+1}$ with $R_{p+1}-\left(R_{p+1}^{\dagger}GR_{i}\right)R_{i}$.
In this way, $R_{p+1}$ will be orthogonal to all the first $p$ row
vectors, and since the dimension of the state space is $p$, $R_{p+1}^{\dagger}GR$
must be zero. Repeating this process for the rest rows, we obtain
the invertible square matrix $R$.

The new bases are 
\[
\brap{i}^{\prime}=R_{ij}\bra{j},\quad1\leq i,j\leq r,
\]
 which satisfy 
\begin{equation}
\dprodp{i}{j}^{\prime}=\left(RGR^{\dagger}\right)_{ij}=\left(\mat{I}_{p}\oplus\mat{O}_{r-p}\right)_{ij},\label{eq:SGS-2}
\end{equation}
 where $\mat{I}_{p}$ is the $p\times p$ identity matrix and $\mat{O}_{r-p}$
is the $\left(r-p\right)\times\left(r-p\right)$ zero matrix. In the
new basis, we define a matrix,
\begin{equation}
\mathbb{H}=RG\mathcal{H}R^{\dagger}=\mat{H}\oplus\mat{O}_{r-p}.\label{eq:ExHam}
\end{equation}
Clearly, if $E$ is an eigenvalue of $\mathbf{H}$ with eigenvector
$W$, it is also an eigenvalue of $\mathbb{H}$, 
\begin{equation}
\left(\mathbb{\mathbb{H}}-E\right)W^{\prime}=0,\label{eq:ExHam-eigenvalue-Eq}
\end{equation}
 with the eigenvector $W^{\prime}$ satisfying 
\begin{equation}
W_{i}^{\prime}=\begin{cases}
W_{i} & ,\quad\text{if }1\leq i\leq p\\
0 & ,\quad\text{if }p<i\leq r
\end{cases}.\label{eq:Exteigenvector1}
\end{equation}
 With relations (\ref{eq:ExHam}) and (\ref{eq:SGS-2}), the left-hand
side of Eq. (\ref{eq:ExHam-eigenvalue-Eq}) can be expressed as 
\begin{eqnarray}
\left(\mathbb{H}-E\right)W^{\prime} & = & RG\mathcal{H}R^{\dagger}W^{\prime}-E\left(\mathbf{I}_{p}\oplus\mathbf{O}_{r-p}\right)W^{\prime}\nonumber \\
 & = & RG\mathcal{H}R^{\dagger}W^{\prime}-ERGR^{\dagger}W^{\prime}\nonumber \\
 & = & R\left(\mathcal{H}^{\dagger}-E\right)GR^{\dagger}W^{\prime}\nonumber \\
 & = & R\left(\mathcal{H}^{\dagger}-E\right)R^{-1}\left(\mathbf{I}_{p}\oplus\mathbf{O}_{r-p}\right)W^{\prime}\nonumber \\
 & = & R\left(\mathcal{H}^{\dagger}-E\right)R^{-1}W^{\prime}\label{eq:ExtHam-LHS1}
\end{eqnarray}
 Since $R$ is invertible, we obtain 
\[
\left(\mathcal{H}^{\dagger}-E\right)R^{-1}W^{\prime}=0.
\]
 $R^{-1}W^{\prime}$ cannot be zero as $R^{-1}$ is invertible and
$W^{\prime}\neq0$. As $E$ is real, $E$ is an eigenvalue of $\mathcal{H}^{\dagger}$
and $\mathcal{H}$.

Conversely, if $E$ is an eigenvalue of $\mathcal{H}$ with eigenvector
$V$, we have 
\[
RG\left(\mathcal{H}-E\right)V=0.
\]
 The right-hand side can be expressed as
\begin{eqnarray*}
RG\left(\mathcal{H}-E\right)V & = & RG\left(\mathcal{H}R^{\dagger}-ER^{\dagger}\right)R^{\dagger-1}V\\
 & = & \left(\mathbb{H}-E\left(\mathbf{I}_{p}\oplus\mathbf{O}_{r-p}\right)\right)R^{\dagger-1}V\\
 & = & \left(\mathbb{H}-E\right)\left(\mathbf{I}_{p}\oplus\mathbf{O}_{r-p}\right)R^{\dagger-1}V,
\end{eqnarray*}
 from which it follows that 
\begin{equation}
\left(\mathbb{H}-E\right)\left(\mathbf{I}_{p}\oplus\mathbf{O}_{r-p}\right)R^{\dagger-1}V=0.\label{eq:ExtHam-Eigenvalue-Eq3}
\end{equation}
To let $E$ be an eigenvalue of $\mathbb{H}$, we need $W^{\prime}\equiv\left(\mathbf{I}_{p}\oplus\mathbf{O}_{r-p}\right)R^{\dagger-1}V$
to be a nonzero vector. By calculating the norm of $W^{\prime}$,
\begin{eqnarray*}
W^{\prime\dagger}W^{\prime} & = & V^{\dagger}R^{\dagger-1}\left(\mathbf{I}_{p}\oplus\mathbf{O}_{r-p}\right)R^{\dagger-1}V\\
 & = & V^{\dagger}GV,
\end{eqnarray*}
we find that $E$ is an eigenvalue of $\mathbb{H}$ if and only if
$V^{\dagger}GV>0$. Under this constraint, as $\mathbb{H}=\mat{H}\oplus\mat{O}_{r-p}$,
$E$ is also an eigenvalue of $\mat{H}$.
\end{proof}

\subsection*{2. Non-positive-semidefinite $G$}

If $G$ is not a positive-semidefinite matrix, at least one of its
eigenvalues is negative. There does not exist an orthonormal basis
in the trace state space. Suppose the $r\times r$ matrix $G$ has
$p$ positive eigenvalues, $q$ negative eigenvalues, and $s=r-p-q$
zero eigenvalues. We can properly choose a unitary matrix $R$ so
that the new basis $\ketp{i}^{\prime}$ satisfies 
\[
\dprodp{i}{j}^{\prime}=\left(RGR^{\dagger}\right)_{ij}=\left(\mat{I}_{p}\oplus-\mat{I}_{q}\oplus\mat{O}_{s}\right)_{ij},
\]
where $\ketp{1}^{\prime},\cdots,\ketp{p}^{\prime}$ are positive norm-square
states, $\ketp{p+1}^{\prime},\cdots,\ketp{p+q}^{\prime}$ are negative
norm-square states, and $\ketp{p+q+1}^{\prime},\cdots,\ketp{r}^{\prime}$
are zero norm states. The negative norm-square states are also called
ghost states. The existence of a ghost state implies the Hamiltonian
is not unitary. 

In analogy with (\ref{eq:Ham-definition}) and (\ref{eq:ExHam}),
we define $\mat{H}$ and $\mathbb{H}$ by 
\[
\mat{H}_{ij}=\brap{i}H\ketp{j},\quad1\leq i,j\leq p+q
\]
 and
\[
\mathbb{H}=RG\mathcal{H}R^{\dagger}=\mat{H}\oplus\mat{O}_{s}.
\]
 We claim:
\begin{itemize}
\item If $E$ is an eigenvalue of $\mat{H}$ with eigenvector $W$, it is
an eigenvalue of $\mathcal{H}$ when $W$ does not couple with any
ghost state.
\item If $E$ is an eigenvalue of $\mathcal{H}$ with eigenvector $V$,
it is an eigenvalue of $\mat{H}$ when $E=0$ or $V^{\dagger}\mathrm{abs}\left(G\right)V=V^{\dagger}GV>0$,
where the function $\mathrm{abs}$ is defined as
\[
\mathrm{abs}\left(G\right)=U^{\dagger}\left(\begin{array}{cccc}
\left|g_{1}\right|\\
 & \left|g_{2}\right|\\
 &  & \ddots\\
 &  &  & \left|g_{n}\right|
\end{array}\right)U,
\]
with 
\[
G=U^{\dagger}\left(\begin{array}{cccc}
g_{1}\\
 & g_{2}\\
 &  & \ddots\\
 &  &  & g_{n}
\end{array}\right)U
\]
 being the eigendecomposition of $G$.
\end{itemize}
The condition $V^{\dagger}\mathrm{abs}\left(G\right)V=V^{\dagger}GV$
implies that, in the basis where $G$ is diagonal, $V$ does not couple
with any ghost state. According to our numerical calculation, $G$
is not positive semidefinite only when $N<M$ and is not integer.
The numerical calculation shows that, except the $E=0$ case, the
condition $V^{\dagger}\mathrm{abs}\left(G\right)V=V^{\dagger}GV$
is usually not satisfied when $G$ is not positive semidefinite. The
proof of the claims is given as follows.
\begin{proof}
\textemdash{}If $E$ is an eigenvalue of $\mat{H}$ with eigenvector
$W$, $E$ is also an eigenvalue of $\mathbb{H}$ with the eigenvector
defined as 
\begin{equation}
W_{i}^{\prime}=\begin{cases}
W_{i} & ,\quad\text{if }1\leq i\leq p+q\\
0 & ,\quad\text{if }p+q<i\leq r
\end{cases}.\label{eq:Exteigenvector2}
\end{equation}
 In analogy with (\ref{eq:ExtHam-LHS1}), we have
\begin{eqnarray*}
\left(\mathbb{H}-E\right)W^{\prime} & = & RG\mathcal{H}R^{\dagger}W^{\prime}-E\left(\mathbf{I}_{p+q}\oplus\mathbf{O}_{s}\right)W^{\prime}\\
 & = & RG\mathcal{H}R^{\dagger}W^{\prime}-E\left(RGR^{\dagger}+2\mathbf{O}_{p}\oplus\mat{I}_{q}\oplus\mathbf{O}_{s}\right)W^{\prime}\\
 & = & R\left(\mathcal{H}-E\right)GR^{\dagger}W^{\prime}-2E\left(\mathbf{O}_{p}\oplus\mat{I}_{q}\oplus\mathbf{O}_{s}\right)W^{\prime}\\
 & = & R\left(\mathcal{H}-E\right)R^{-1}\left(\mat{I}_{p}\oplus-\mat{I}_{q}\oplus\mat{O}_{s}\right)W^{\prime}-2E\left(\mathbf{O}_{p}\oplus\mat{I}_{q}\oplus\mathbf{O}_{s}\right)W^{\prime}.
\end{eqnarray*}
 If the following conditions are satisfied, $E$ is an eigenvalue
of $\mathcal{H}$: 
\[
\begin{cases}
\left(\mathbf{O}_{p}\oplus\mat{I}_{q}\oplus\mathbf{O}_{s}\right)W^{\prime} & =0\\
\left(\mat{I}_{p}\oplus-\mat{I}_{q}\oplus\mat{O}_{s}\right)W^{\prime} & \neq0
\end{cases}.
\]
With (\ref{eq:Exteigenvector2}) and $W\neq0$, it implies that, if
\begin{equation}
\left(\mathbf{O}_{p}\oplus\mat{I}_{q}\right)W=0,\label{eq:Noghost-condition}
\end{equation}
$E$ is an eigenvalue of $\mathcal{H}$. Equation (\ref{eq:Noghost-condition})
is a constraint under which the eigenvector does not couple with the
ghost states.

Conversely, if $E$ is an eigenvalue of $\mathcal{H}$ with eigenvector
$V$, 
\[
RG\left(\mathcal{H}-E\right)V=0.
\]
 The left-hand side of the equation can be expressed as
\begin{eqnarray*}
RG\left(\mathcal{H}-E\right)V & = & RG\mathcal{H}R^{\dagger}R^{\dagger-1}V-ERGR^{\dagger}R^{\dagger-1}V\\
 & = & \mathbb{H}R^{\dagger-1}V-E\left(\mat{I}_{p}\oplus-\mat{I}_{q}\oplus\mat{O}_{s}\right)R^{\dagger-1}V\\
 & = & \left(\mathbb{H}-E\right)\left(\mat{I}_{p+q}\oplus\mat{O}_{s}\right)R^{\dagger-1}V+2E\left(\mat{O}_{p}\oplus\mat{I}_{q}\oplus\mat{O}_{s}\right)R^{\dagger-1}V\\
 & = & \left(\mathbb{H}-E\right)W^{\prime}+2EW^{\prime\prime},
\end{eqnarray*}
 where we have defined 
\begin{eqnarray*}
W^{\prime} & \equiv & \left(\mat{I}_{p+q}\oplus\mat{O}_{s}\right)R^{\dagger-1}V,\\
W^{\prime\prime} & \equiv & \left(\mat{O}_{p}\oplus\mat{I}_{q}\oplus\mat{O}_{s}\right)R^{\dagger-1}V.
\end{eqnarray*}
If $E=0$, or $W^{\prime}\neq0$ and $W^{\prime\prime}=0$, $E$ is
an eigenvalue of $\mathbb{H}$. $W^{\prime}\neq0$ implies 
\begin{eqnarray}
W^{\prime\dagger}W^{\prime} & = & V^{\dagger}R^{-1}\left(\mat{I}_{p+q}\oplus\mat{O}_{s}\right)R^{\dagger-1}V\nonumber \\
 & = & V^{\dagger}R^{-1}\left(\mat{I}_{p}\oplus-\mat{I}_{q}\oplus\mat{O}_{s}\right)R^{\dagger-1}V\nonumber \\
 & = & V^{\dagger}R^{-1}RGR^{\dagger}R^{\dagger-1}V\nonumber \\
 & = & V^{\dagger}GV>0,\label{eq:W-prime-norm}
\end{eqnarray}
 where we use the constraint $W^{\prime\prime}=0$ in the second equality.
With the equation 
\[
2\left(\mat{O}_{p}\oplus\mat{I}_{q}\oplus\mat{O}_{s}\right)=\left(RGR^{\dagger}\right)^{2}-RGR^{\dagger},
\]
 $W^{\prime\prime}=0$ is equivalent to 
\begin{equation}
V^{\dagger}\left(GR^{\dagger}RG-G\right)V=0.\label{eq:V-norm-constraint3}
\end{equation}
 Combining constraints (\ref{eq:W-prime-norm}) and (\ref{eq:V-norm-constraint3}),
we findthat, if 
\[
V^{\dagger}GR^{\dagger}RGV=V^{\dagger}GV>0,
\]
 $E$ is an eigenvalue of $\mat{H}$. 

The matrix $GR^{\dagger}RG$ seems to be dependent on $R$, but actually
it only depends on $G$. Indeed, any unitary transformation $R\to UR$
does not change $GR^{\dagger}RG$. In general, if the eigendecomposition
of $G$ is 
\[
G=U^{\dagger}\left(\begin{array}{cccc}
g_{1}\\
 & g_{2}\\
 &  & \ddots\\
 &  &  & g_{n}
\end{array}\right)U,\quad U^{\dagger}U=\mat{I},
\]
 we can choose $R$ as 
\[
R_{i}=\begin{cases}
\frac{1}{\sqrt{\left|g_{i}\right|}}U_{i}, & \quad\text{if }g_{i}\neq0\\
U_{i}, & \quad\text{if }g_{i}=0
\end{cases}.
\]
 Then, we obtain 
\[
GR^{\dagger}RG=U^{\dagger}\left(\begin{array}{cccc}
\left|g_{1}\right|\\
 & \left|g_{2}\right|\\
 &  & \ddots\\
 &  &  & \left|g_{n}\right|
\end{array}\right)U,
\]
 which clearly only depends on $G$. 
\end{proof}

\section{Algorithms}

The numerical computation is performed by C++ and the matlab program.
We use the C++ program to generate the norm matrices and $\mathcal{H}$
matrices and then use matlab to find eigenvalues and eigenstates.
Here we introduce the algorithms for generating trace states, calculating
norm matrices, and building $\mathcal{H}$ matrices.

\paragraph*{1. Generate trace states}

Trace states are represented by integer numbers. The bosonic and fermionic
creation operators are mapped to $0$ and 1, respectively. Then, an
$M$-bit single trace state is mapped as an $M$-bit binary number,
and a multiple trace state is an array of integers. Because of the
cyclic symmetry, a single trace state corresponds to several integers.
Among these integers we choose the smallest integer. For example,
$\tr\bar{a}\bar{b}\bar{b}\ket{0}$ is mapped to $\left(011\right)_{2}=3$
rather than $\left(110\right)_{2}=6$. We then go through all integers
between $0$ and $2^{M}-1$. A number is a single trace state only
when it meets two conditions:
\begin{itemize}
\item There is no cyclic rotation on this integer producing a smaller integer.
\item The corresponding trace state is nonvanishing. A trace state is vanishing
if it can be partitioned into an even number of identical consecutive
parts, each of which has an odd number of $\bar{b}$. For example,
$\tr\bar{b}\bar{b}\bar{b}\bar{b}\ket{0}$ vanishes as it can be partitioned
into four $\bar{b}$s. 
\end{itemize}
After generating all single trace states, we can build multiple trace
states out of single trace states. The procedure is similar to the
recursive relation (\ref{eq:Multiple-trace-states-2}) for calculating
the number of trace states.

\paragraph*{2. Calculate norm matrices}

To build a norm matrix, we need to calculate $\dprod{i}{j}$ for each
pair of states $i,\, j$. The norm can be calculated as follows. If
two $M$-bit states $i,j$ do not have the same number of $\bar{b}$,
then $\dprod{i}{j}=0$. Otherwise, if both have $n$ fermionic operators,
there are $n!\left(M-n\right)!$ ways to contract their color indices.
Take $\tr\bar{a}\bar{a}\bar{b}\bar{b}\ket{0}$ and $\tr\bar{a}\tr\bar{a}\bar{b}\bar{b}\ket{0}$
as an example. We first write the states as
\begin{eqnarray*}
\tr\bar{a}\bar{a}\bar{b}\bar{b}\ket{0} & = & \bar{a}_{\alpha}^{\beta}\bar{a}_{\beta}^{\gamma}\bar{b}_{\gamma}^{\rho}\bar{b}_{\rho}^{\alpha}\ket{0},\\
\tr\bar{a}\tr\bar{a}\bar{b}\bar{b}\ket{0} & = & \bar{a}_{i}^{i}\bar{a}_{j}^{k}\bar{b}_{k}^{l}\bar{b}_{l}^{j}\ket{0}.
\end{eqnarray*}
 Using the commutation and anticommutation relations, we can expand
the norm into $2!\times2!=4$ terms, 
\begin{eqnarray*}
\bra{0}\tr bbaa\tr\bar{a}\tr\bar{a}\bar{b}\bar{b}\ket{0} & = & \bra{0}b_{\alpha}^{\rho}b_{\rho}^{\gamma}a_{\gamma}^{\beta}a_{\beta}^{\alpha}\bar{a}_{i}^{i}\bar{a}_{j}^{k}\bar{b}_{k}^{l}\bar{b}_{l}^{j}\ket{0}\\
 & = & \delta_{j}^{\beta}\delta_{\gamma}^{k}\delta_{i}^{\alpha}\delta_{\beta}^{i}\left(\delta_{l}^{\rho}\delta_{\alpha}^{j}\delta_{k}^{\gamma}\delta_{\rho}^{l}-\delta_{k}^{\rho}\delta_{\alpha}^{l}\delta_{l}^{\gamma}\delta_{\rho}^{j}\right)\\
 &  & +\delta_{i}^{\beta}\delta_{\gamma}^{i}\delta_{j}^{\alpha}\delta_{\beta}^{k}\left(\delta_{l}^{\rho}\delta_{\alpha}^{j}\delta_{k}^{\gamma}\delta_{\rho}^{l}-\delta_{k}^{\rho}\delta_{\alpha}^{l}\delta_{l}^{\gamma}\delta_{\rho}^{j}\right).
\end{eqnarray*}
 The sign of each term is determined by how many times a swap occurs
among $b$ and $\bar{b}$: and odd (even) number of swaps produces
a negative (positive) sign. The first term can be written as
\[
\delta_{j}^{\beta}\delta_{\gamma}^{k}\delta_{i}^{\alpha}\delta_{\beta}^{i}\delta_{l}^{\rho}\delta_{\alpha}^{j}\delta_{k}^{\gamma}\delta_{\rho}^{l}=\left(\delta_{j}^{\beta}\delta_{\beta}^{i}\delta_{i}^{\alpha}\delta_{\alpha}^{j}\right)\left(\delta_{\gamma}^{k}\delta_{k}^{\gamma}\right)\left(\delta_{l}^{\rho}\delta_{\rho}^{l}\right),
\]
 where Kronecker delta functions are put into three groups. The contraction
of the indices in each group produces a factor of $N$, which implies
the first term is equal to $N^{3}$. Repeating the procedure, we obtain
\[
\bra{0}\tr bbaa\tr\bar{a}\tr\bar{a}\bar{b}\bar{b}\ket{0}=2N^{3}-2N.
\]
 Finally, the result is normalized by multiplying $1/N^{4}$, which
yields $2/N-2/N^{3}$. 

Our algorithm simply simulates the procedure and hence has $\mathcal{O}\left(M!\right)$
time complexity to calculate each entry of a norm matrix. For numerical
computation of higher $M$, we need to improve time complexity significantly.

\paragraph*{3. Build $\mathcal{H}$ matrices}

To build $\mathcal{H}$ matrices, we need to calculate the action
of trace operators on trace states. Let us take an example that the
trace operator is $\tr Aab$, where $A$ is any creation operator
chain. To calculate $\tr Aab\tr S\ket{0}$, we need to find all possible
ways to partition $S$ into the form $B\bar{a}C\bar{b}D$ or $B\bar{b}C\bar{a}D$,
where $B$, $C$, $D$ are any creation operator chains. Each partition
corresponds to one way to contract the indices among annihilation
and creation operators. The results of these two contraction schemes
are 
\begin{eqnarray}
\tr Aab\tr B\bar{b}C\bar{a}D\ket{0} & \to & \left(-1\right)^{\pi\left(AB\bar{b}CD\to A\bar{b}DBC\right)}\tr ADB\tr C\ket{0},\label{eq:Alg-1}\\
\tr Aab\tr B\bar{a}C\bar{b}D\ket{0} & \to & \left(-1\right)^{\pi\left(ABC\bar{b}D\to A\bar{b}CDB\right)}\tr AC\tr DB\ket{0},\label{eq:Alg-2}
\end{eqnarray}
 where $\pi\left(AB\bar{b}CD\to A\bar{b}DBC\right)$ denotes the number
of swaps occurring among the fermionic operators as the chain being
reordered from$AB\bar{b}CD$ to $A\bar{b}DBC$. Let $f\left(A\right)$
denote the number of $\bar{b}$ in $A$; then, 
\[
\pi\left(AB\bar{b}CD\to A\bar{b}DBC\right)=f\left(B\right)+f\left(D\right)f\left(BC\right).
\]
 The complete result of $ $$\tr Aab\tr S\ket{0}$ can be written
as 
\begin{eqnarray*}
\tr Aab\tr S\ket{0} & = & \sum_{B\bar{b}C\bar{a}D=S}\left(-1\right)^{\pi\left(AB\bar{b}CD\to A\bar{b}DBC\right)}\tr ADB\tr C\ket{0}\\
 &  & +\sum_{B\bar{a}C\bar{b}D=S}\left(-1\right)^{\pi\left(ABC\bar{b}D\to A\bar{b}CDB\right)}\tr AC\tr DB\ket{0}.
\end{eqnarray*}
In analogy with (\ref{eq:Alg-1}) and (\ref{eq:Alg-2}), for two trace
states, we have 
\begin{eqnarray*}
\tr Aab\tr B\bar{a}C\tr D\bar{b}E\ket{0} & \to & \left(-1\right)^{\pi\left(ABCD\bar{b}E\to A\bar{b}CBED\right)}\tr ACBED\ket{0},\\
\tr Aab\tr B\bar{b}C\tr D\bar{a}E\ket{0} & \to & \left(-1\right)^{\pi\left(AB\bar{b}CDE\to A\bar{b}EDCB\right)}\tr AEDCB\ket{0}.
\end{eqnarray*}

The algorithm takes $\mathcal{O}\left(M^{2}\right)$ to calculate
one row of the $\mathcal{H}$ matrix. Since there are about $2^{M}$
trace states, it takes $\mathcal{O}\left(M^{2}2^{M}\right)$ to build
an $\mathcal{H}$ matrix, which is much faster than building a norm
matrix.

\bibliographystyle{plain}
\nocite{*}
\bibliography{ref}

\begin{thebibliography}{10}

\bibitem{Thorn:1991fv}
C.~B. Thorn.
\newblock {Reformulating string theory with the 1/N expansion}.
\newblock In {\em {The First International A.D. Sakharov Conference on Physics
  Moscow, USSR, May 27-31, 1991}}, 1991.

\bibitem{'tHooft:1990eb}
G.~'t~Hooft.
\newblock {Quantization of Discrete Deterministic Theories by Hilbert Space
  Extension}.
\newblock {\em Nucl. Phys.}, B342:471--485, 1990.

\bibitem{'tHooft:1987}
G.~'t~Hooft.
\newblock {On the Quantization of Space and Time}.
\newblock In V.A.~Berezin Eds. M.A.~Markov and V.P. Frolov, editors, {\em
  {{Proc. of the 4th Seminar on Quantum Gravity}, May 25-29, 1987, Moscow,
  USSR.}}, pages 551--567. World Scientific Press, 1988.

\bibitem{'tHooft:1993gx}
G.~'t~Hooft.
\newblock {Dimensional reduction in quantum gravity}.
\newblock In {\em {Salamfest 1993:0284-296}}, pages 0284--296, 1993.

\bibitem{Mandelstam:1973jk}
S.~Mandelstam.
\newblock {Interacting String Picture of Dual Resonance Models}.
\newblock {\em Nucl. Phys.}, B64:205--235, 1973.

\bibitem{Mandelstam:1974hk}
S.~Mandelstam.
\newblock {Interacting String Picture of the Neveu-Schwarz-Ramond Model}.
\newblock {\em Nucl. Phys.}, B69:77--106, 1974.

\bibitem{Goddard:1972ky}
P.~Goddard, C.~Rebbi, and C.~B. Thorn.
\newblock {Lorentz covariance and the physical states in dual resonance
  models}.
\newblock {\em Nuovo Cim.}, A12:425--441, 1972.

\bibitem{Goddard:1973qh}
P.~Goddard, J.~Goldstone, C.~Rebbi, and C.~B. Thorn.
\newblock {Quantum dynamics of a massless relativistic string}.
\newblock {\em Nucl. Phys.}, B56:109--135, 1973.

\bibitem{'tHooft:1973jz}
G.~'t~Hooft.
\newblock {A Planar Diagram Theory for Strong Interactions}.
\newblock {\em Nucl. Phys.}, B72:461, 1974.

\bibitem{Thorn:1979gu}
C.~B. Thorn.
\newblock {A Fock Space Description of the 1/$N_c$ Expansion of Quantum
  Chromodynamics}.
\newblock {\em Phys. Rev.}, D20:1435, 1979.

\bibitem{Giles:1977mpa}
R.~Giles and C.~B. Thorn.
\newblock {A Lattice Approach to String Theory}.
\newblock {\em Phys. Rev.}, D16:366, 1977.

\bibitem{Gliozzi:1976qd}
F.~Gliozzi, J.~Scherk, and D.~I. Olive.
\newblock {Supersymmetry, Supergravity Theories and the Dual Spinor Model}.
\newblock {\em Nucl. Phys.}, B122:253--290, 1977.

\bibitem{Ramond:1971gb}
P.~Ramond.
\newblock {Dual Theory for Free Fermions}.
\newblock {\em Phys. Rev.}, D3:2415--2418, 1971.

\bibitem{Neveu:1971rx}
A.~Neveu and J.~H. Schwarz.
\newblock {Factorizable dual model of pions}.
\newblock {\em Nucl. Phys.}, B31:86--112, 1971.

\bibitem{Neveu:1971iw}
A.~Neveu, J.~H. Schwarz, and C.~B. Thorn.
\newblock {Reformulation of the Dual Pion Model}.
\newblock {\em Phys. Lett.}, B35:529--533, 1971.

\bibitem{Thorn:1971jc}
C.~B. Thorn.
\newblock {Embryonic Dual Model for Pions and Fermions}.
\newblock {\em Phys. Rev.}, D4:1112--1116, 1971.

\bibitem{Neveu:1971iv}
A.~Neveu and J.~H. Schwarz.
\newblock {Quark Model of Dual Pions}.
\newblock {\em Phys. Rev.}, D4:1109--1111, 1971.

\bibitem{Green:1980zg}
M.~B. Green and J.~H. Schwarz.
\newblock {Supersymmetrical Dual String Theory}.
\newblock {\em Nucl. Phys.}, B181:502--530, 1981.

\bibitem{Bardakci:1970nb}
K.~Bardakci and M.~B. Halpern.
\newblock {New dual quark models}.
\newblock {\em Phys. Rev.}, D3:2493, 1971.

\bibitem{Sun:2014dga}
S.~Sun and C.~B. Thorn.
\newblock {Stable String Bit Models}.
\newblock {\em Phys. Rev.}, D89(10):105002, 2014.

\bibitem{Green:1983hw}
M.~B. Green, J.~H. Schwarz, and L.~Brink.
\newblock {Superfield Theory of Type II Superstrings}.
\newblock {\em Nucl. Phys.}, B219:437--478, 1983.

\bibitem{Chen:2015GitHub}
G.~Chen.
\newblock {String bit project} source code.
\newblock \url{https://github.com/gaolichen/stringbit}.
\newblock Accessed: 2016-01-23.

\bibitem{Thorn:2015wli}
Charles~B. Thorn.
\newblock {1/N Perturbations in Superstring Bit Models}.
\newblock {\em Phys. Rev.}, D93(6):066003, 2016.

\bibitem{rotman1999introduction}
J.~Rotman.
\newblock {\em An Introduction to the Theory of Groups}.
\newblock Graduate Texts in Mathematics. Springer New York, 1999.

\bibitem{Thorn:1996fa}
C.~B. Thorn.
\newblock {Substructure of string}.
\newblock In {\em {Strings 96: Current Trends in String Theory Santa Barbara,
  California, July 15-20, 1996}}, 1996.

\bibitem{Bergman:1995wh}
O.~Bergman and C.~B. Thorn.
\newblock {String bit models for superstring}.
\newblock {\em Phys. Rev.}, D52:5980--5996, 1995.

\end{thebibliography}

\end{document}